\documentclass[useAMS,usenatbib]{mn2e}
\usepackage{graphicx}
\usepackage{subfigure}
\usepackage{ amssymb }
\usepackage{ gensymb }
\usepackage{rotating}
\usepackage{pdflscape}
\usepackage[total={17.8cm,24.0cm},centering]{geometry}
\usepackage{times}

\title[LAMP: Variability of accretion indicators]{LAMP: The Long-term Accretion Monitoring Program of T Tauri stars in Chamaeleon-I}
\author[Costigan, Scholz, Stelzer, Ray, Vink, Mohanty ]{G. Costigan$^{1,2,3}$\thanks{E-mail:
gcostigan@cp.dias.ie}, A. Scholz$^{1}$, B. Stelzer$^{4}$, T. Ray$^{1}$, J. S. Vink$^{2}$, S. Mohanty$^{5}$ \\
$^{1}$ School of Cosmic Physics, Dublin Institute for Advanced Studies, 31 Fitzwilliam Place, Dublin 2, Ireland \\
$^{2}$ Armagh Observatory, College Hill, Armagh, BT61 9DG, Northern Ireland \\
$^{3}$ European Southern Observatory, Karl-Schwarzschild-Str. 2, 85748 Garching bei M\"unchen, Germany \\
$^{4}$ INAF - Osservatorio Astronomico di Palermo, Piazza del Parlamento 1, 90134 Palermo, Italy \\
$^{5}$ Department of Physics, Imperial College London, London SW7 2AZ, United Kingdom }

\begin{document}

\date{ }

\pagerange{\pageref{firstpage}--\pageref{lastpage}} \pubyear{2012}

\maketitle

\label{firstpage}

\begin{abstract}
We present the results of a variability study of accreting young stellar objects in the Chameleon I star-forming region which is based on $\sim$~300 high resolution optical spectra from the multi-object fibre spectrograph FLAMES/GIRAFFE at the ESO/VLT. Twenty five objects with spectral types from G2\,-\,M5.75 were observed 12 times over the course of 15 months. Using the emission lines H$\alpha$~(6562.81\,\AA) and Ca\,II (8662.1\,\AA) as accretion indicators we found 10 accreting and 15 non-accreting objects.

We derived accretion rates for all accretors in the sample using the H$\alpha$ equivalent width, H$\alpha$ 10\% width and the Ca\,II (8662.1\,\AA) equivalent width. We found the H$\alpha$ equivalent width of accretors varied by $\sim$ 7\,-\,100 \AA~over the 15 months. This corresponds to a mean amplitude of variations in derived accretion rate of $\thicksim$~0.37\,dex. The amplitude of variations in derived accretion rate from Ca\,II equivalent width were $\thicksim$~0.83 dex and from H$\alpha$ 10\% width were $\thicksim$~1.11 dex. Based on the large amplitude of variations in accretion rates derived from the H$\alpha$ 10\% width with respect to the other diagnostics, we do not consider it to be a reliable accretion rate estimator.
Taking the variations in H$\alpha$ EW and Ca\,II EW accretion rates to be closer to the true value, they suggest that the spread which has been found around the accretion rate to stellar mass relation is not due to the variability of individual objects on time-scales of weeks to $\thicksim$~1 year. From these variations we can also infer that the accretion rates are stable within \textless~0.37\,dex over time-scales of less than 15 months.

A major portion of the accretion variability was found to occur on less than the shortest time-scales in our observations, 8\,-\,25 days, which is comparable with the rotation periods of these young stellar objects. This could be an indication that what we are probing is spatial structure in the accretion flows, and also suggests that observations on time-scales of $\thicksim$~a couple of weeks are sufficient to limit the total extent of accretion rate variations in typical young stars. 

No episodic accretion was observed, all 10 accretors continuously accreted for the entire period of observations, and though they may have undetected, low accretion rates, the non-accretors never showed any large changes in their emission that would imply a jump in their accretion rate.

\end{abstract}


\section{Introduction}
\label{s1}

Accretion is a vital feature in the process of star formation as it controls the flow of matter and angular momentum from the circumstellar environment on to the young stellar object (YSO). For low-mass YSOs, the current paradigm for this process is magnetospheric accretion \citep{1990RvMA....3..234C,1991ApJ...370L..39K}: In short, the strong magnetic field of the young star truncates the inner part of the disc and channels the accretion flow from the inner edge of the disc to the star \citep[see review by][]{2007prpl.conf..479B}.

\indent Accretion causes a continuous excess flux from the ultra-violet (UV) to the near-infrared as well as a spectrum of emission lines, including the H Balmer series, the Ca\,II infrared triplet (Ca IRT), and He\,I lines. Given the relevance of accretion, it is somewhat worrisome that our understanding of the observational indicators for accretion is poor, in particular their relation to the mass accretion rate $\dot{M}$. The analysis of the lines and continuum relies on simplified assumptions about the geometry and structure of the accretion shock \citep[e.g.][]{2008ApJ...681..594H}. 
Moreover, many emission lines have additional components due to strong winds/outflows, either from the star or from the disc \citep{2007prpl.conf..231R}.

\indent One promising way to improve our understanding is to study accretion spectroscopically in the time-domain. The majority of the spectroscopic monitoring studies of YSOs have focused on a relatively small number of targets, specifically selected for their known variability. Some of these studies have led to a detailed close-up view on the disc-accretion system \citep[e.g,][] {1993ApJS...89..321G,1999A&A...349..619B} and provided unique insights into the magnetospheric accretion/ejection scenario \citep[e.g.][]{1995AJ....109.2800J,2007A&A...463.1017B}. In other cases the changes in the variability patterns are highly complex and difficult to interpret \citep[e.g.][]{2005A&A...440..595A}.

Various models have been put forward to explain accretion variability. Misalignment between rotation and stellar magnetic field axes will result in a non-uniform accretion flow. Different parts of the flow will then come into view as the star rotates, changing the emission signature we observe on the time-scale of the rotation period \citep{2008MNRAS.385.1931K}. Non-axisymmetric fields will also introduce asymmetry into the system \citep{2008ApJ...687.1323M}. Additionally inhomogeneities in the disc will cause the observed accretion rate to vary, as is the case with an accreting T Tauri star found by \citet{1999A&A...349..619B} with roughly constant brightness, broken by a quasi-periodic fading. This was attributed to a wall in the inner circumstellar disc which obscures the light from the central source periodically as it rotates. In the same way, such a wall could be a source of apparent accretion variability on a much longer time-scale if placed further out in the circumstellar disc. \cite{2010ApJ...719.1896V} showed that for young embedded objects, gravitational instabilities in the disc and disc fragmentation will lead to large scale accretion variations, such as FUor and EXor events where the accretion rates can vary between $10^{-4}$M$_{\sun}$yr$^{-1}$ and $10^{-8}$M$_{\sun}$yr$^{-1}$ over very long periods of thousands of years. Another source of variability is differential rotation between the disc and the star. This can lead to expanding field lines, which could radically reduce the mass accretion rate as the magnetic field lines open, and will then cause it to increase again upon reconnection \citep{1998AIPC..431..533G}. These magnetic reconnection led variations are expected to occur on the time-scales on the order of the rotation period of the circumstellar disc \citep{1996ApJ...468L..37H}. 

Apart from a few benchmark objects, observations have yet to distinguish between these scenarios. Each of these cases will show different amplitudes and time-scales of variations. In particular variability studies can provide information on the spatial structure and dynamics of the accretion flow \citep{2003A&A...409..169B}. They can also help to understand the origin of the scatter in $\dot{M}$ at a given stellar mass $M_{*}$, and refine the $\dot{M} \sim M_{*}^{2}$ correlation \citep{2003ApJ...592..266M}. The scatter within this relation can be up to two orders of magnitude for a given mass, and it is not clear if this scatter comes from time-varying accretion rates of individual objects, differences in the average accretion rates of objects of the same mass or both. 

\cite{Nguyen09} found that the variability in accretion rates of a sample of YSOs is dominated by variations on time-scales of a few days, and also the amplitude of variations is not enough to explain the full spread in the $\dot{M}$-$M_{*}$ relation. This is in agreement with previous studies \citep{2004A&A...424..603N}, and suggests that initial conditions or environmental differences are more likely to be the cause of this spread. However long term variations could also play a role \citep{2006ApJ...645L..69D}. Our observations are designed to investigate the variability on longer time-scales than previous studies and better define the spread associated with the $\dot{M}$-$M_{*}$ relation.

\indent In this paper, we set out to investigate the extent and characteristics of accretion-related variability in YSOs in an unbiased sample of objects, not chosen for their variability. We observed 25 low-mass YSOs in the Chamaeleon-I star forming region (age $\sim 2$\,Myr) using the multi-object fibre spectrograph FLAMES at the ESO/VLT. For each target, we obtained 12 epochs, covering time-scales from $\sim 1$ week to 15 months. Rather than focusing on individually selected targets, our main goal was to constrain long-term accretion variability for {\it typical} YSOs. We use information from three accretion-related emission lines (H$\alpha$ 6563\,\AA, He-I 6678\,\AA, Ca-II 8662\,\AA) to investigate possible accretion within these systems. 

\indent The paper is structured as follows: After this introduction, we describe the observations and the data in Sect. \ref{s2}, we discuss the accretion indicators and the measurements of emission lines in Sect. \ref{section:emissionlines} and Sect. \ref{sec:OriginEmission}. Methods used to derived accretion rates are presented in Sect. \ref{sec:accretion_rate}. The results are discussed in Sect. \ref{sec:discusion} and conclusions are drawn in Sect. \ref{sec:conclusion}. In the Appendix we discuss the possible presence of veiling in the spectra of the accretors.

\section{Observations and data preparation}
\label{s2}

\begin{figure}
\centering
 \includegraphics[width=0.47\textwidth]{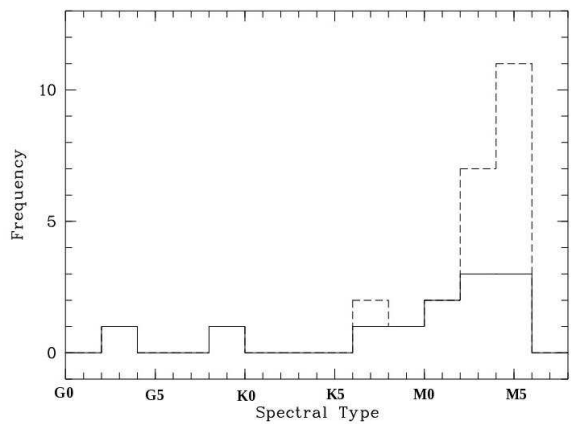}
 \caption{Histogram of spectral type distribution within sample. The dashed line indicates all objects and the solid line indicates the objects with discs.}
\label{fig:hist_spec}
\end{figure}

\indent The observations were carried out using the multi-object fibre spectrograph FLAMES/GIRAFFE at the ESO/VLT for ESO programmes 082.C-0005(A) and 084.C-0094(A). Our targets are confirmed members of the $\sim$2\,Myr old Cha-I star forming region, selected from the census by \citet{2007ApJS..173..104L} which contains 215 objects. Whenever possible we used the coordinates from the UCAC2 catalogue for the fibre positioning \citep{2004AJ....127.3043Z}. The faintest objects (spectral type later than $\sim$M4) are not registered by UCAC; here we relied on the less accurate coordinates from 2MASS. We observed one field (25 arcmin in diameter) focused on the area with the highest density of YSOs in the molecular cloud in the southern part of Cha-I with central coordinates $\alpha~11^h08^m20.0^s$, $\delta$\,-$77^o36'10''$ (J2000). When configuring the fibres priority was given to objects with a) evidence for accretion from emission lines and b) evidence for a circumstellar disc from mid-infrared excess (e.g. Class II objects or classical T Tauri stars (CTTs), see Sect. \ref{sec:OriginEmission}). Apart from that, no other selection criteria were used. The final sample of targets contained 50 objects, however twenty five of these objects were dropped from the analysis because no measurable stellar continuum was observed above the sky background emission. Two objects which have no continuum were kept as they both showed strong H$\alpha$ emission and other spectral lines. The spectral type distribution for the objects with and without discs in this sample is shown in Fig. \ref{fig:hist_spec}.

\indent We observed this sample 6 times in each of the two campaigns. The observations cover the time span from January 2009 to March 2009 and from December 2009 to April 2010. The minimum separation between consecutive observations is $\sim 1$\,week. The exact dates of the observations are given in Table \ref{tab:observations}.

\begin{table}
  \caption{Observations }
  \begin{tabular}{@{}cccccc@{}}
\hline
Obs. Period 1 & Time   & Obs. Period 2 & Time & Grism  \\
Obs.  Date    &  UTC  &  Obs.  Date    &  UTC &        \\
\hline
04-Jan-2009 & 05:10 &  09-Dec-2009 & 07:06 & HR21   \\
04-Jan-2009 & 05:26 &  09-Dec-2009 & 07:21 & HR15   \\
16-Jan-2009 & 04:35 &  03-Jan-2010 & 06:30 & HR21   \\
16-Jan-2009 & 04:51 &  03-Jan-2010 & 06:46 & HR15   \\
03-Feb-2009 & 04:40 &  15-Jan-2010 & 07:26 & HR21   \\
03-Feb-2009 & 04:56 &  15-Jan-2010 & 07:42 & HR15   \\
11-Feb-2009 & 05:30 &  27-Jan-2010 & 04:11 & HR21   \\
11-Feb-2009 & 05:46 &  27-Jan-2010 & 04:26 & HR15   \\
19-Feb-2009 & 05:38 &  10-Feb-2010 & 06:30 & HR21   \\
19-Feb-2009 & 05:54 &  10-Feb-2010 & 06:45 & HR15   \\
01-Mar-2009 & 02:42 &  07-Mar-2010 & 06:20 & HR21   \\
01-Mar-2009 & 02:59 &  07-Mar-2010 & 06:36 & HR15   \\
\hline
\end{tabular}
\label{tab:observations}
\end{table}

\indent At each epoch two spectra of each source were taken with two high-resolution GIRAFFE grisms, HR15N and HR21, with a wavelength coverage of 6470\,-\,6790\,\AA~and 8480\,-\,9000\,\AA~respectively. These settings were selected because they cover the accretion-related emission lines H$\alpha$, He\,I, and the Ca IRT (see Sect. \ref{section:emissionlines}). The nominal resolution for these grisms is in the range of $R\sim$ 16000 - 17000. For each epoch and setting we obtained a long exposure (800\,s for HR21 and 1000\,s for HR15N) and in addition a short exposure (30\,s) in case of saturation. As saturation was not an issue even for the brightest objects; we use the short exposures only for consistency checks.

\indent For our analysis, we rely on the fully processed science frames as provided by the ESO/GIRAFFE pipeline (`SRBS' files). The pipeline reduction includes bias correction, extraction, flatfielding, transmission correction and wavelength calibration. The spectral response is not corrected, because no standard stars were observed, i.e. absolute flux measurements are not feasible with our data set. Note that in the standard pipeline reduction, sky subtraction and scattered light subtraction does not take place. 

\begin{figure}
\centering
 \includegraphics[width=0.34\textwidth,angle=270]{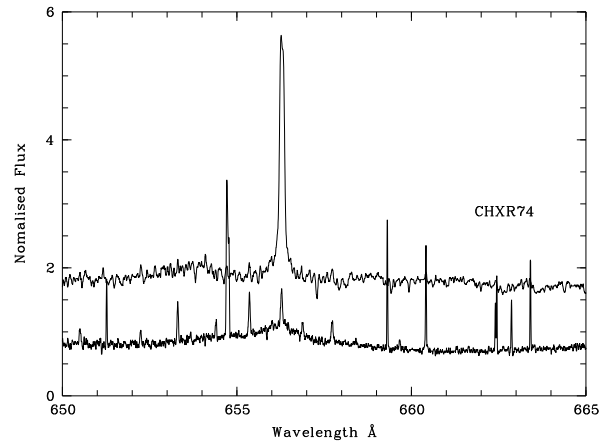}
 \caption{A spatially averaged spectrum of the sky background H$\alpha$ emission, and a single epoch spectrum from a non-accretor in the sample, CHXR74, which is offset by 1 for clarity.}
\label{fig:av_sky}
\end{figure}

\indent Spectra taken in molecular cloud regions are affected by emission line flux from the nebular background \citep[e.g.][]{2005AJ....129..363S}. For fibre spectrographs this can be problematic, because a measurement of the background close to the target is usually not feasible. To be able to assess the background contamination, we placed a number of additional fibres on positions without any source (35 and 58 sky fibres for the first and second observation periods).
Fig. \ref{fig:av_sky} shows an averaged spectra from eight sky fibres, with a single epoch spectra from an object in the sample, CHXR74. Here we see that the amount the sky emission contributes to the H$\alpha$ flux is small (here CHXR74 has a H$\alpha$ EW $\sim$ 8.6\,\AA).

We used two different methods to remove the sky background continuum levels from the object spectra. For the majority of objects, a spatially averaged sky continuum value was derived for each epoch, and was removed from their spectra. To increase our spatial resolution and the reliability of our background  subtraction, in addition to the sky fibres, we used 18 target spectra that showed no continuum emission above the sky background. 
Using the fluctuations in our emission line measuements to test how reliable this sky subtraction is, we found that this method was insufficient for four object's shorter wavelength spectra (6470\,-\,6790\,\AA), which are discussed in Sect.\ref{section:emissionlines}. 

\section{Emission Lines}
\label{section:emissionlines}
 The dominant emission line in our spectra is H$\alpha$~(6562.81\,\AA) and is present in almost all objects. In addition, both the He\,I (6678.2\,\AA) and Ca\,II triplet emission lines (8498, 8542, 8662.1\,\AA) are covered and are present in the spectra of a few objects. 

\subsection{H$\alpha$ emission}\label{section:Ha_emission}
\begin{figure*}
\begin{tabular}{ccccc}
\includegraphics[scale=0.180]{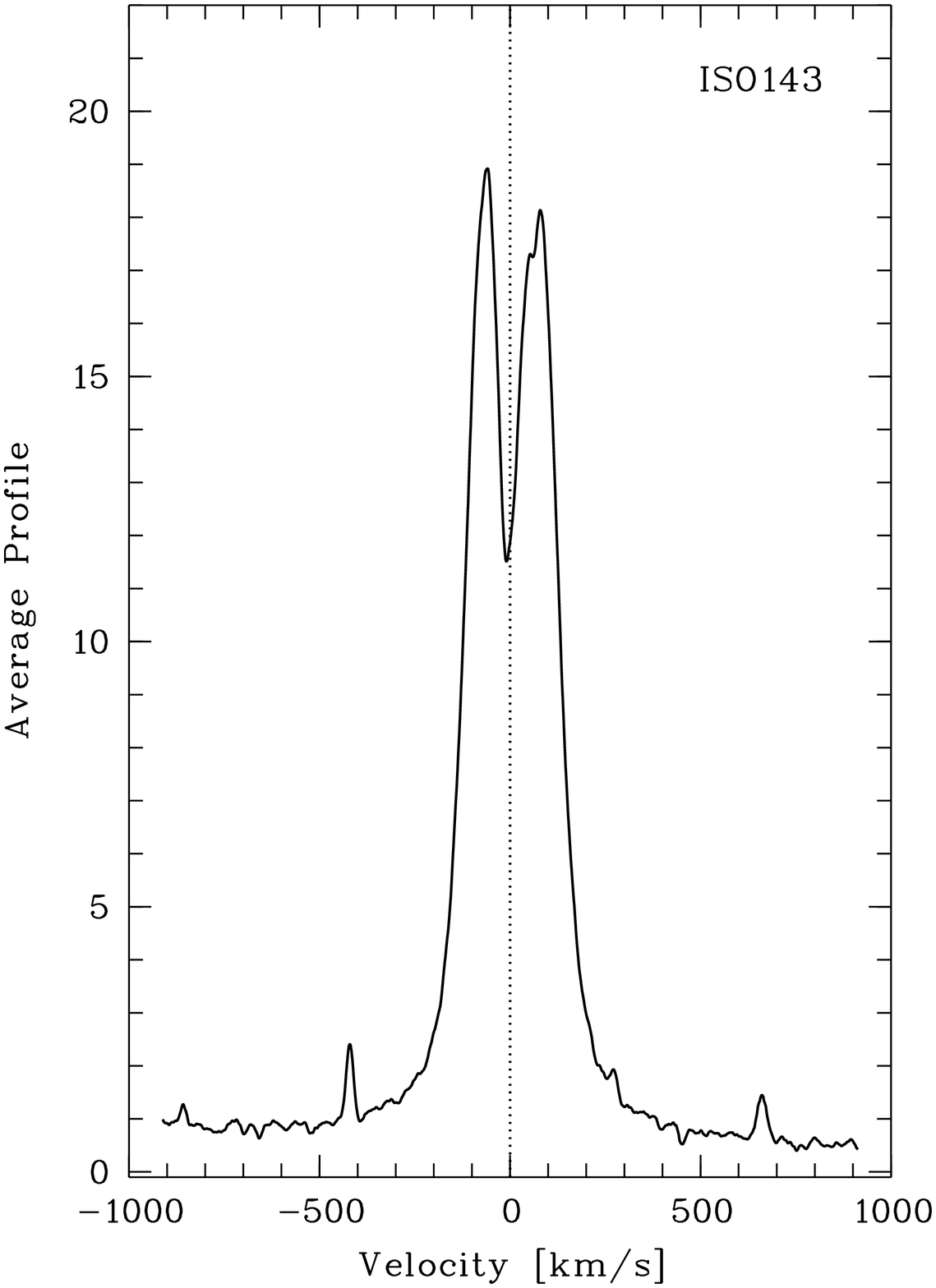}
\includegraphics[scale=0.180]{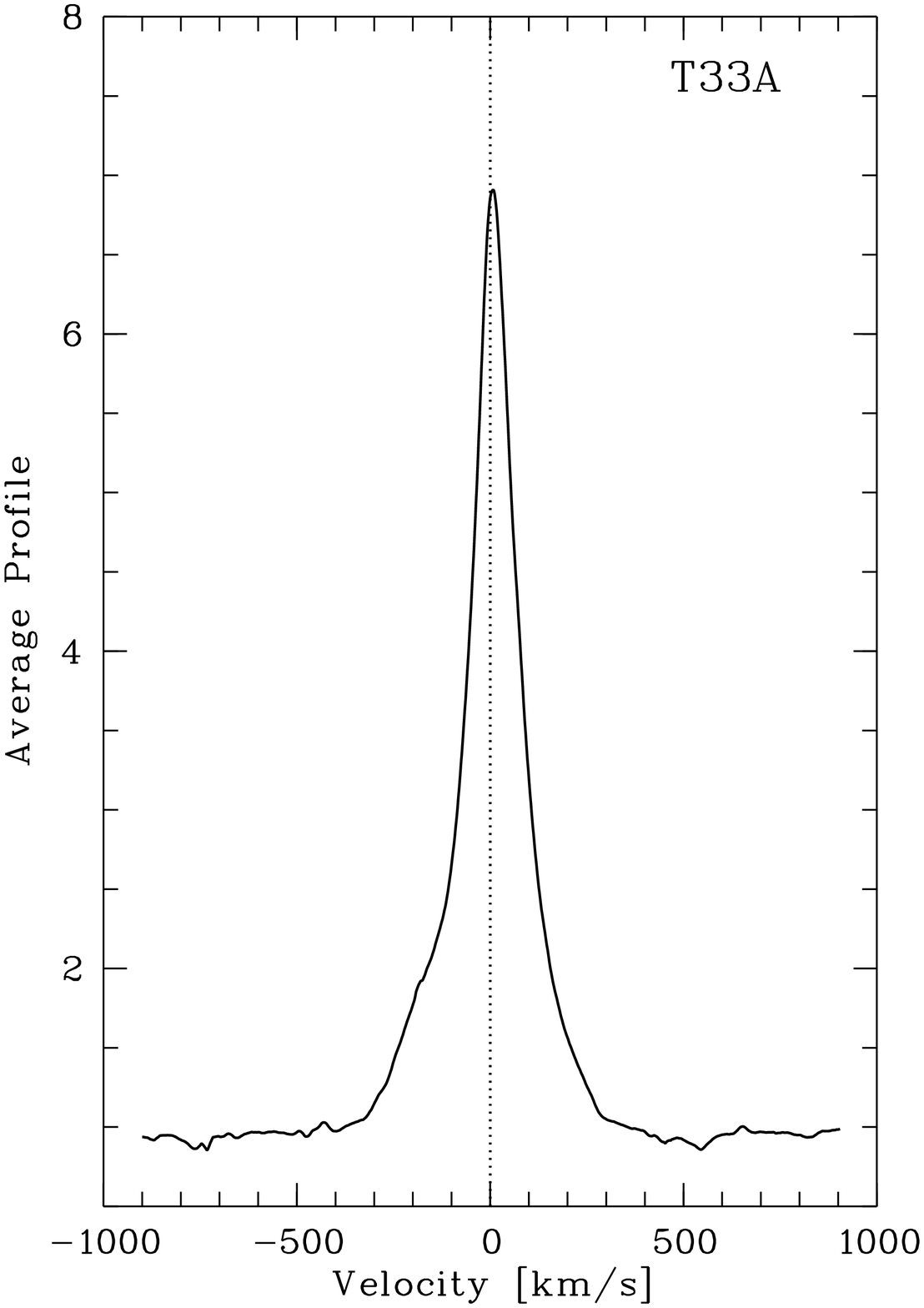} 
\includegraphics[scale=0.180]{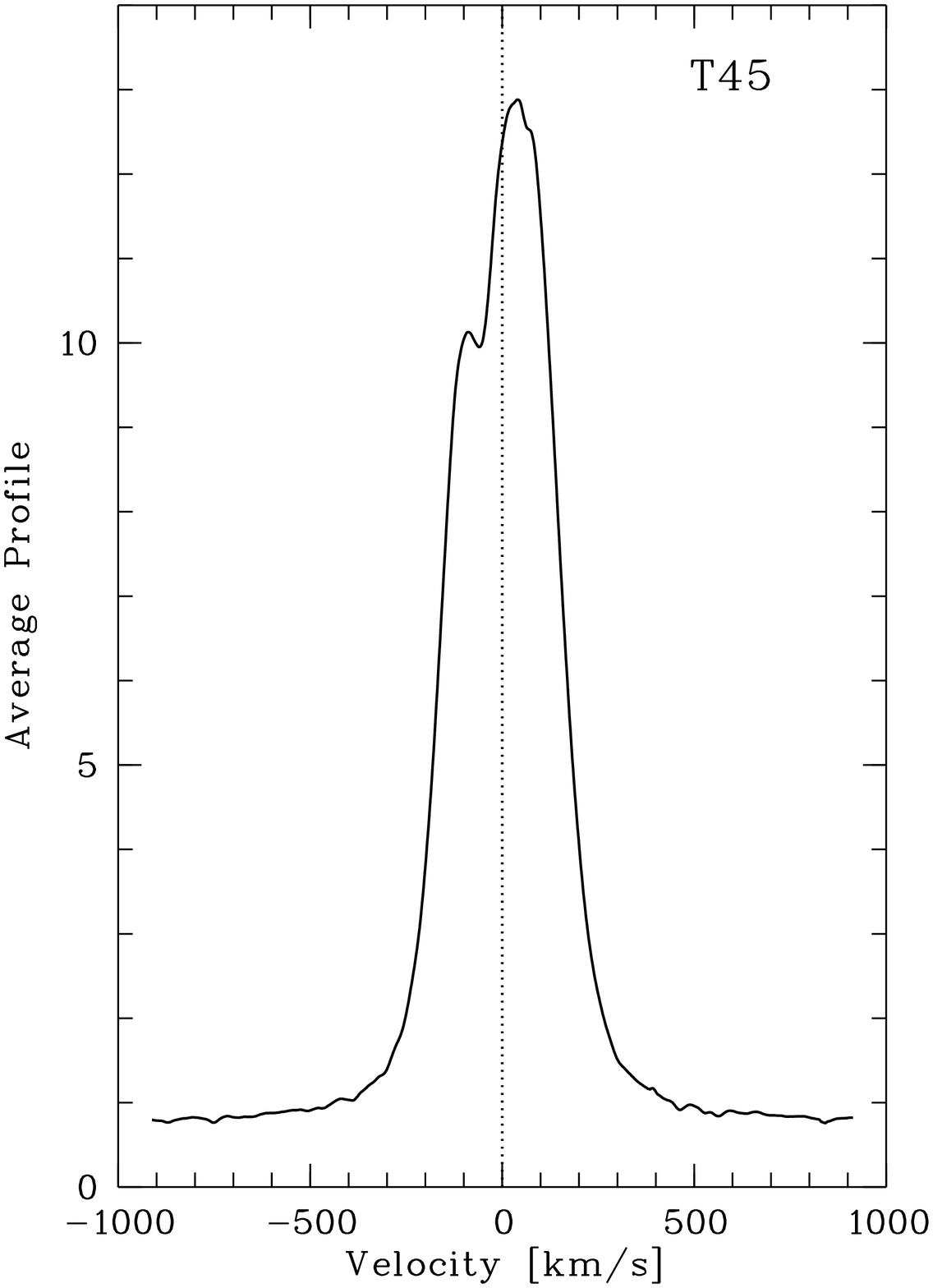} 
\includegraphics[scale=0.180]{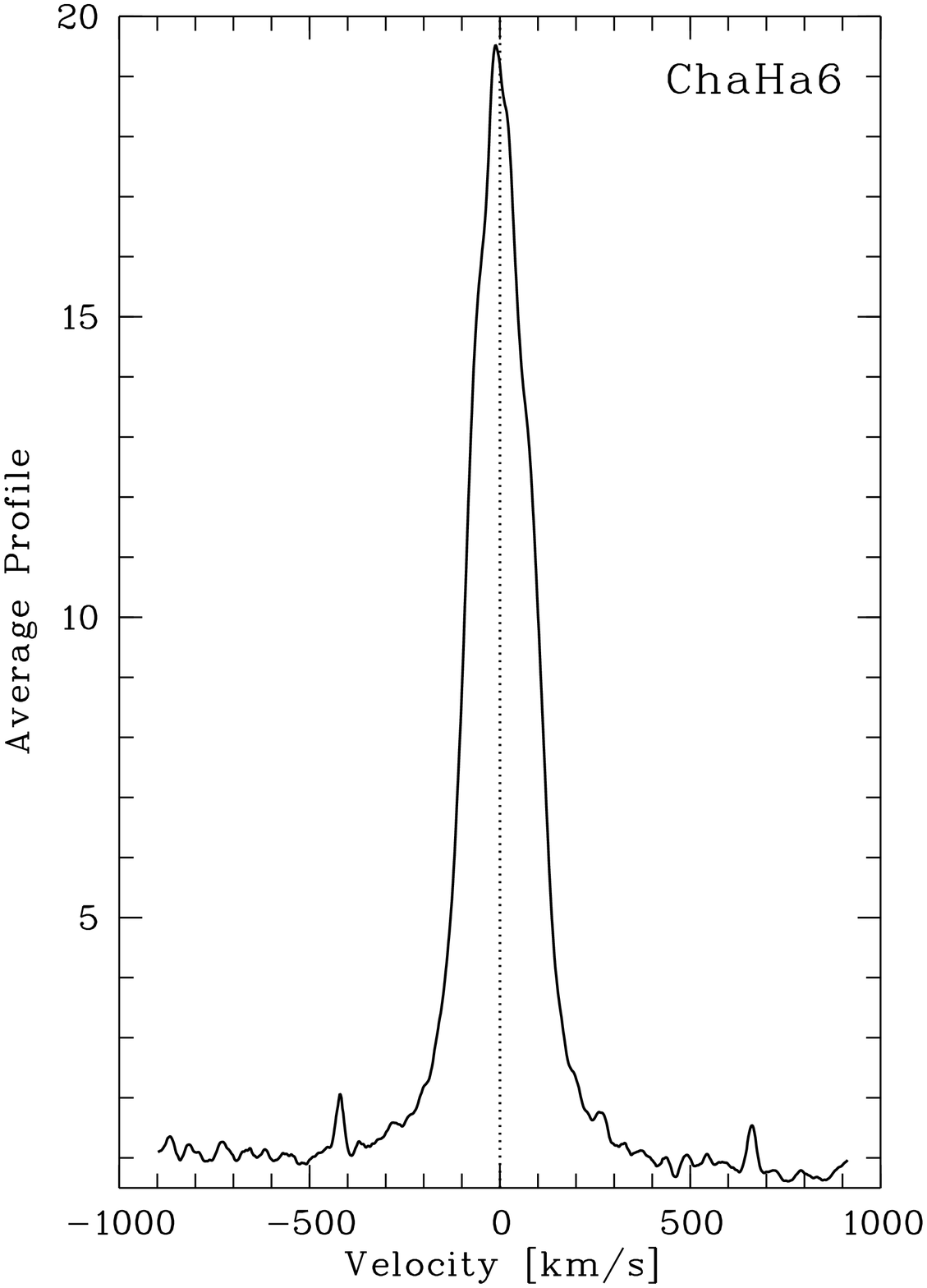}
\includegraphics[scale=0.180]{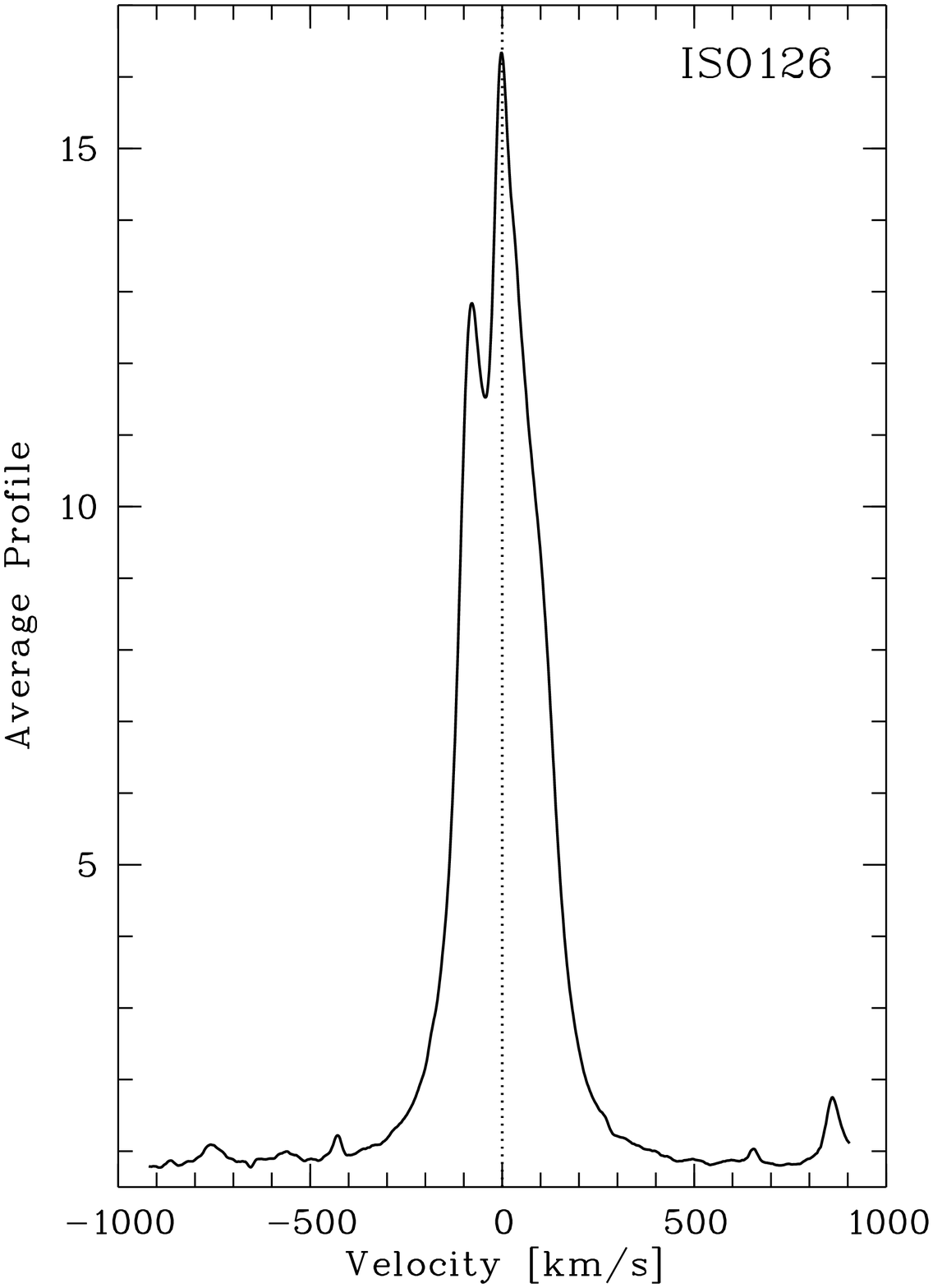}\\
\includegraphics[scale=0.180]{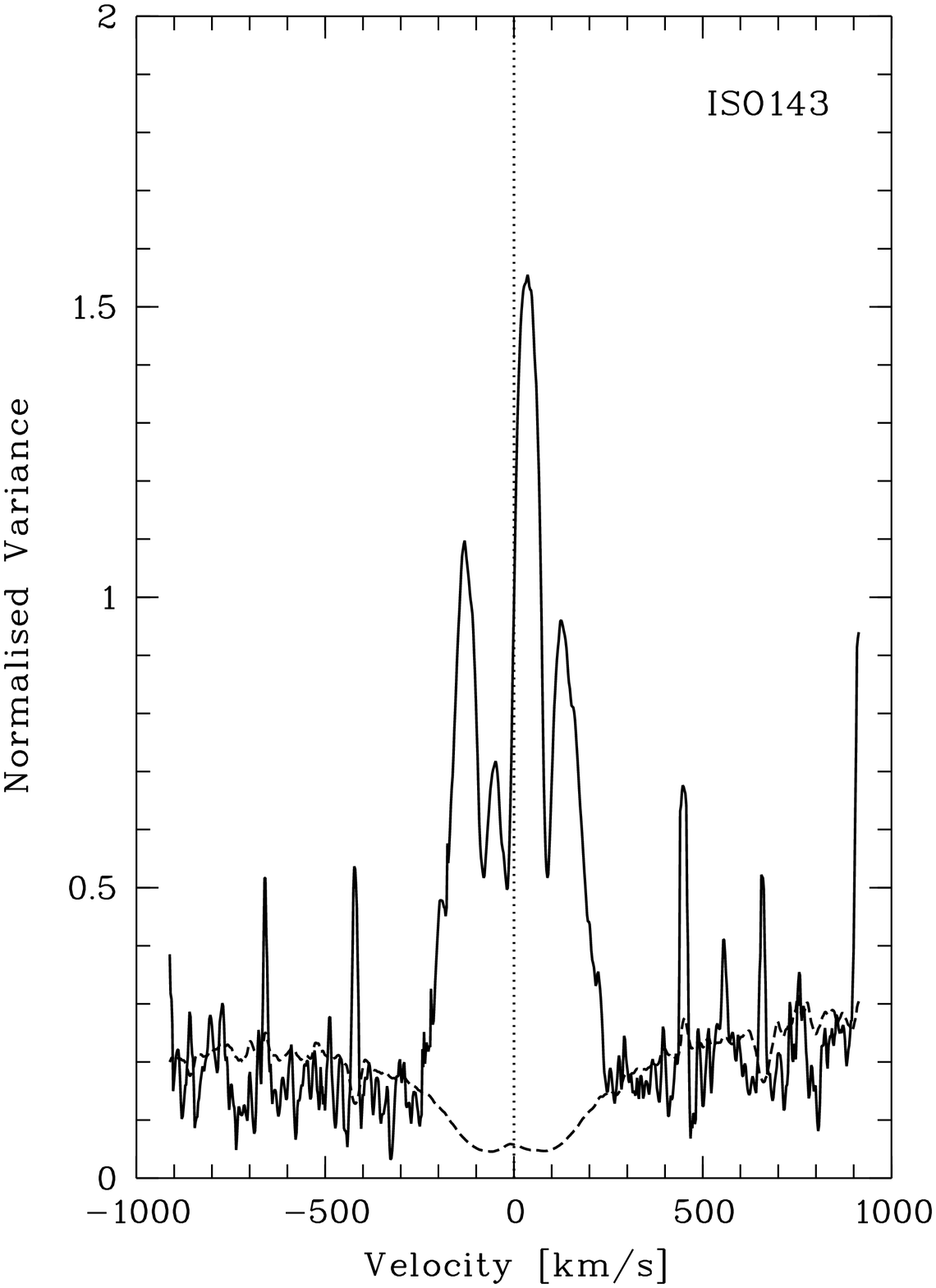}
\includegraphics[scale=0.180]{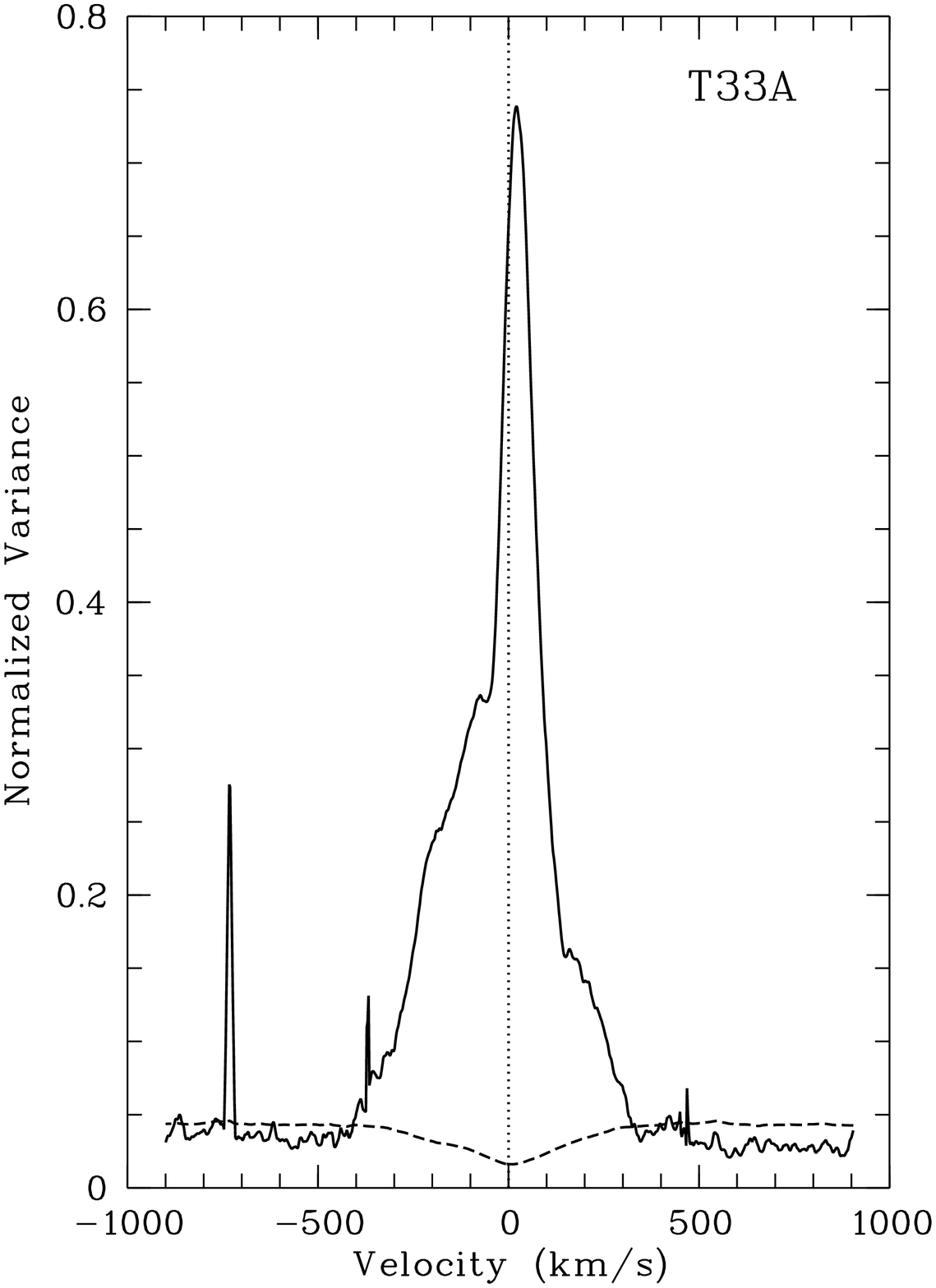}
\includegraphics[scale=0.180]{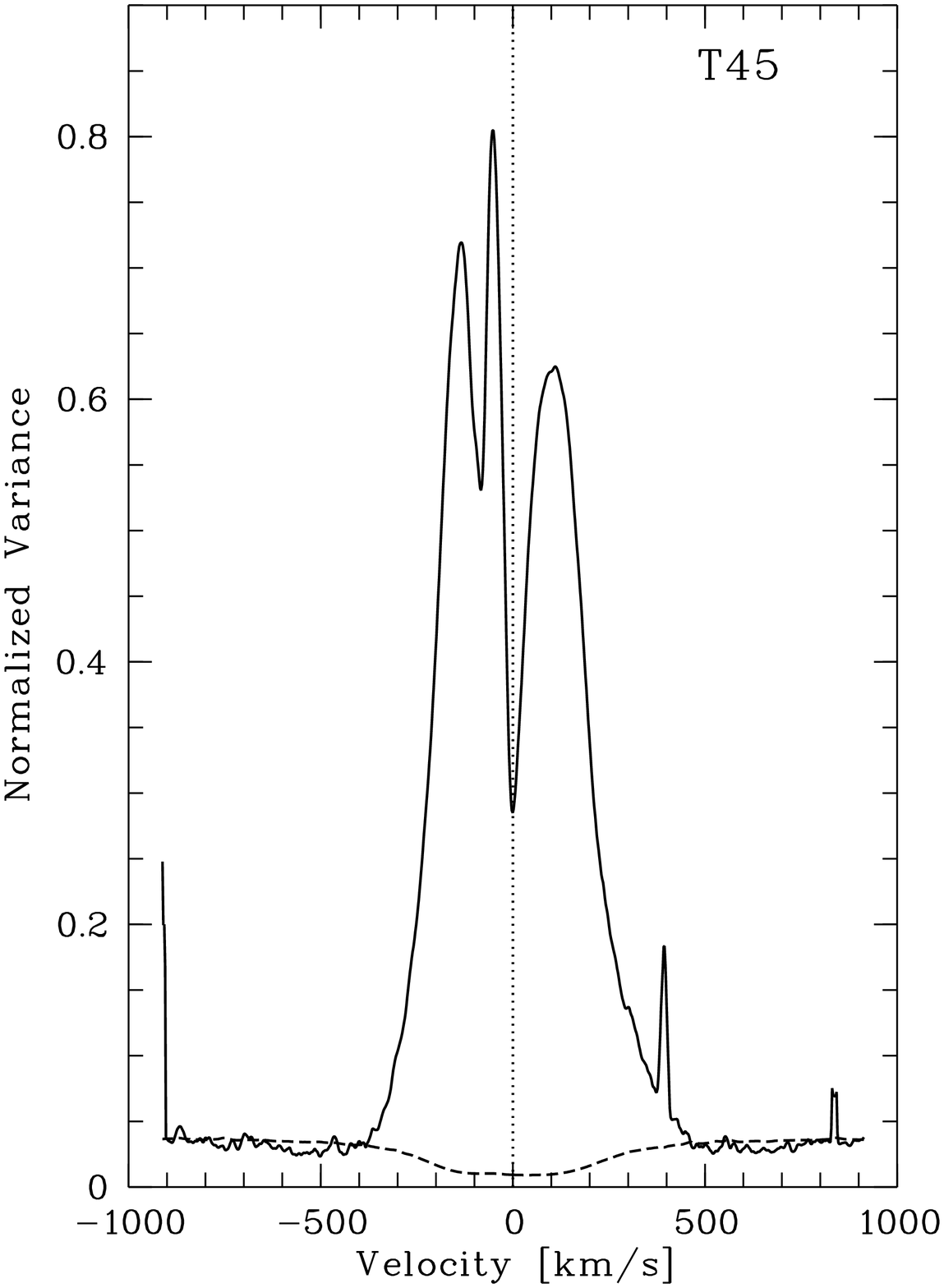}
\includegraphics[scale=0.180]{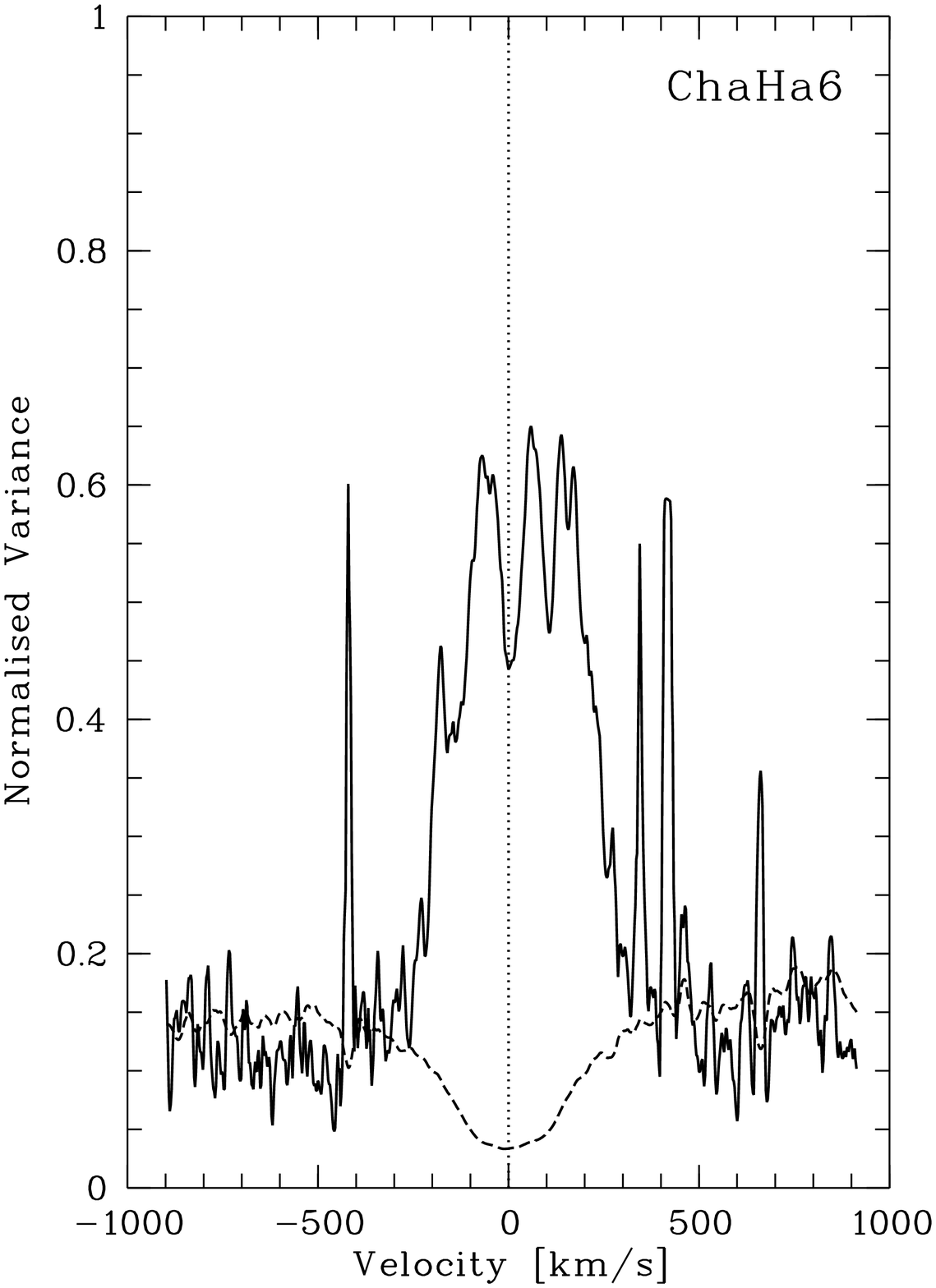}
\includegraphics[scale=0.180]{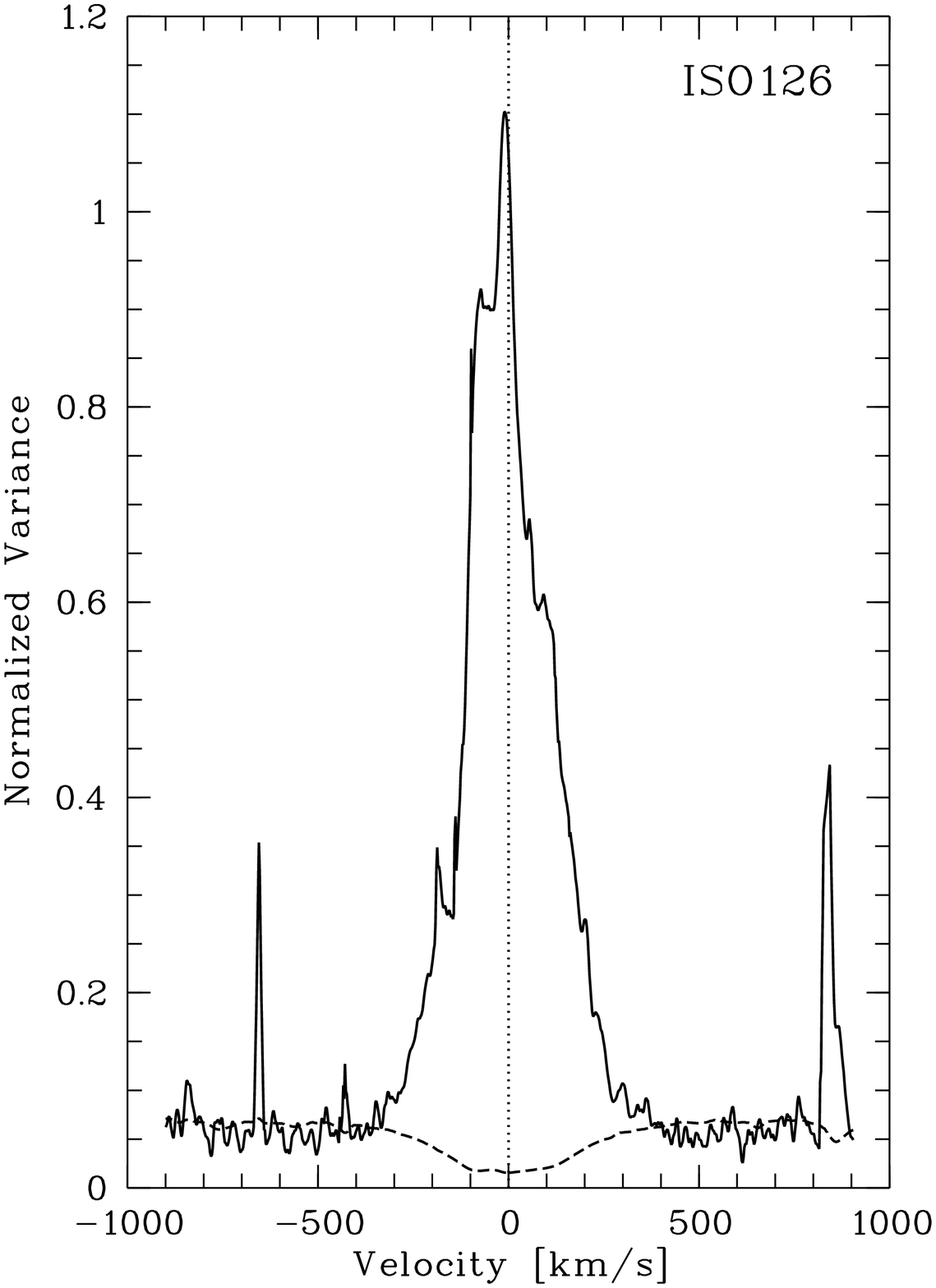}\\

\includegraphics[scale=0.180]{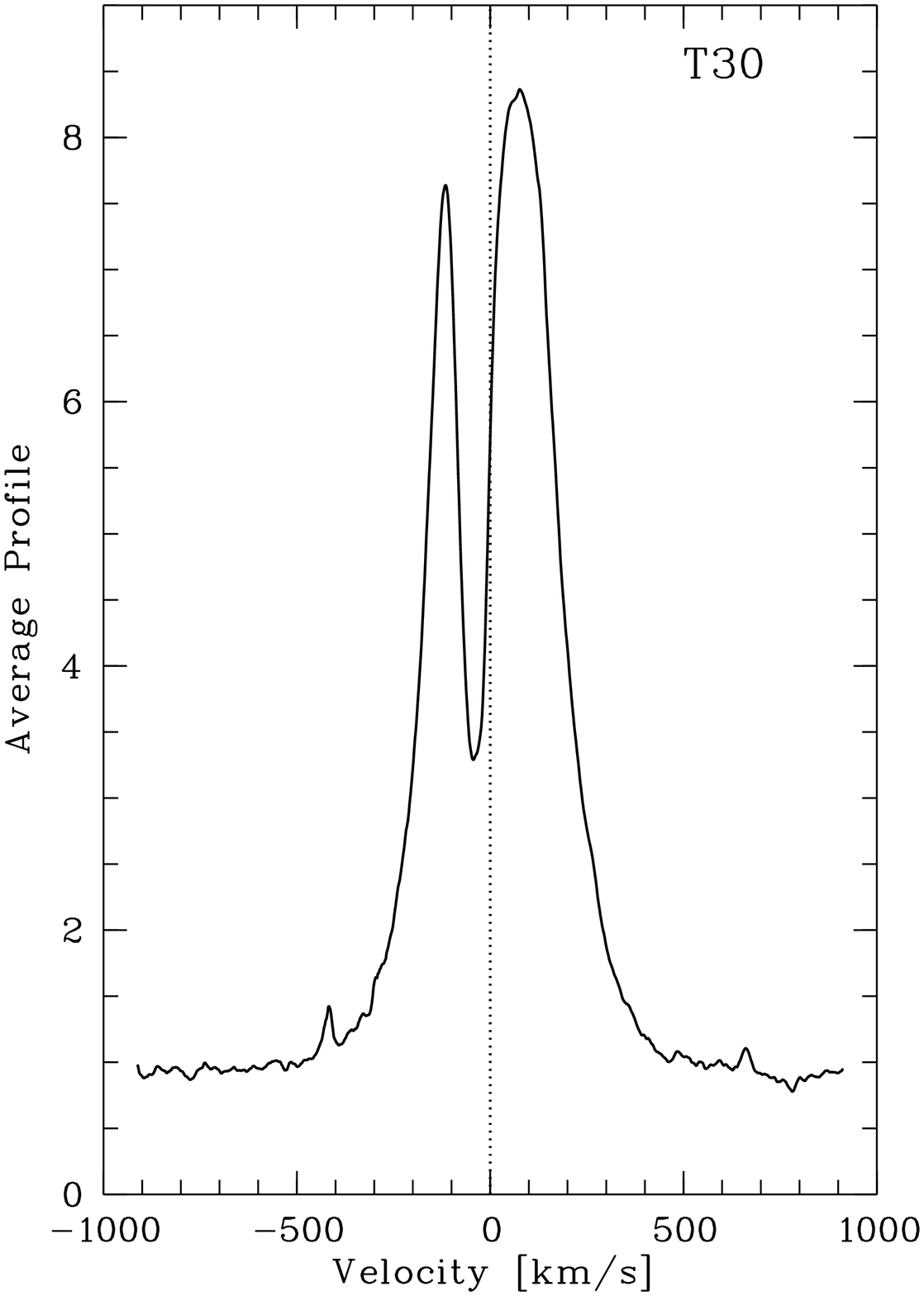}
\includegraphics[scale=0.180]{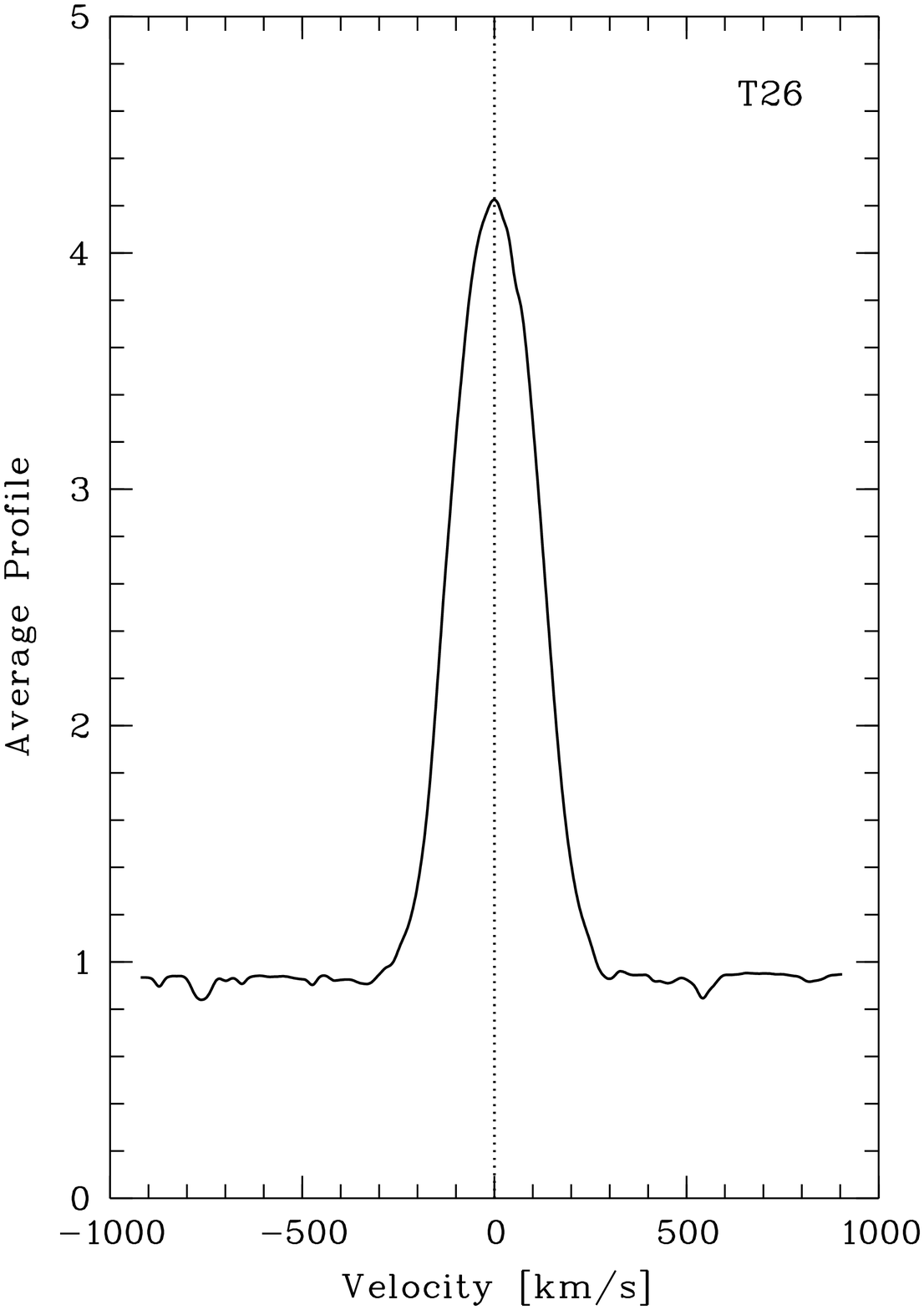}
\includegraphics[scale=0.180]{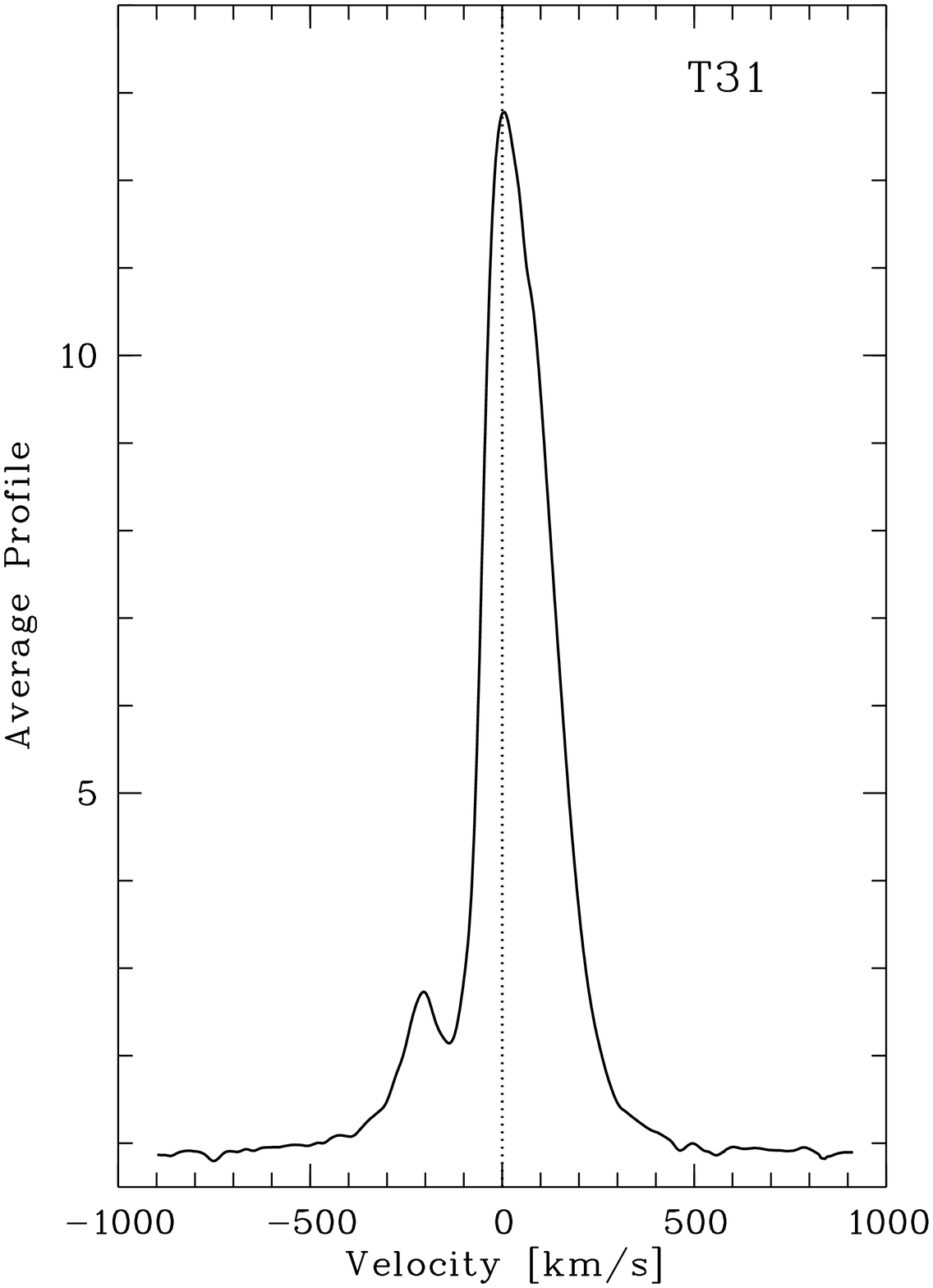} 
\includegraphics[scale=0.180]{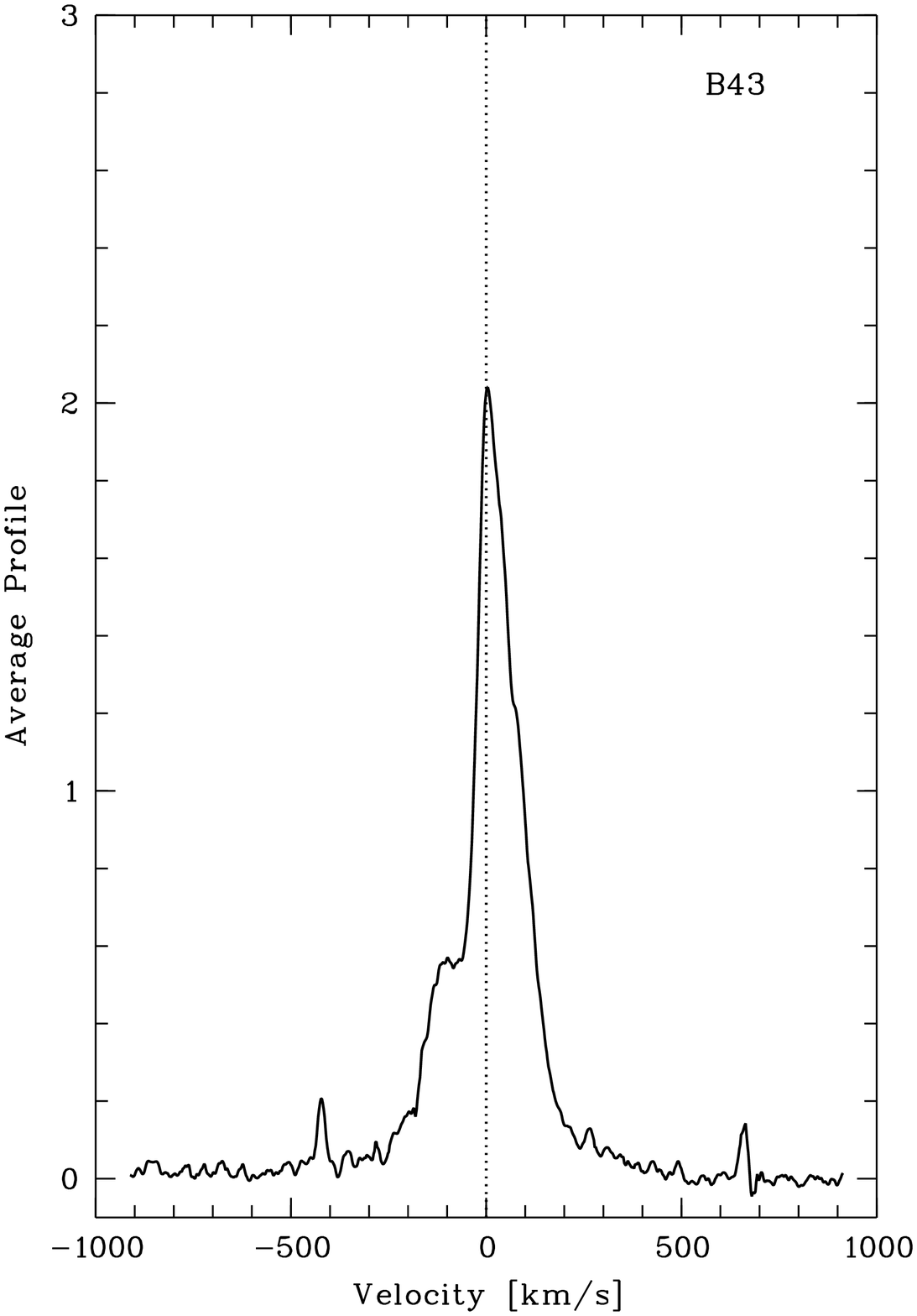}

\includegraphics[scale=0.180]{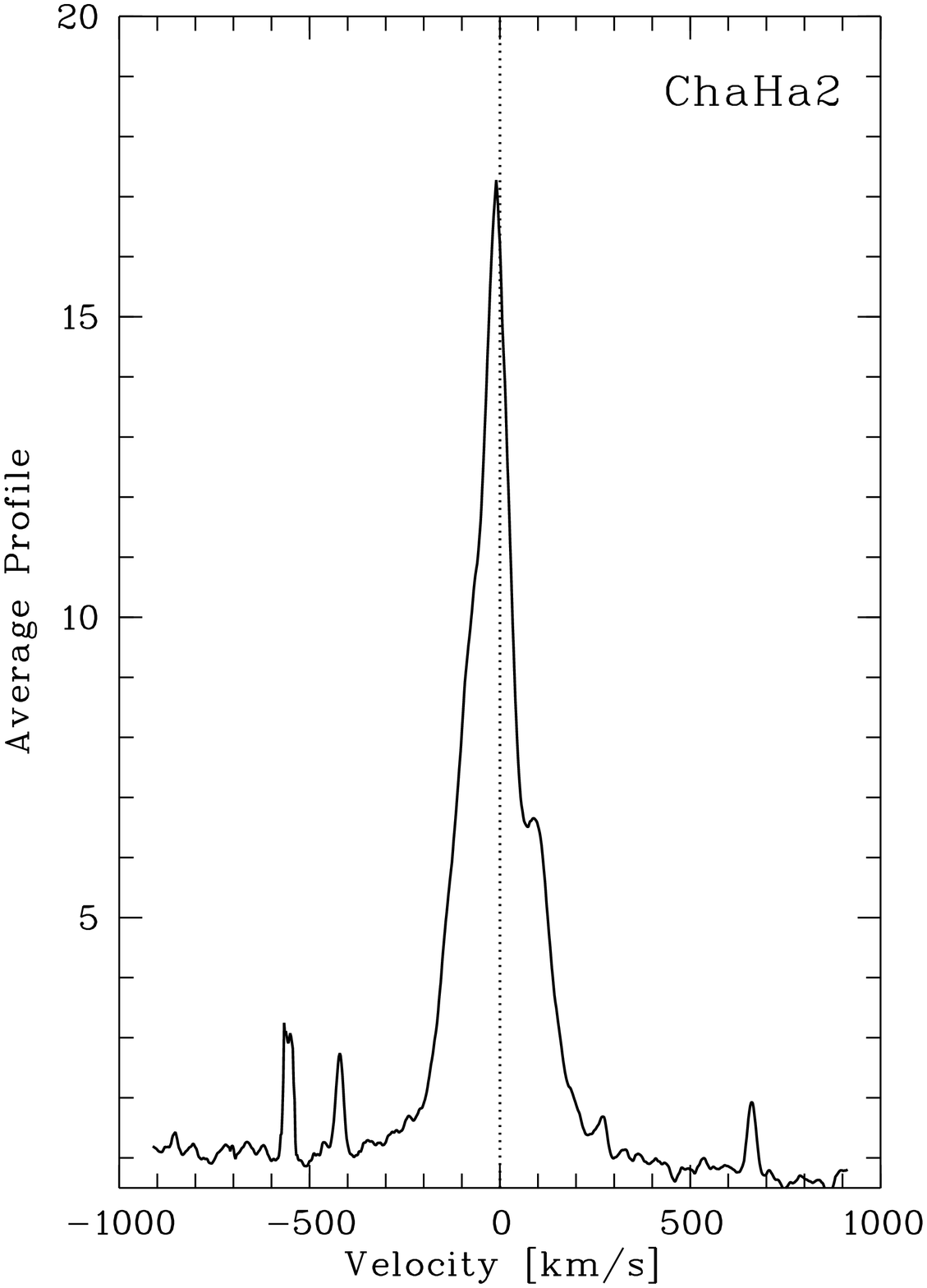}  \\

\includegraphics[scale=0.180]{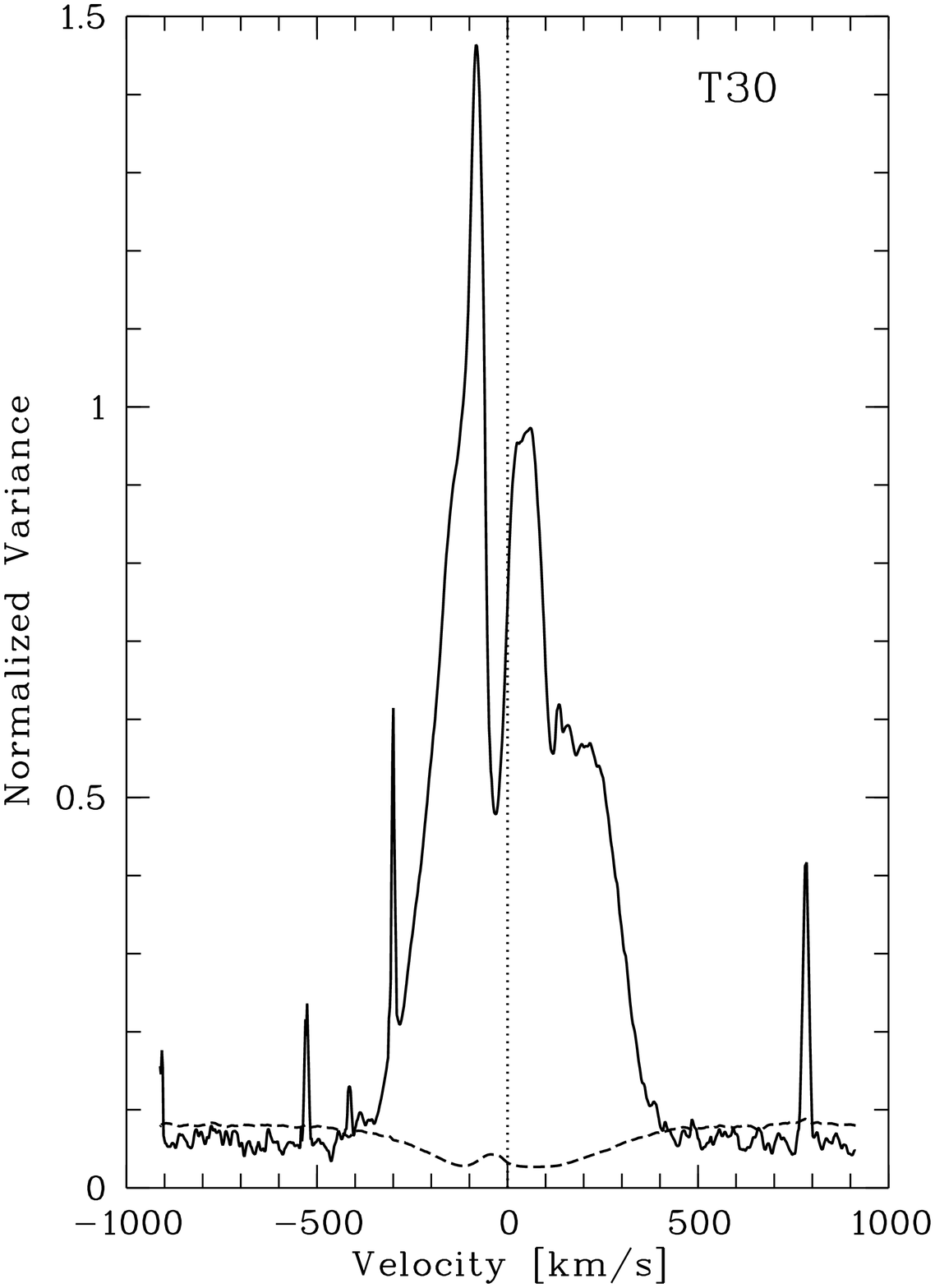}
\includegraphics[scale=0.180]{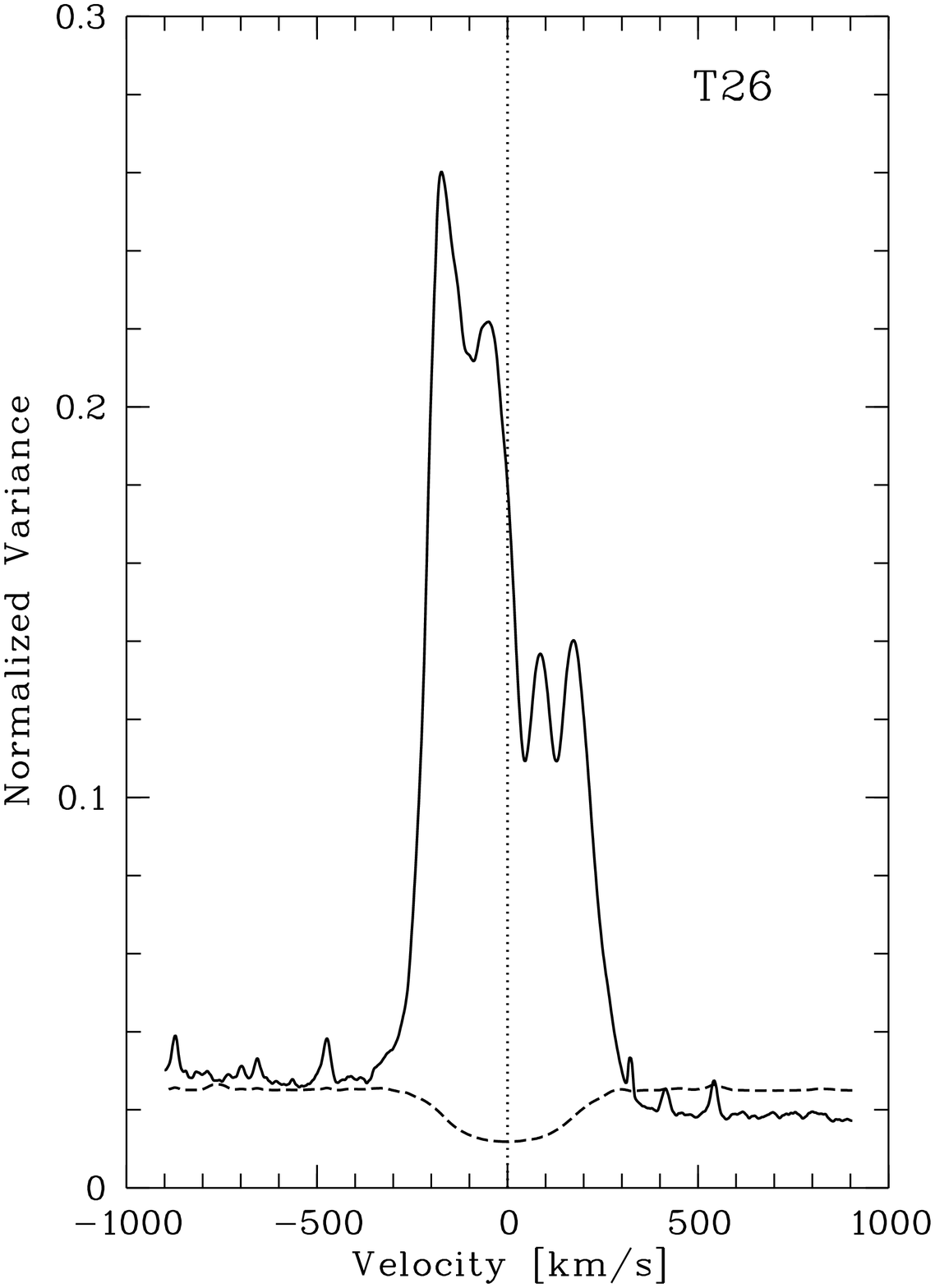}
\includegraphics[scale=0.180]{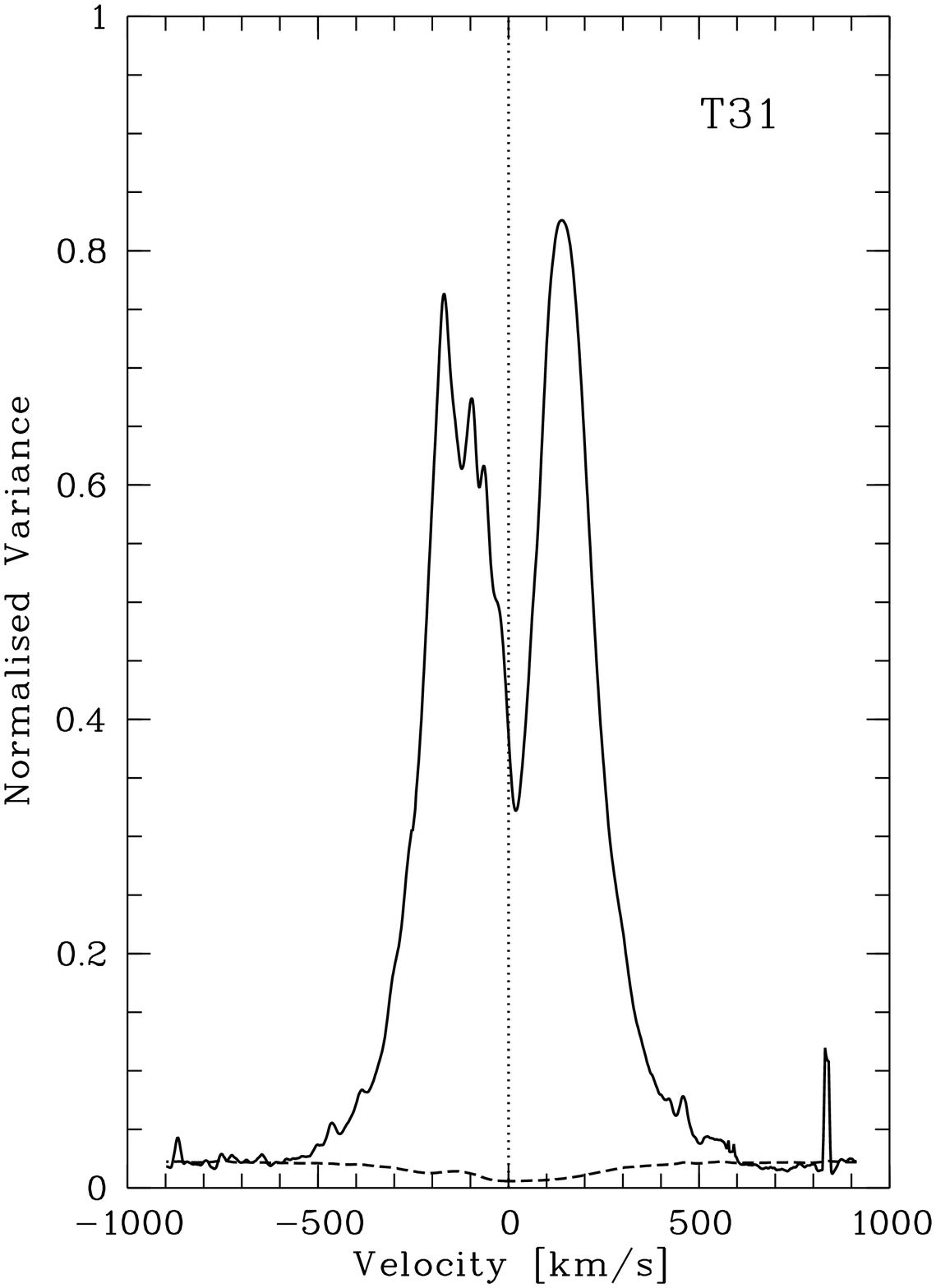}
\includegraphics[scale=0.180]{figure_3_u.ps}
\includegraphics[scale=0.180]{figure_3_u.ps}\\
\end{tabular}
\caption{Average and variance profiles for H$\alpha$ emission feature for the 10 objects with strongest H$\alpha$ emission (see Sect. \ref{section:Ha_emission}). The name of the object is given in each frame. Vertical lines mark the central wavelength for H$\alpha$ emission (6562.81\AA). The horizontal dashed lines in the variance profile plots represent the zero variability level allowing for signal to noise. B43 and ChaH$\alpha$2 are shown without variance profiles, as their continua were not measureable. }
\label{fig:Accretion_Profiles}
\end{figure*}

\begin{figure*}
\centering
\begin{tabular}{cc}
    \includegraphics[width=0.19\textwidth ]{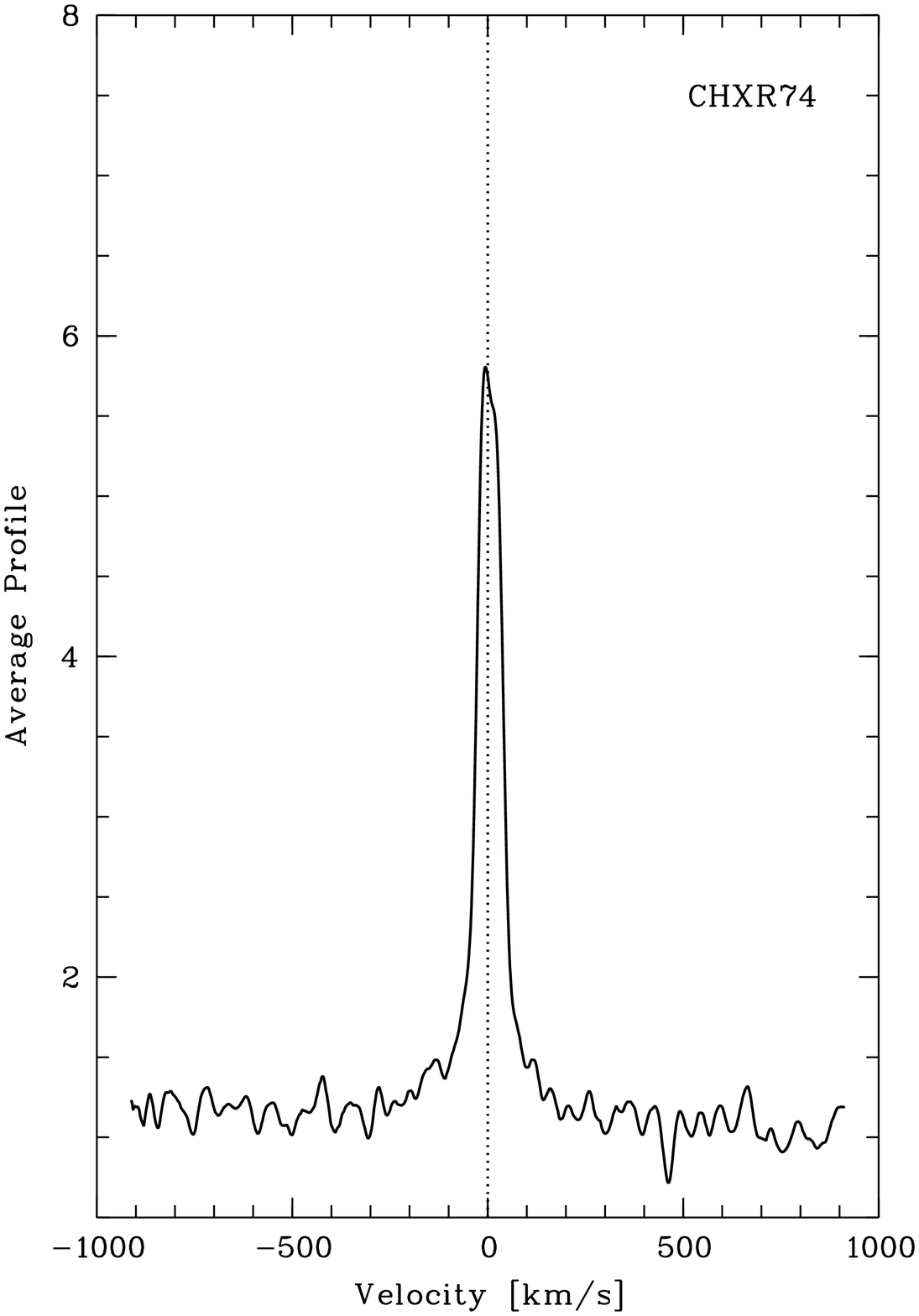}
    \includegraphics[width=0.19\textwidth]{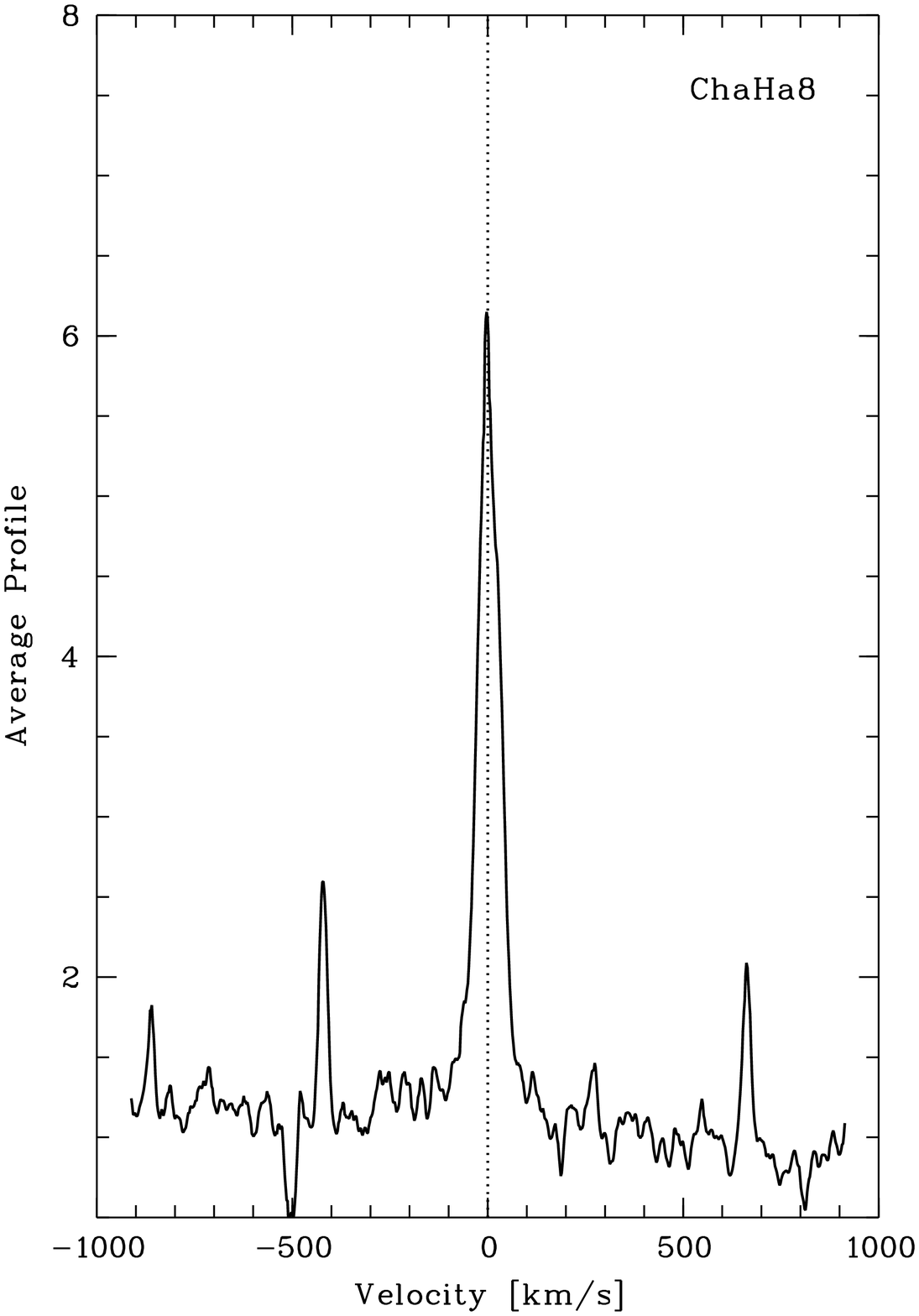}
    \includegraphics[width=0.19\textwidth]{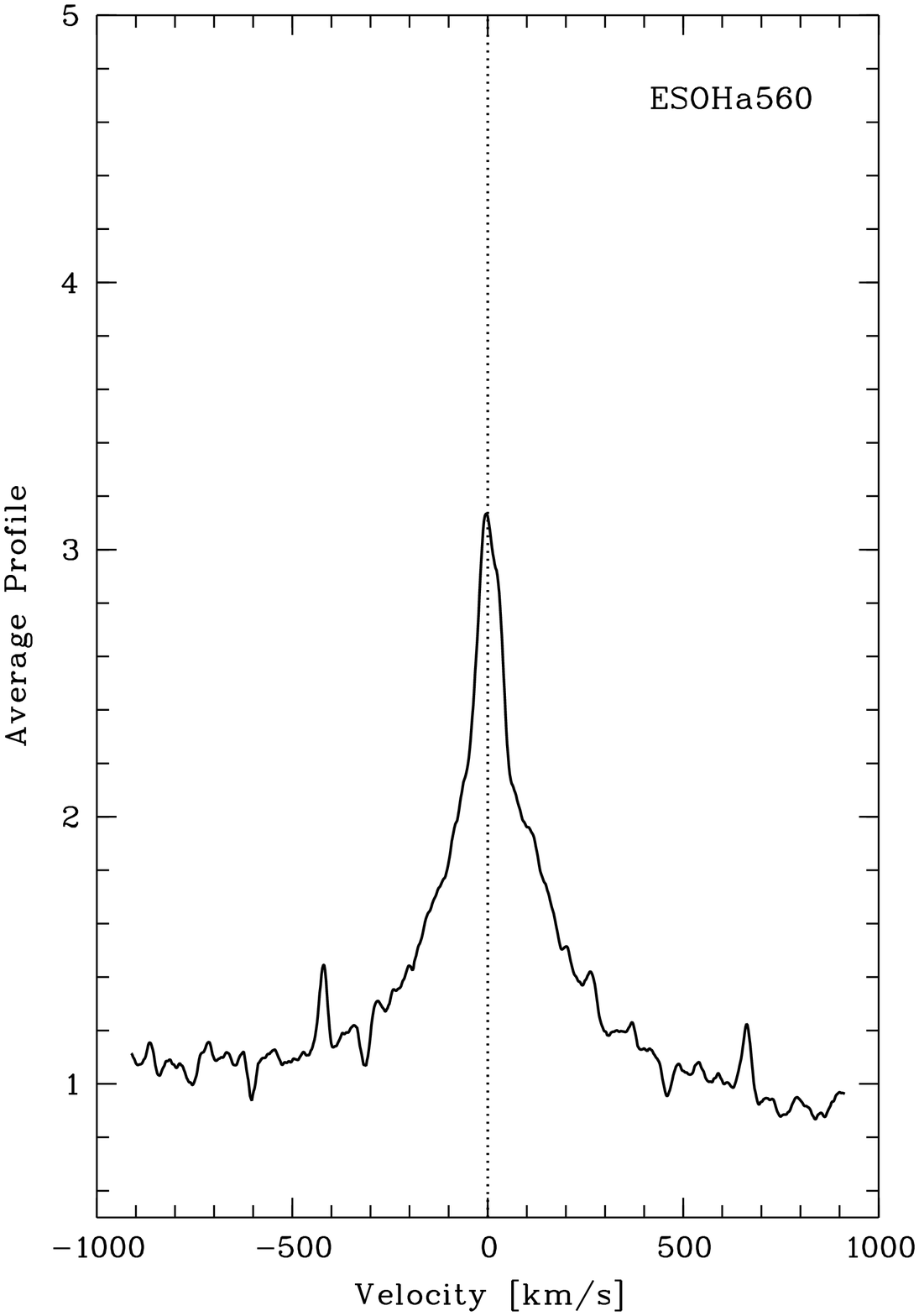}
    \includegraphics[width=0.19\textwidth]{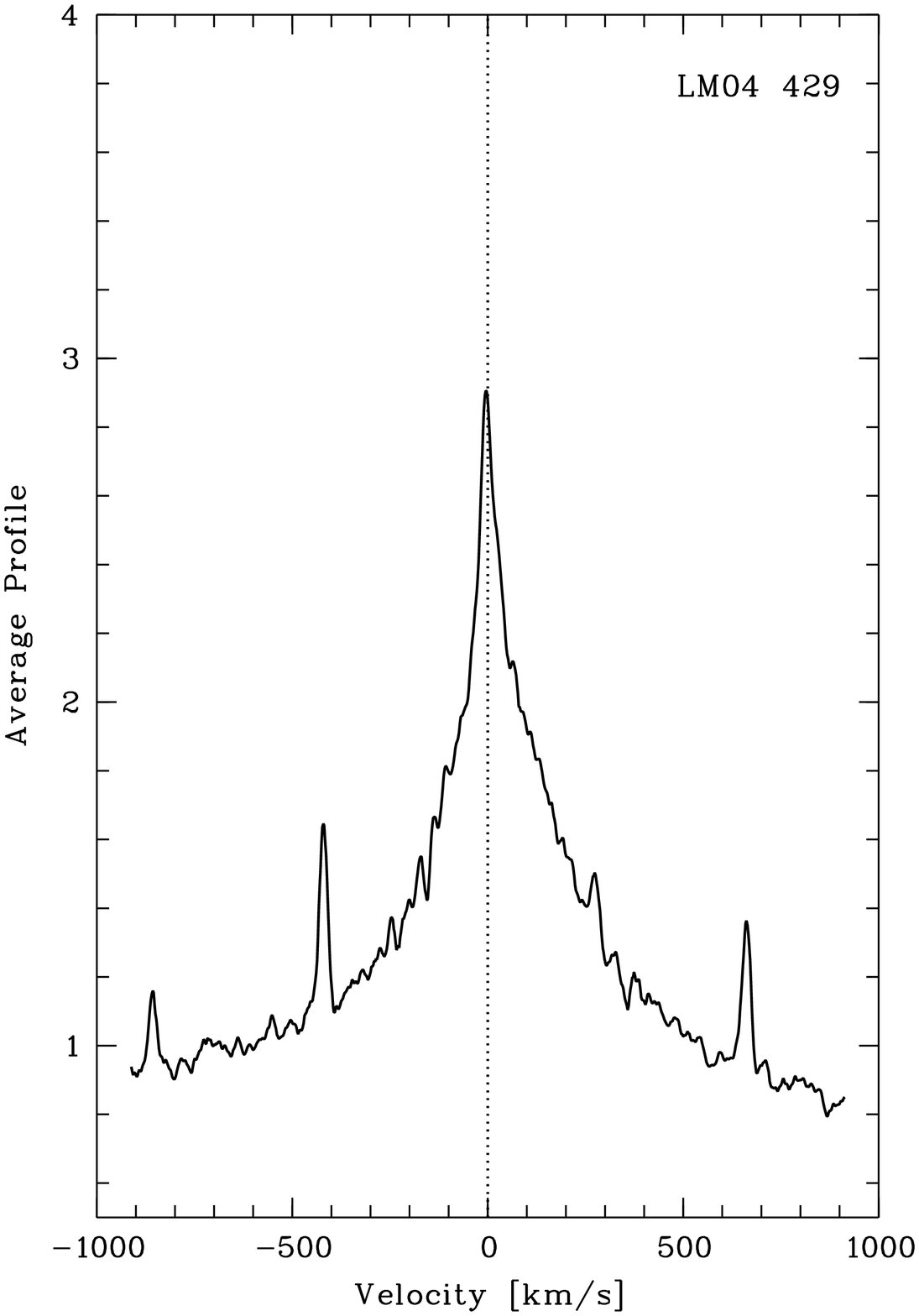}\\
    \includegraphics[width=0.19\textwidth]{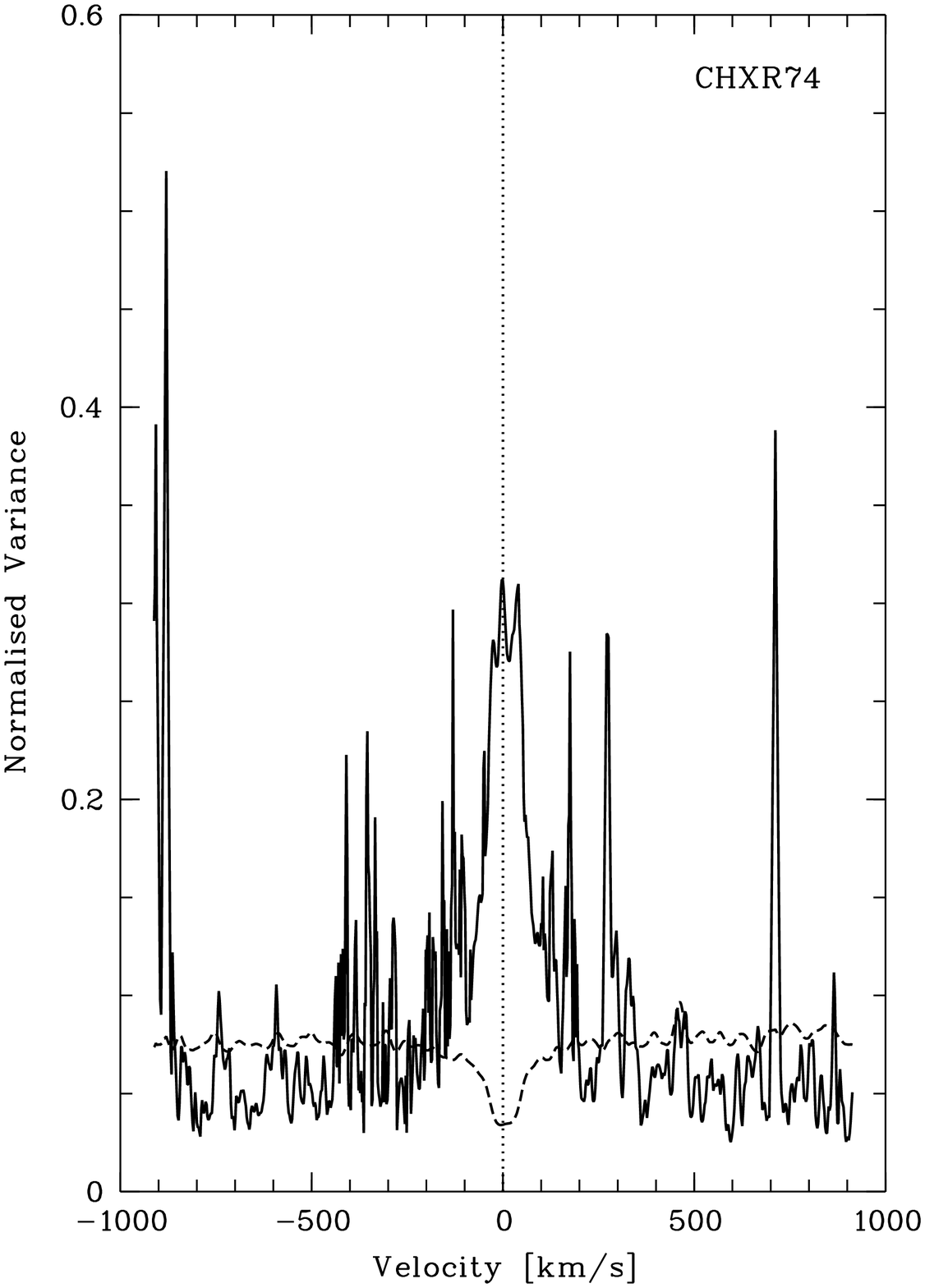} 
    \includegraphics[width=0.19\textwidth]{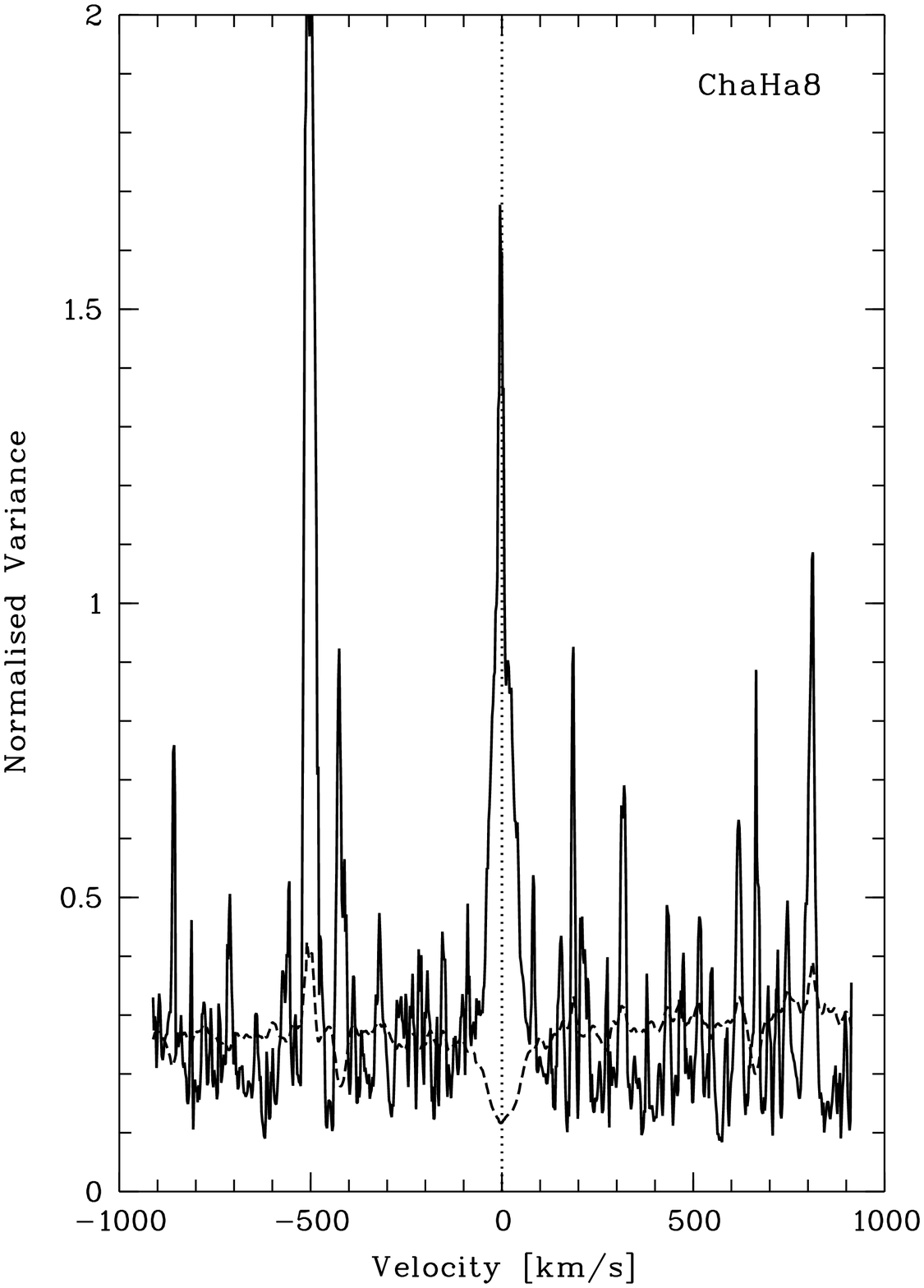}
    \includegraphics[width=0.19\textwidth]{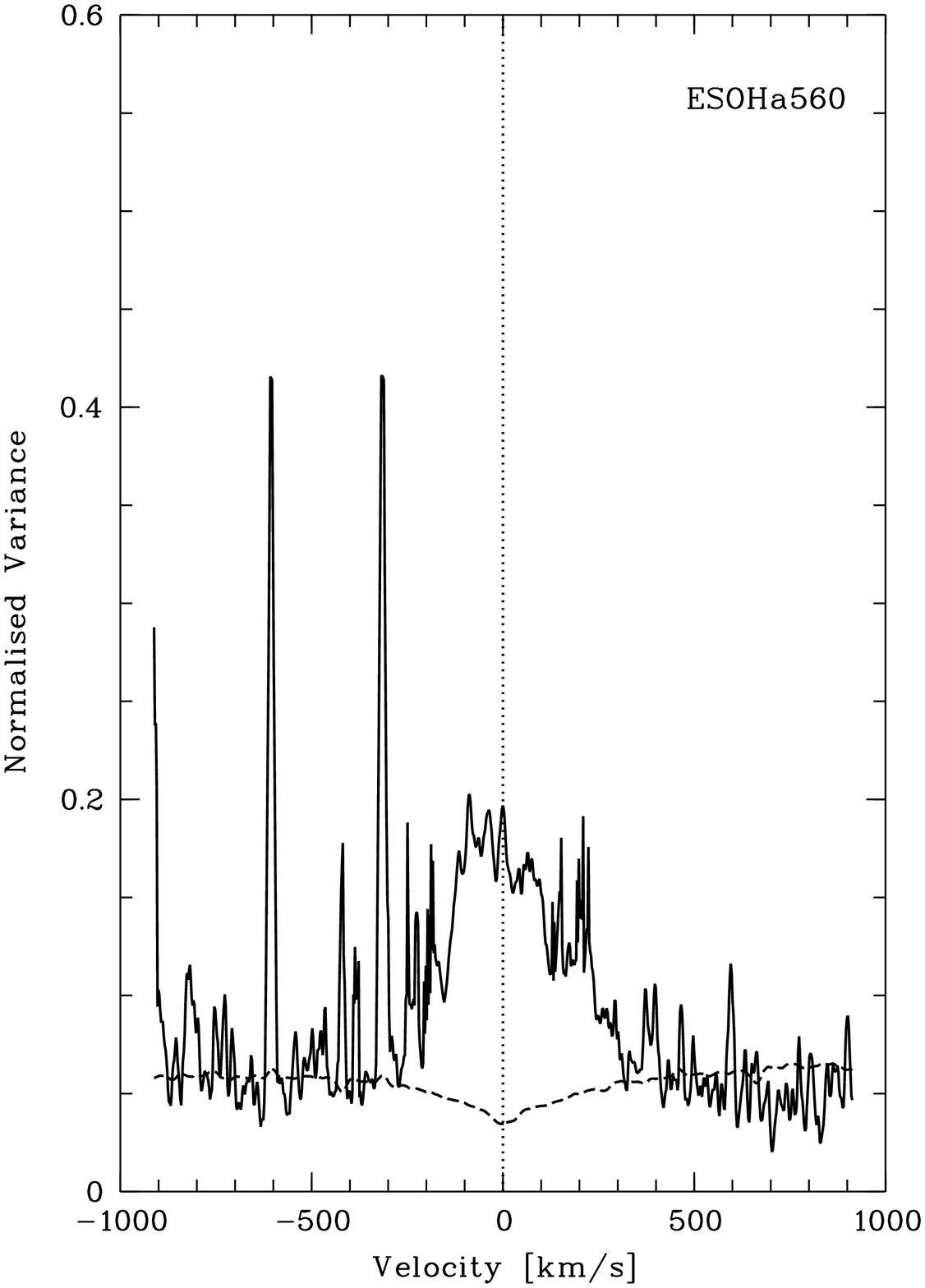}
    \includegraphics[width=0.19\textwidth]{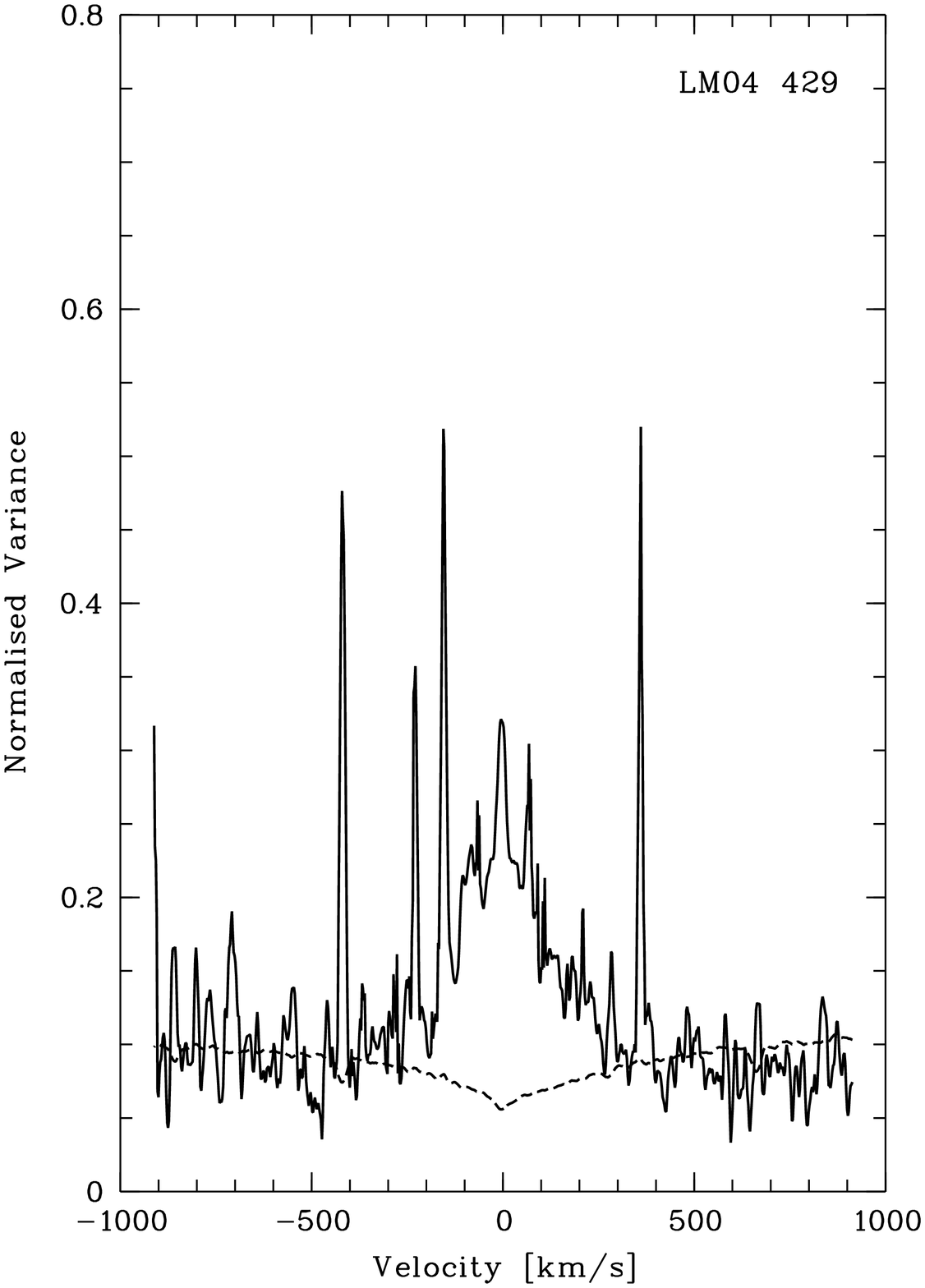}\\
\end{tabular}
 \caption{Average H$\alpha$ profiles for two objects with weak emission (left, CHXR74, ChaH$\alpha$8) and for two objects with weak emission but extended wings (right, ESOHa560 and LM04 429). Below each is their variance profile. The horizontal dashed lines in the variance profile plots represent the zero variability level allowing for signal to noise. }
\label{fig:Non-Accretor_Prof}
\end{figure*}

In Fig. \ref{fig:Accretion_Profiles} and \ref{fig:Non-Accretor_Prof} we show average line profiles for a selection of objects in our sample. Based on the shape of the H$\alpha$ profiles, the objects clearly fall in two groups. Ten of them show strong, broad H$\alpha$ emission (see Fig.~\ref{fig:Accretion_Profiles}), often asymmetric and accompanied by absorption features. In this case 7 out of 10 show obvious absorption superimposed on the emission line, 5 have blueshifted and 1 has redshifted absorption. Most of the remaining 15 objects have weak, symmetric H$\alpha$~emission as seen in the first two upper panels of Fig. \ref{fig:Non-Accretor_Prof}. A small subgroup of these objects, show broad wings in their H$\alpha$ profiles as seen in the last two upper panels of of Fig. \ref{fig:Non-Accretor_Prof}.

\indent Under each average profile in Fig. \ref{fig:Accretion_Profiles} and Fig. \ref{fig:Non-Accretor_Prof} the normalised variance profile is plotted, which can be used to assess the variability as a function of wavelength. The variance at a certain wavelength is given by:

\begin{equation}
 \sigma^{2}({\lambda}) =\frac{1}{n-1} \displaystyle\sum_{i=1}^{n}(I_{\lambda,i} - \bar{I}_{\lambda})^{2}
\end{equation}
\noindent where $I_{\lambda,i} $ is the flux at wavelength $\lambda$ for spectrum number $i$, $\bar{I}_{\lambda}$ is the average flux at that wavelength and $n$ is the total number of spectra for a given object \citep{1995AJ....109.2800J}. The normalised variance is then given by $ \sigma_{N}^{2}(\lambda) =  \sigma^{2}(\lambda)/\bar{I}_{\lambda} $.

\indent The horizontal dashed line in these plots indicates the zero variability level, which is given by
\begin{equation}
 \sigma_{N}^{2}(\lambda) = \sigma_{N,0}^{2}\left(\frac{\sqrt{\bar{I_{\lambda}}}}{\bar{I_{\lambda}}}\right)^{2}
\end{equation}
where $\sigma_{N,0}^{2}$ is the normalised variance in the continuum \citep{2007ApJ...671..842S,2006ApJ...638.1056S}.

The variance profiles in Fig. \ref{fig:Accretion_Profiles}, reveal that the variability is not uniform across the line profile. We can see that seven of these objects show multiple peaks of enhanced variations. (Two objects B43 and ChaH$\alpha$2 are shown without variance profiles, as they have weak continuum emission compared to the sky background emission).

\indent In all spectra we measured H$\alpha$ equivalent widths (EW) and 10\% widths. The 10\% width is simply the full width of the emission line at 10\% of the peak height, and is measured in km\,s$^{-1}$. The equivalent width of a line is given by 
\begin{equation}
 EW = \sum \frac{ F_{c} - F_{\lambda}}{F_{c}} \Delta\lambda
\end{equation}
where $F_{c}$ is the continuum flux and $F_{\lambda}$ is flux integrated over the line and the units of EW are \AA. We used three different integration window sizes, depending on the emission line width (25, 15, or 7.5\,\AA). The continuum was found by averaging measurements from two 10\,\AA~windows either side of the integration window. To determine the 10\% width of the H$\alpha$ line, we used the same continuum as for the EW to find the peak height above the continuum. In Table \ref{tab:Objects} we list the average and the spread in EW and 10\% width for all objects as well as the window sizes and an estimate of the measurement errors. (A brief explanation of how the measurement errors where estimated is given in Appendix \ref{sec:appendix_errors}).

Due to the low signal to noise ratio in some spectra, a number of the 10\% width measurements are considered to be upper limits. In these cases the H$\alpha$ wings are not clearly distinguishable from the continuum due to the noise. This does not affect the EW measurements in the same way, as to calculate the EW we integrate over the full line. For the objects with more measurements than upper limits, we only take the measurements in calculating the average 10\% width given in Table \ref{tab:Objects}.

As previously mentioned in Sect.\ref{s2}, for four objects in the sample, a further step in the sky subtraction was taken. When a range of sky continuum values from across the field of view was removed from these objects, it was found that their EW measurements for a single epoch varied by 20\,-\,50\AA, causing their derived accretion rate change of up to an order of magnitude. For these four objects (ISO143, ChaH$\alpha$2, ChaH$\alpha$6 and B43) we used 3\,-\,4 nearest neighbour fibres to estimate the local sky background continuum. Again by varying this local sky continuum value within the range given by the neighbouring sky fibres, we find the EW measurements stay stable for ISO143 and ChaH$\alpha$6. However objects B43 and ChaH$\alpha$2 still show very large variations in EW, and so for these objects we quote only 10\% width measurements which are not affected by low continuum counts.

\begin{table*}
 \centering
  \caption{Data for observed sample. Spectral types from \protect\cite{2007ApJS..173..104L}, SED classes from \protect\cite{2008ApJ...675.1375L}. Coordinates are for epoch J2000. EW and 10\% width measurements listed are the averages from all 12 epochs. The $\Delta$EW and $\Delta$10\% width are the maximum - minimum measurements for the 12 epochs, unless otherwise indicated. Integration Window is the window over which the EW and 10\% widths are calculated. Error estimates are given for EW and 10\% width  measurements (See Appendix \ref{sec:appendix_errors} for more). Sky subtraction method is indicated in the second last column, M indicates that the the mean sky continuum value was removed for that object, NN indicates that the nearest neighbour fibre method was used to estimate the local sky continuum for removal, see Sect.\ref{s2} for details. Note: Sky subtraction was not possible for B43 or ChaH$\alpha$2 so measurements are not given.  }
  \begin{tabular}{@{}lccclllllllll@{}}
  \hline
          & Spectral& SED &$\alpha$&$\delta$                &  H$\alpha$ EW & $\Delta$EW&  Error   &H$\alpha$ 10\%& $\Delta$10\% &Error & Sky & Integration     \\
   Object &  Type   & Class &  [hh:mm:ss]&  [$\degree$ ' '']&   (\AA)       &         &             &(km\,s$^{-1}$) &             &      & Sub.   &Window  (\AA)   \\
\hline 	                         
CHXR28		&	K6	 & III   & 11:07:55.8	&-77 27 25.5	&   0.6  &   0.8 & (0.1) & \textless194&-         & (36) & M & 7.5         \\
CHXR20		&	K6	 & II    & 11:06:45.1	&-77 27 02.3 	&   0.3 *&   0.7 & (0.1) &\textless255 &  -       & (24) & M & 7.5         \\
T22		&	M3	 & III   & 11:06:43.5 	&-77 26 34.5	&   3.0  &   1.3 & (0.1) & 419$^{*}$   &  384     & (29) & M & 15         \\
ISO143		&	M5	 & II    & 11:08:22.4	&-77 30 27.7	& 138.2* & 172.2 &(13.6) & 395         &   66     & (5)  &NN & 25         \\
T33A 		&	G7	 & II    &  11:08:15.2	&-77 33 53.1	&  25.9  &  19.7 &(0.4)  & 433   $^{*}$&  158     & (1)  & M & 15         \\
T39A		&	M2	 & III   & 11:09:11.6	&-77 29 12.7	&   5.3  &   4.2 &(0.1)  & 197         &   86     & (3)  & M & 15         \\
ChaH$\alpha$2		&	M5.25	 & II    & 11:07:42.5   &-77 33 59.3	&  -     &  -    & -     & 349         &   74     & (6)  &NN & 25          \\ 
B43		&	M3.25	 & II    & 11:09:47.4	&-77 26 29.0	&     -  & -     &   -   & 398 *       &   68     & (5)  &NN & 25           \\
T45		&	M1.25	 & II    & 11:09:58.7	&-77 37 09.0	& 104.4  &  64.8 &(1.3)  & 506         &   69     & (1)  & M & 25          \\
ChaH$\alpha$6		&	M5.75	 & II    & 11:08:39.5	&-77 34 16.6	&  81.7  &  55.7 &(3.8)  & 350         &  118     & (6)  & NN& 25          \\
ISO126		&	M1.25	 & II    & 11:08:2.97	&-77 38 42.5	&  79.9  &  72.1 &(4.0)  & 386         &   68     & (1)  & M & 25          \\
ChaH$\alpha$8		&	M5.75	 & III   & 11:07:46.1	&-77 40 08.9	&   6.7  &   9.1 &(1.3)  & 170*        &  108     & (24) & M & 15          \\
ChaH$\alpha$5		&	M5.5	 & III   & 11:08:24.1	&-77 41 47.3	&  7.3*  &  4.5  &(1.4)  & 168*        &  116     & (12) & M & 15          \\
ESOHa566	&	M5.75	 & III   & 11:09:45.2	&-77 40 33.2	&  7.5*  &  3.1  &(1.6)  & 158*        &   75     & (3)  & M & 15          \\
T30		&	M2.5	 & II    & 11:07:58.1	&-77 42 41.3	&  59.0  &  53.9 &(2.2)  & 561         &  213     & (1)  & M & 25          \\
T34		&	M3.75	 & III   & 11:08:16.5	&-77 44 37.2	&   3.0  &   2.6 &(0.2)  & 194*        &   78     & (11) & M & 15          \\
ChaH$\alpha$3		&	M5.5	 & III   & 11:07:52.3	&-77 36 56.9	&   6.9  &  10.6 &(0.5)  & 222*        &  146     & (12) & M & 15          \\
T26		&	G2	 & II    & 11:07:20.7	&-77 38 07.3	&   20.7 &   7.6 &(0.2)  & 430         &   90     & (1)  & M & 25          \\
ESOHa560	&	M4.5	 & III   & 11:07:38.3	&-77 47 16.8	&   10.2 &   8.0 &(0.7)  & 526*        &  179     & (7)  & M & 15          \\
LM04 429	&	M5.75	 & III   & 11:07:24.4	&-77 43 48.9	&   11.7*&  11.2 &(1.6)  &\textless618 & -        & (23) & M & 25          \\
CHXR22E		&	M3.5	 & II    & 11:07:13.3	&-77 43 49.8	&   9.5* &   8.1 &(0.7)  & 412*        &  357     & (37) & M & 25          \\
CHXR21		&	M3	 & III   & 11:07:11.5	&-77 46 39.4	&   9.1* &   7.5 &(0.2)  & 425*        &  211     & (4)  & M & 25          \\
T31		&	K8	 & II    & 11:08:01.5	&-77 42 28.8	&  68.9  &  36.7 &(0.4)  & 470         &  214     & (1)  & M & 25          \\
CHXR76		&	M4.25	 & III   & 11:07:35.2	&-77 34 49.3	&  12.6  &   6.4 &(1.3)  & 153         &   58     & (2)  & M & 15          \\
CHXR74		&	M4.25	 & III   & 11:06:57.3	&-77 42 10.6	&   8.6  &   3.5 &(0.4)  & 174*        &   74     & (1)  & M & 15          \\
                                                                                            
\hline
\multicolumn{13}{l}{$^{*}$ In these cases the mean is not based on 12 measurements. See Tables \ref{tab:All_EW_values} and \ref{tab:All_Obs_10width_values}) for more.}
\end{tabular}
\label{tab:Objects}
\end{table*}

\begin{figure*}
\centering
\begin{tabular}{ll}
\includegraphics[scale=0.33,angle=270]{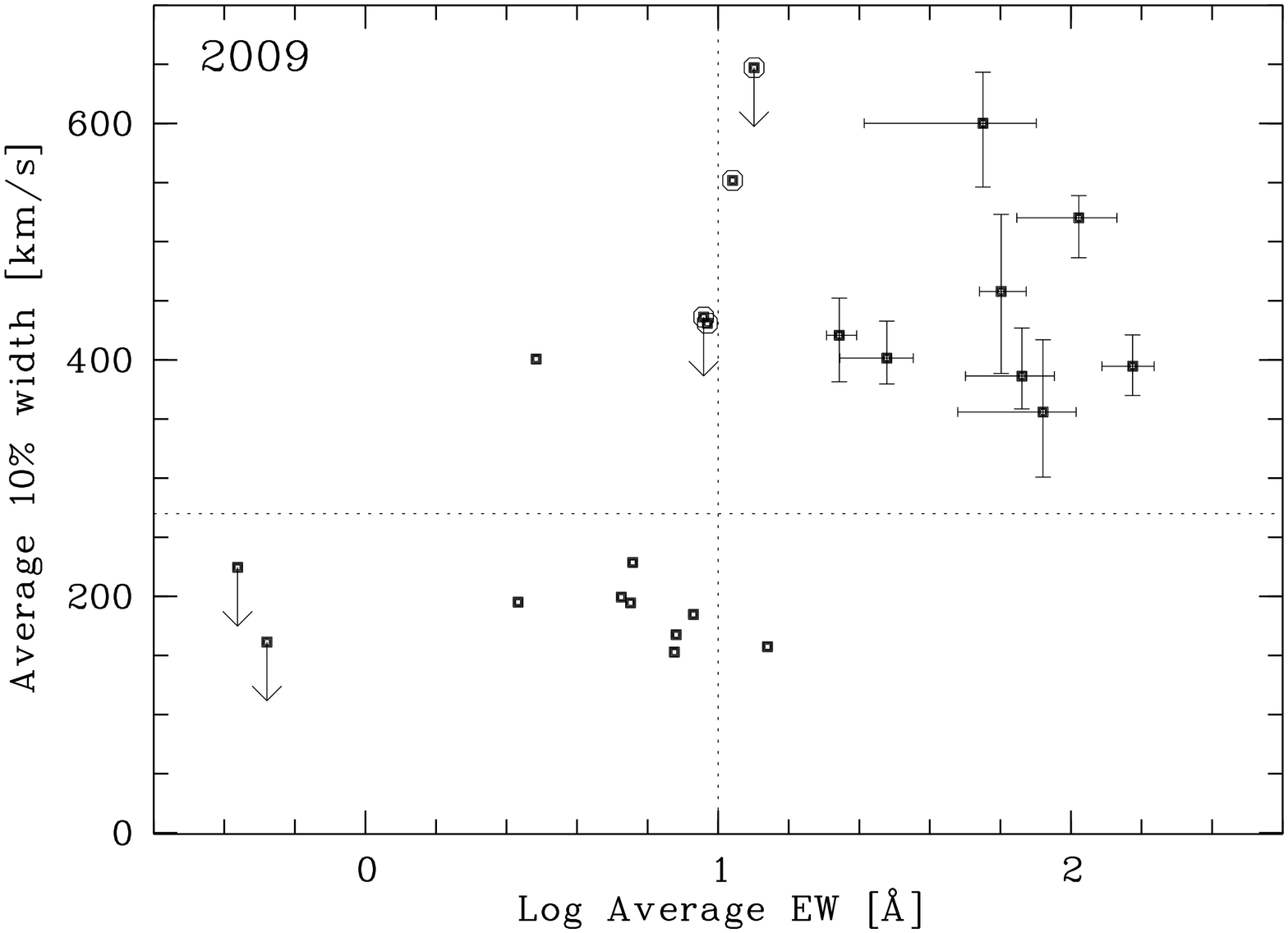}
\includegraphics[scale=0.33,angle=270]{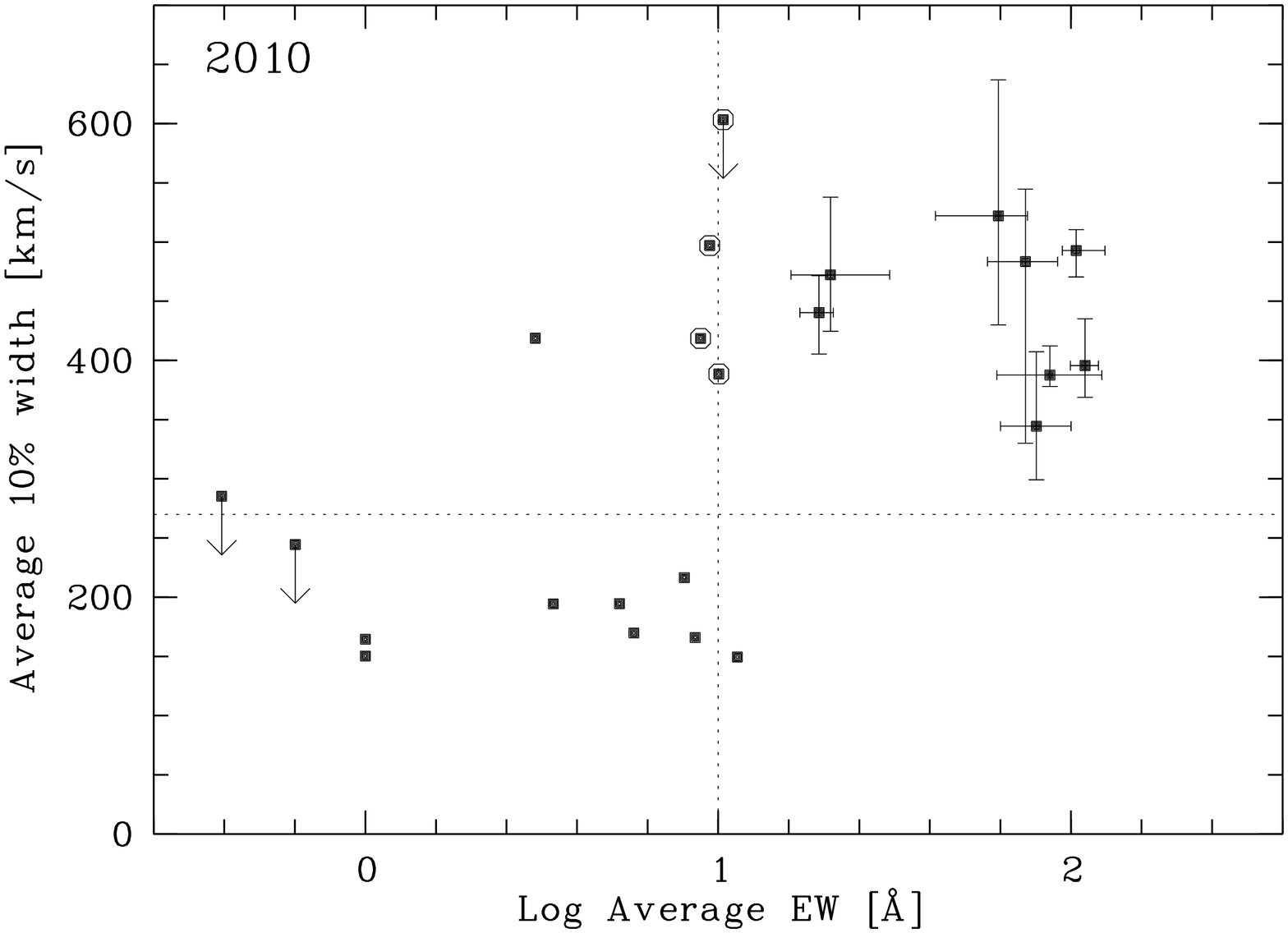}\\
\includegraphics[scale=0.33,angle=270]{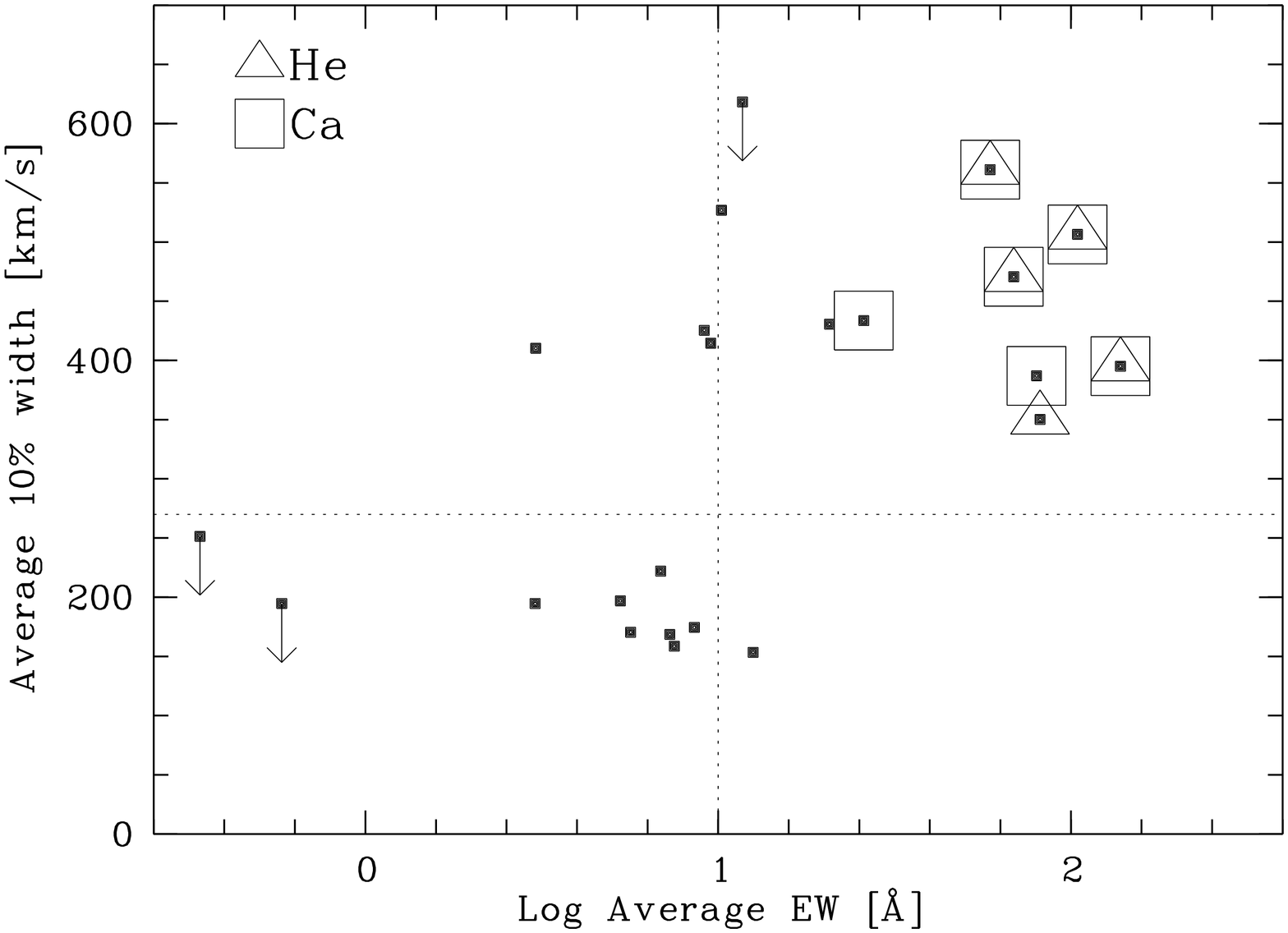}
\includegraphics[scale=0.33,angle=270]{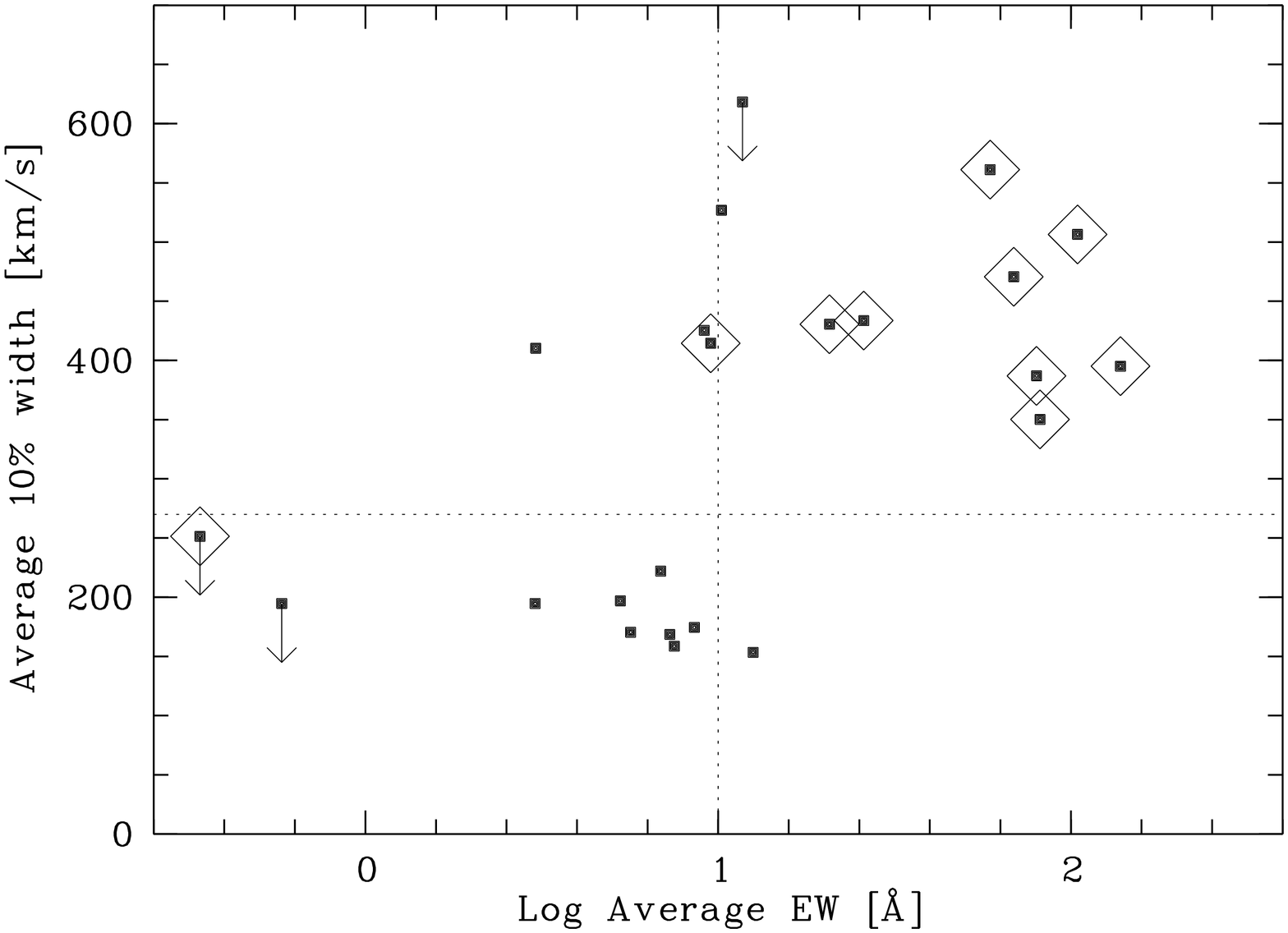} \\
\end{tabular}
 \caption{\textbf{Two Upper Panels:} Average H$\alpha$ 10\% width versus H$\alpha$ EW for the first and second observation periods on the left and right respectively. The horizontal line is at 270 km\,s$^{-1}$ and the vertical line is at 10\,\AA, which are the standard limits to differentiate between WTTs and CTTs. The spread (max - min) in measurements for objects with the broadest H$\alpha$ profiles are indicated by the over-plotted horizontal and vertical bars. The arrows pointing down indicate upper limits in 10\% width. The ringed objects have weak but borad H$\alpha$ emission and are dicussed in Sect. \ref{sec:subsection_accretion}. \textbf{Lower Left Panel:} Average H$\alpha$ 10\% width and EW for all observations, showing objects with Helium emission (triangles) and Calcium emission (squares). \textbf{Lower Right Panel:} Average H$\alpha$ 10\% width and EW for all observations where objects marked with diamonds are Class II YSOs according to \citet{2008ApJ...675.1375L}.}
\label{fig:10EW}
\end{figure*}

\indent In the top two panels of Fig. \ref{fig:10EW} we plot the average 10\% width against EW for all 23 well measured objects, one panel for each observing period. 

 Over-plotted on these two panels is the often adopted cut off between CTTs and Weak lined T Tauri stars (WTTs), an EW of 10\,\AA~and a 10\%~width of 270\,km\,s$^{-1}$ \citep{2003ApJ...582.1109W}. \cite{2003AJ....126.2997B} have introduced a more precise CTTs/WTTs division based on spectral type: for K6 stars it is 5.9\,\AA, for M1 stars it is 10.1\,\AA~and for M6 objects 24.1\,\AA. The mean spectral type in our sample is M1, making the 10\,\AA~a reasonable value to take, as the three K and two G type stars in our sample, either have very high EW or an EW well below the cut off for their spectral type.

Eight objects stand out with EW $>20$\,\AA~and 10\% width $>350$\,km\,s$^{-1}$ (two objects, B43 and ChaH$\alpha$2 have 10\% width $>270$\,km\,s$^{-1}$ but no EW measurements). They appear in the upper right quadrant and coincide with the objects found to have highly variable, mostly asymmetric line profiles (see above). Defining the spread as the maximum - minimum value measured during one observation season, the spread in H$\alpha$ EW and 10\% width measurements are indicated by the over-plotted horizontal and vertical bars. Although variability is clearly present, the objects in the right upper quadrant have continuous strong H$\alpha$ emission over both observation periods and remain a separate group. Objects with 10\% width $<270$\,km\,s$^{-1}$ also show some variations but are not shown in the plot for clarity.

\indent The subgroup of objects mentioned above with weak H$\alpha$ emission but broad wings, are ringed in the upper plots of Fig. \ref{fig:10EW}, with EW$<20$\,\AA~and 10\% width\,$>270$\,km\,s$^{-1}$. These objects show significant variability over the course of the observations; in two cases their 10\% widths fall below $270$\,km\,s$^{-1}$ in some epochs. 

It is worth noting that the changes in EW and 10\% width both show erratic behaviour, with no recognisable pattern across the time period of our observations. However it is possible that the temporal resolution of our observations was not high enough to resolve any periodic or quasi-periodic variations in these quantities.

\subsection{Other emission lines}

In the spectra for which He\,I (6678.2\,\AA) and Ca\,II (8662.1\,\AA) emission lines were detected, we measured the EW of these lines. For the He\,I line we used a 6\,\AA~integration window and the average of two 10\,\AA~windows on either side of the integration window to measure the continuum. For Ca\,II we use an integration window of either 12\,\AA~or 20\,\AA~for the stronger lines, and the average of three different continuum measurements, two 10\,\AA~windows centred at 8605 and 8585\,\AA~and a 20\,\AA~window centred at 8670\,\AA. Table \ref{tab:HaHeCa} shows the EW measurements for both lines. Estimates of the measurement errors for both lines are also given in Table \ref{tab:HaHeCa}. (See Appendix \ref{sec:appendix_errors} for an explanation of how these were derived).
\begin{table*}
 \centering
  \caption{Emission Line Measurements for objects with either He or Ca emission in their spectra. EW measurements listed are the averages from each observation period, $\Delta$EW is the maximum - minimum measurement in an observation period. Estimate of measurement errors are given in parentheses (See Appendix \ref{sec:appendix_errors} for more). }
  \begin{tabular}{@{}lcccccccccc@{}}
  \hline
&\multicolumn{5}{|c|}{Ca\,II Emission} &  & \multicolumn{3}{|c|}{He\,I Emission} \\
   Object & Obs.1 EW & $\Delta$Ca EW &  Obs. 2 EW &  $\Delta$ EW &Error  & & Obs.1 EW &  Obs.2 EW & Error &  \\
          & [\AA] &  & [\AA] &  & & & [\AA] &  [\AA] &  & \\

\hline
ISO143    &      4.6   &  5.0    &     3.2   &   3.5  & (0.30)&  &   7.7   &    10.0 &  (3.00)  \\  
T33A      &      2.0*  &  2.5    &     -     &   -    & (0.07)&  &   -     &    -    &   -      \\
ChaH$\alpha$2    &      -     &  -      &      -    &     -  & -     &  &     -   &     -   &  -       \\
B43       &      12.4  &  15.8   &     19.3  &   7.0  & (0.90)&  &     X   &     -   &   -      \\
T45       &      21.6  &  24.7   &     19.8  &   16.3 & (0.05)&  &    1.2  &    1.1  &  (0.05)  \\
ChaH$\alpha$6    &       -    &   -     &     -     &   -    & -     &  &   1.4   &    1.3  &  (0.20)   \\
ISO126    &      6.2   &  5.6    &     6.7   &   4.3  & (0.10)&  &   -     &     -   &   -       \\
T30       &      11.4  &  17.1   &     7.4   &   12.4 & (0.10)&  &   1.7   &   1.5   &  (0.20)   \\
T26       &        -   &    -    &     -     &    -   & -     &  &   -     &     -   &       -   \\
T31       &      3.3   &  6.2    &     7.4   &   8.4  & (0.01)&  &   0.6   &   1.0   &  (0.06)  \\
                                                           
\hline

\multicolumn{8}{l}{$^{*}$ : Ca\,II in emission in only four epochs for this object.} \\
\multicolumn{8}{l}{X : He\,II present in spectra, but no continuum for measurements.} \\
\end{tabular}
\label{tab:HaHeCa}
\end{table*}

\begin{figure*}
\begin{tabular}{ccccc}
\includegraphics[scale=0.180]{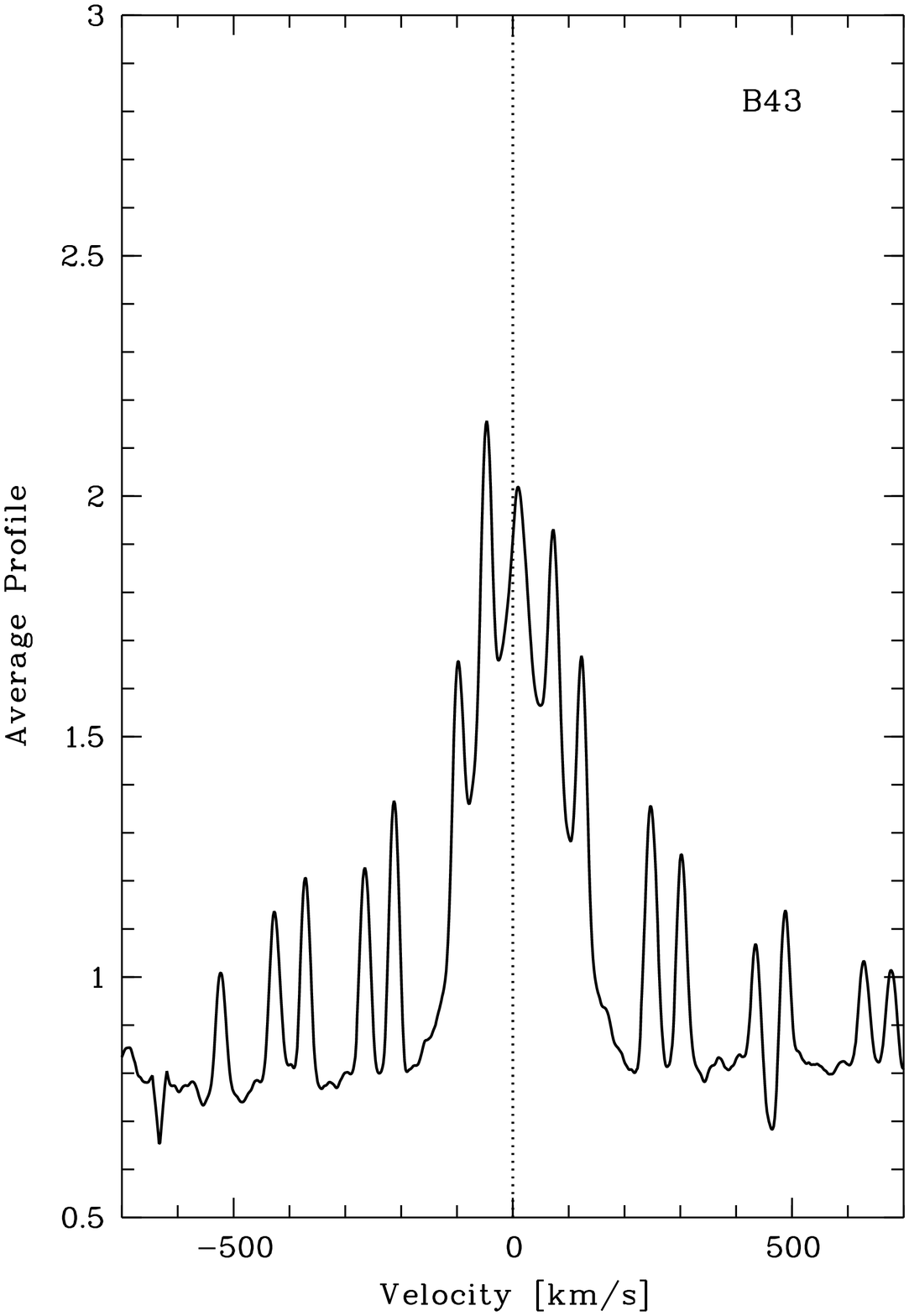}
\includegraphics[scale=0.180]{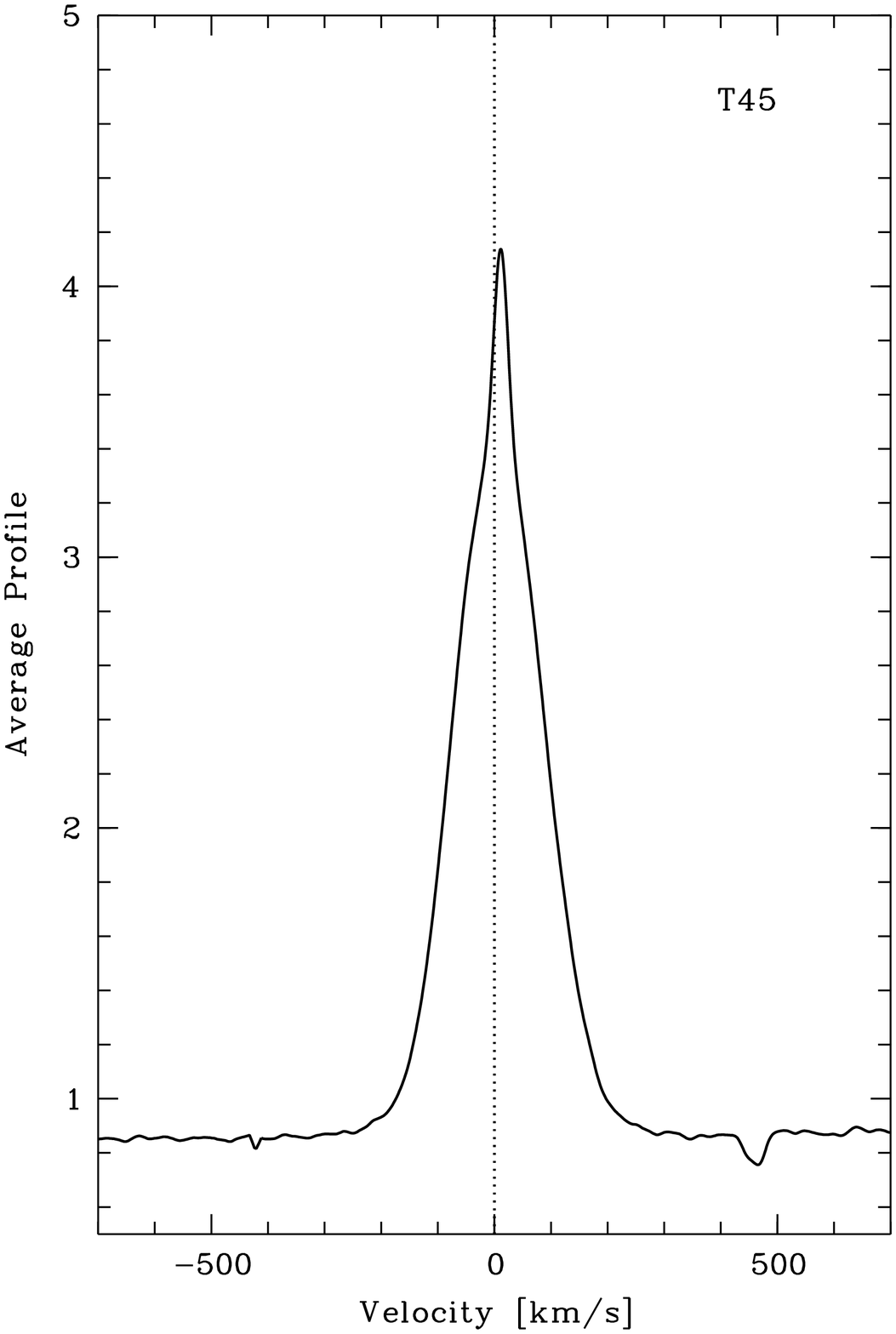}
\includegraphics[scale=0.180]{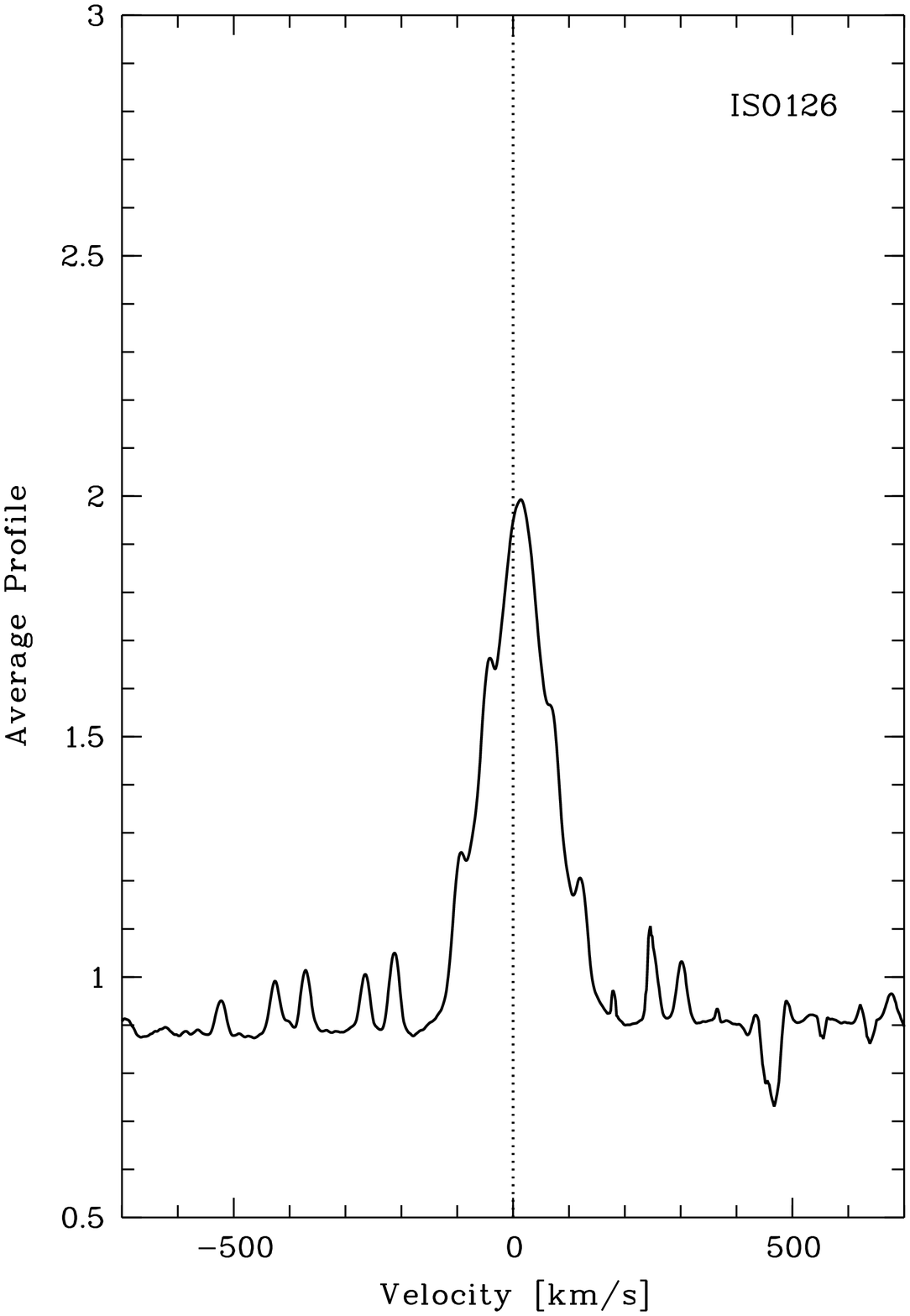}
\includegraphics[scale=0.180]{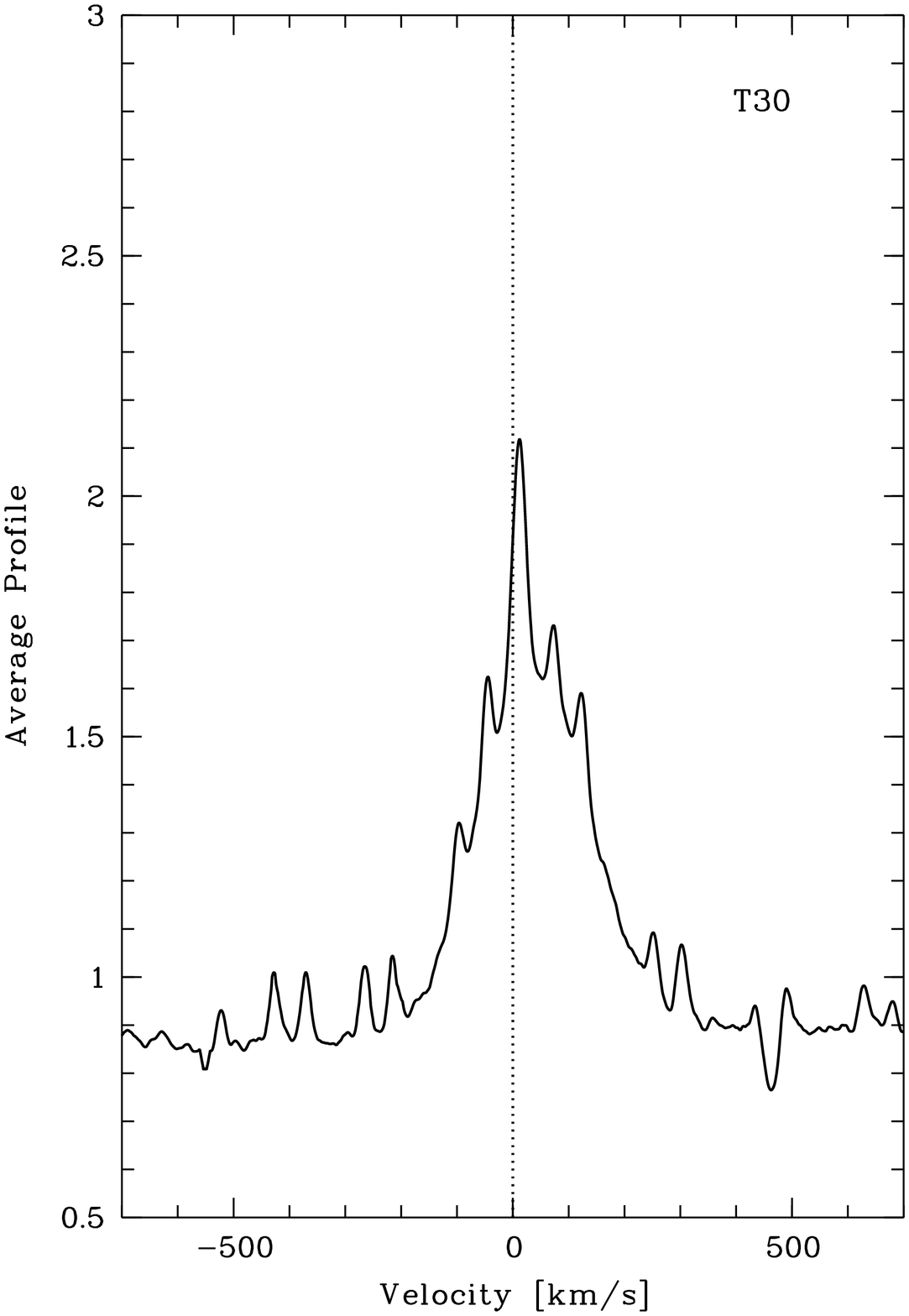}
\includegraphics[scale=0.180]{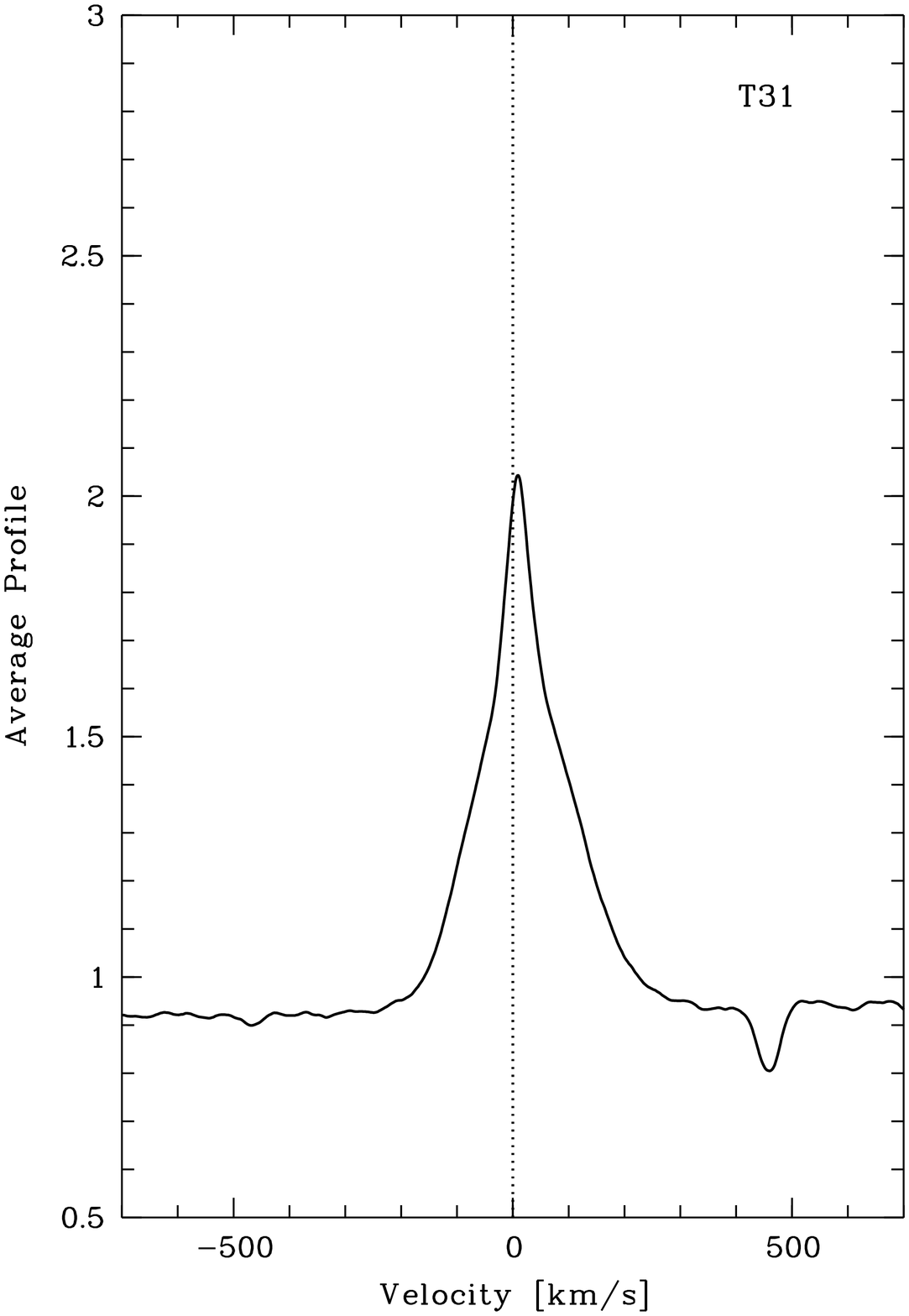}\\

\includegraphics[scale=0.180]{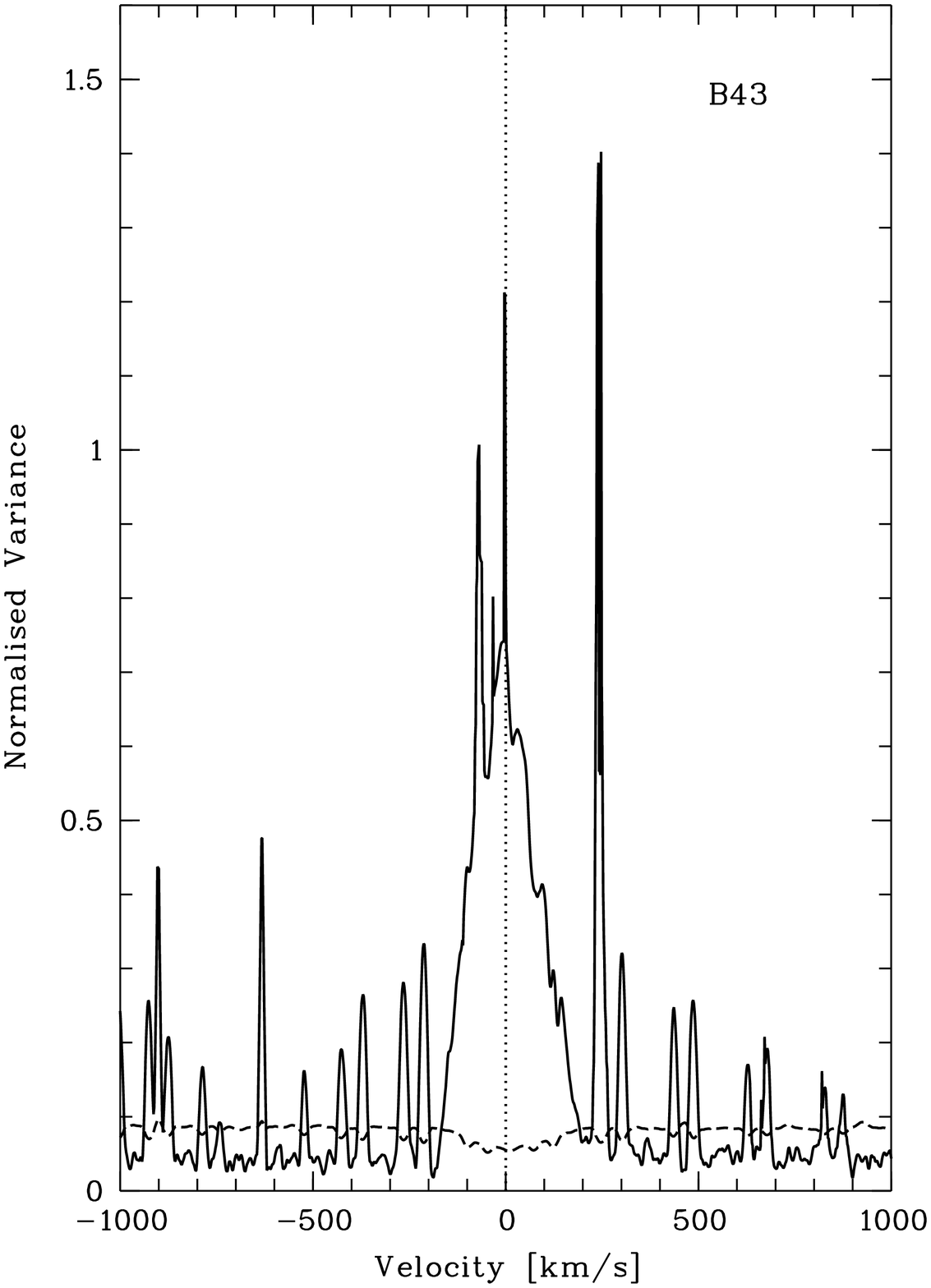}
\includegraphics[scale=0.180]{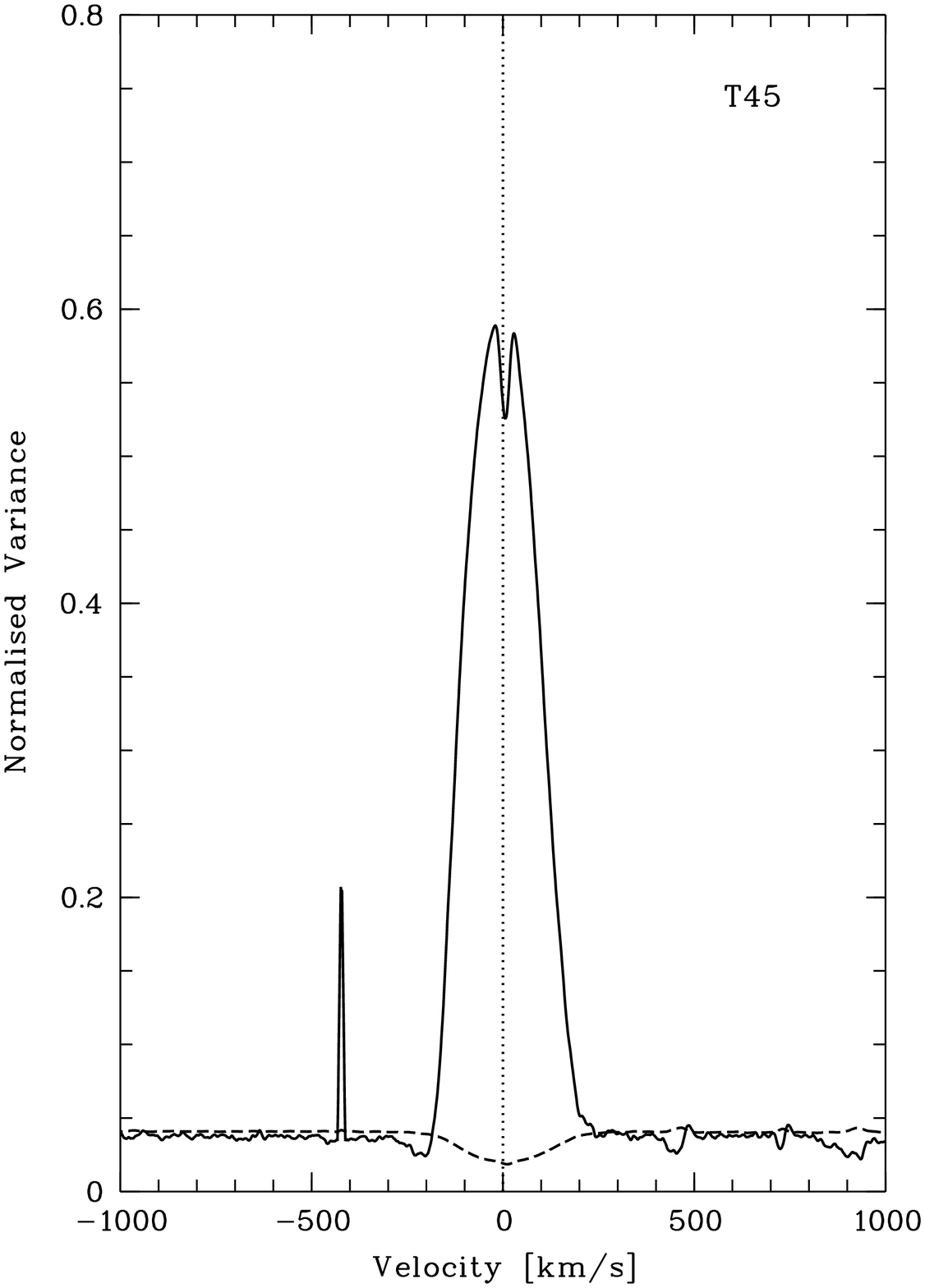}
\includegraphics[scale=0.180]{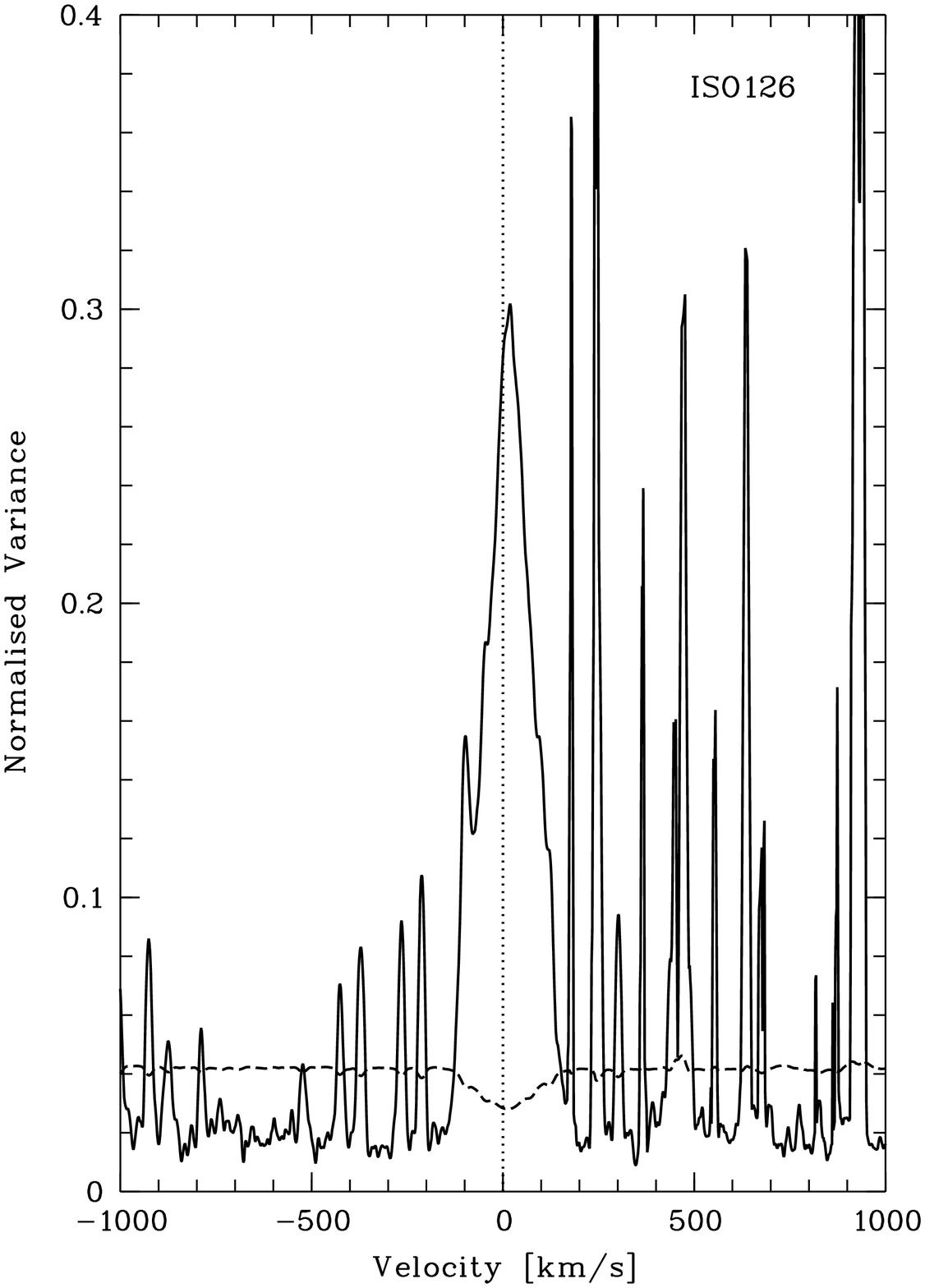}
\includegraphics[scale=0.180]{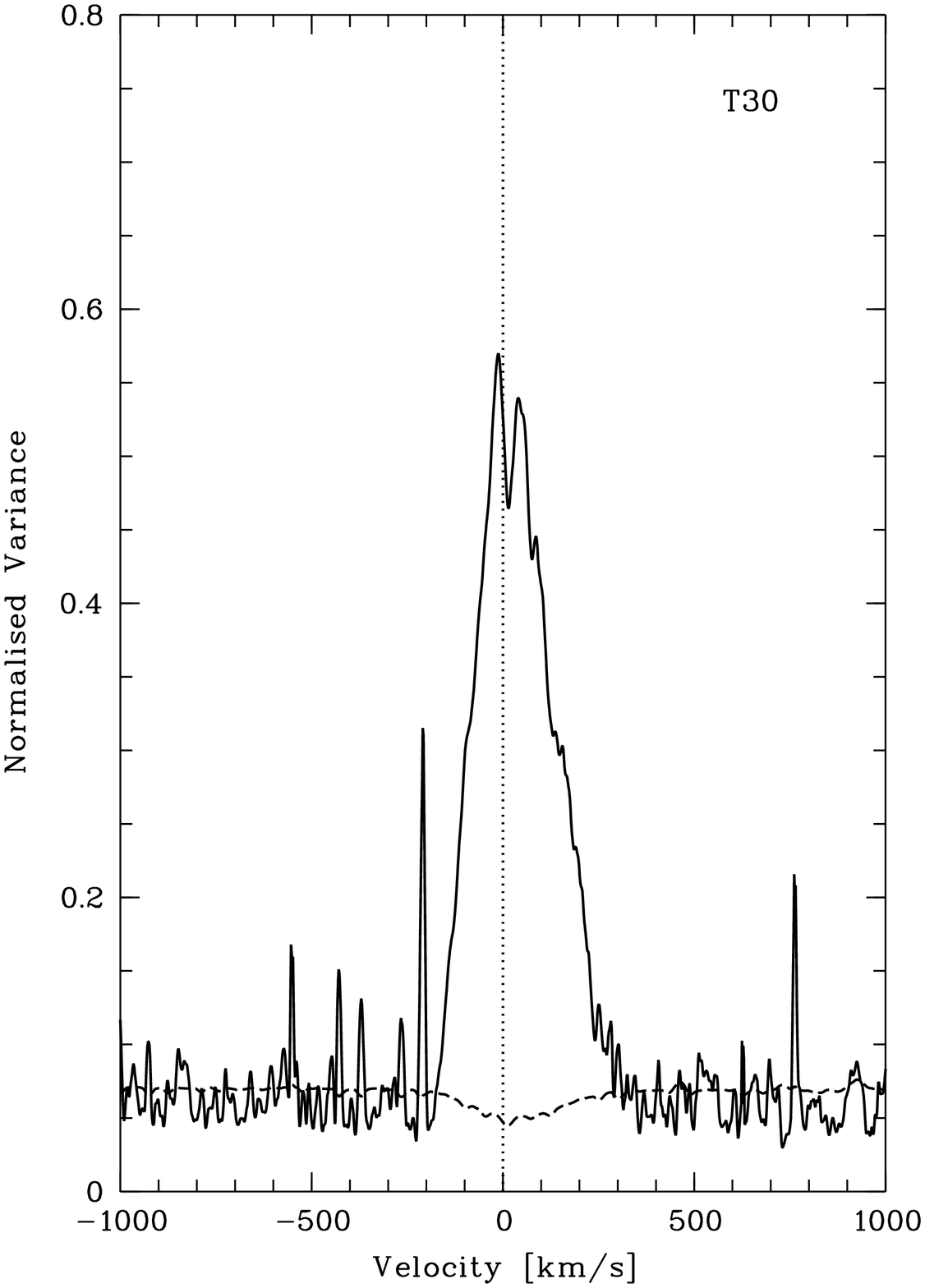}
\includegraphics[scale=0.180]{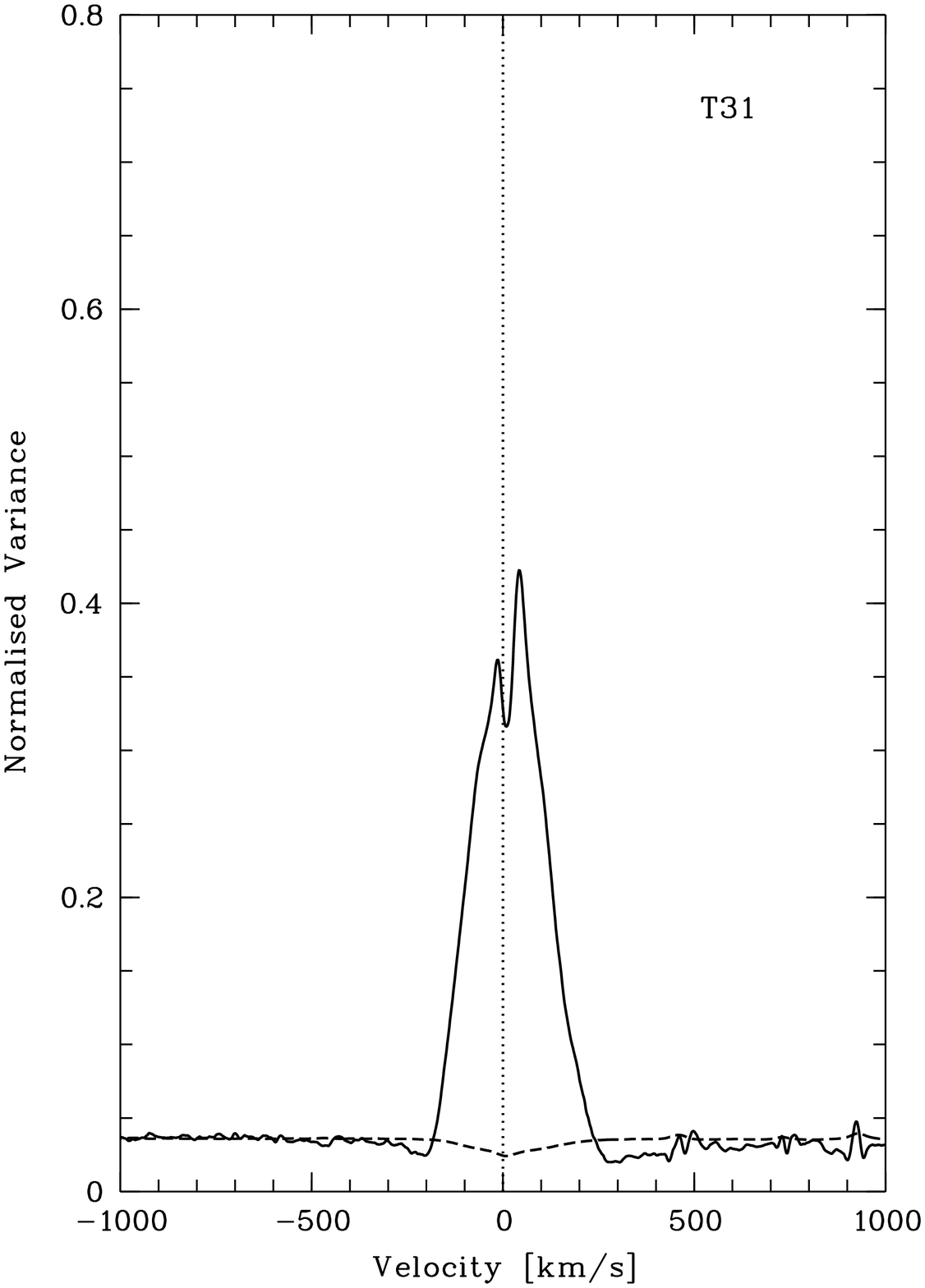}\\

\includegraphics[scale=0.180]{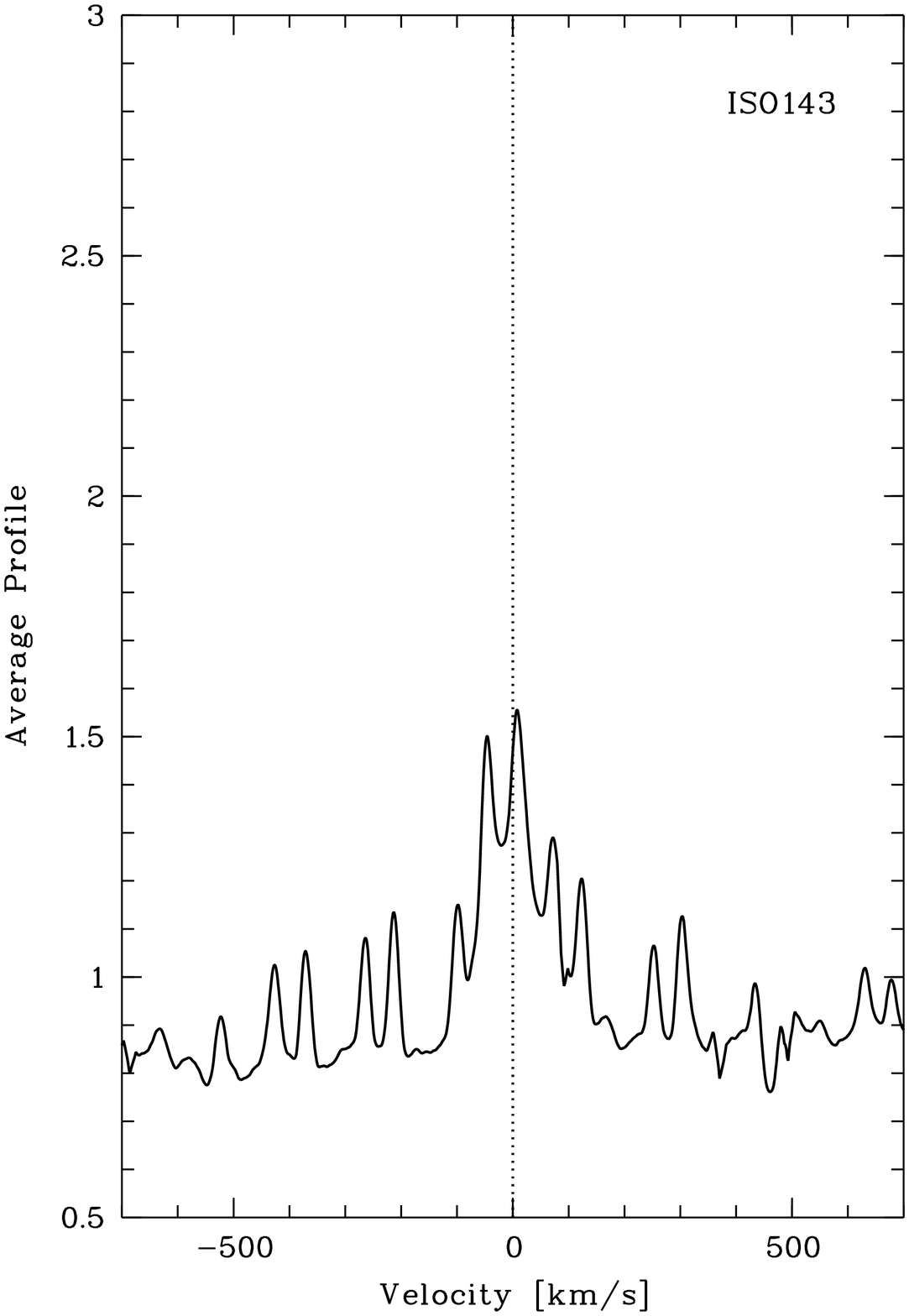} 
\includegraphics[scale=0.180]{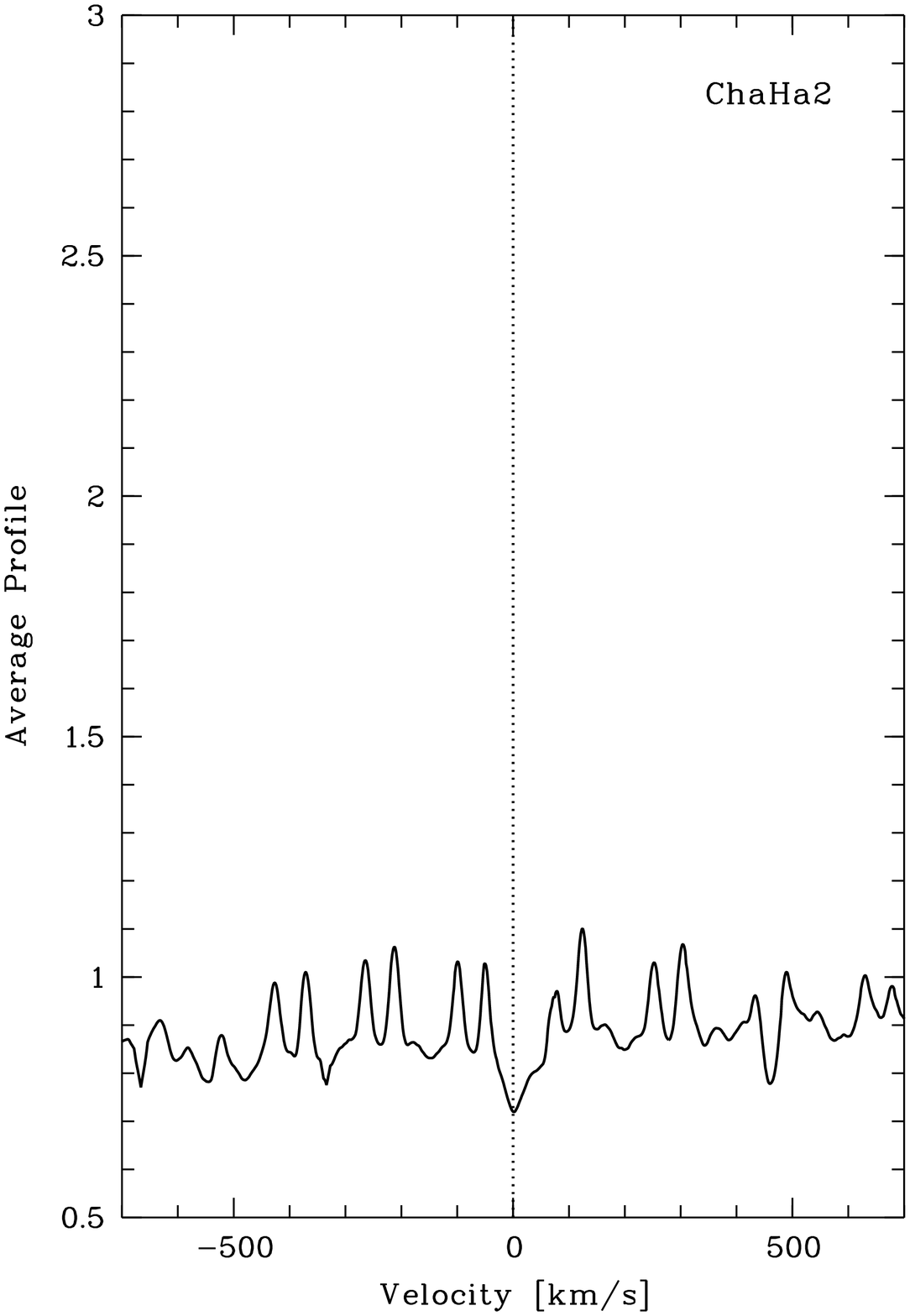} 
\includegraphics[scale=0.180]{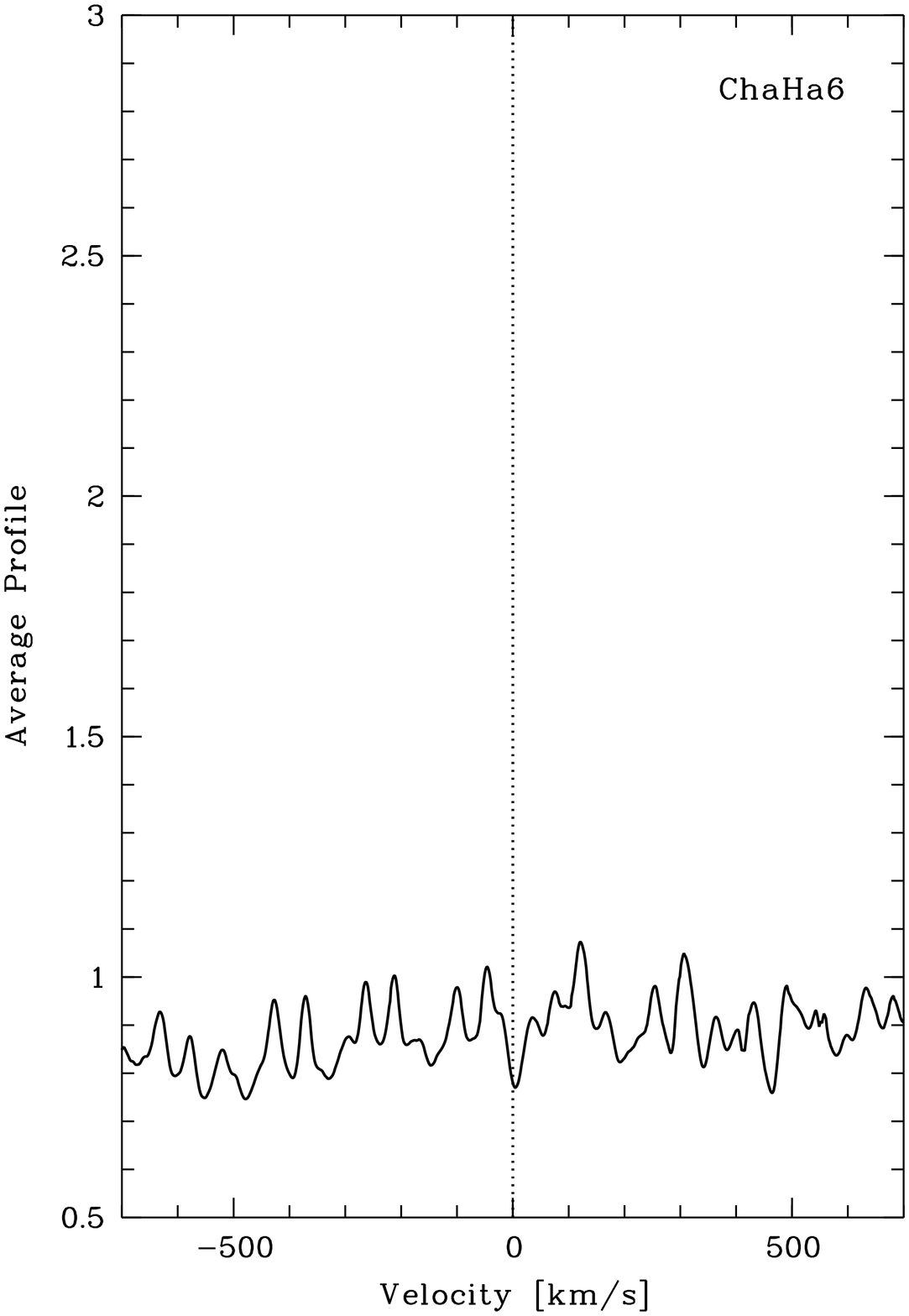}
\includegraphics[scale=0.180]{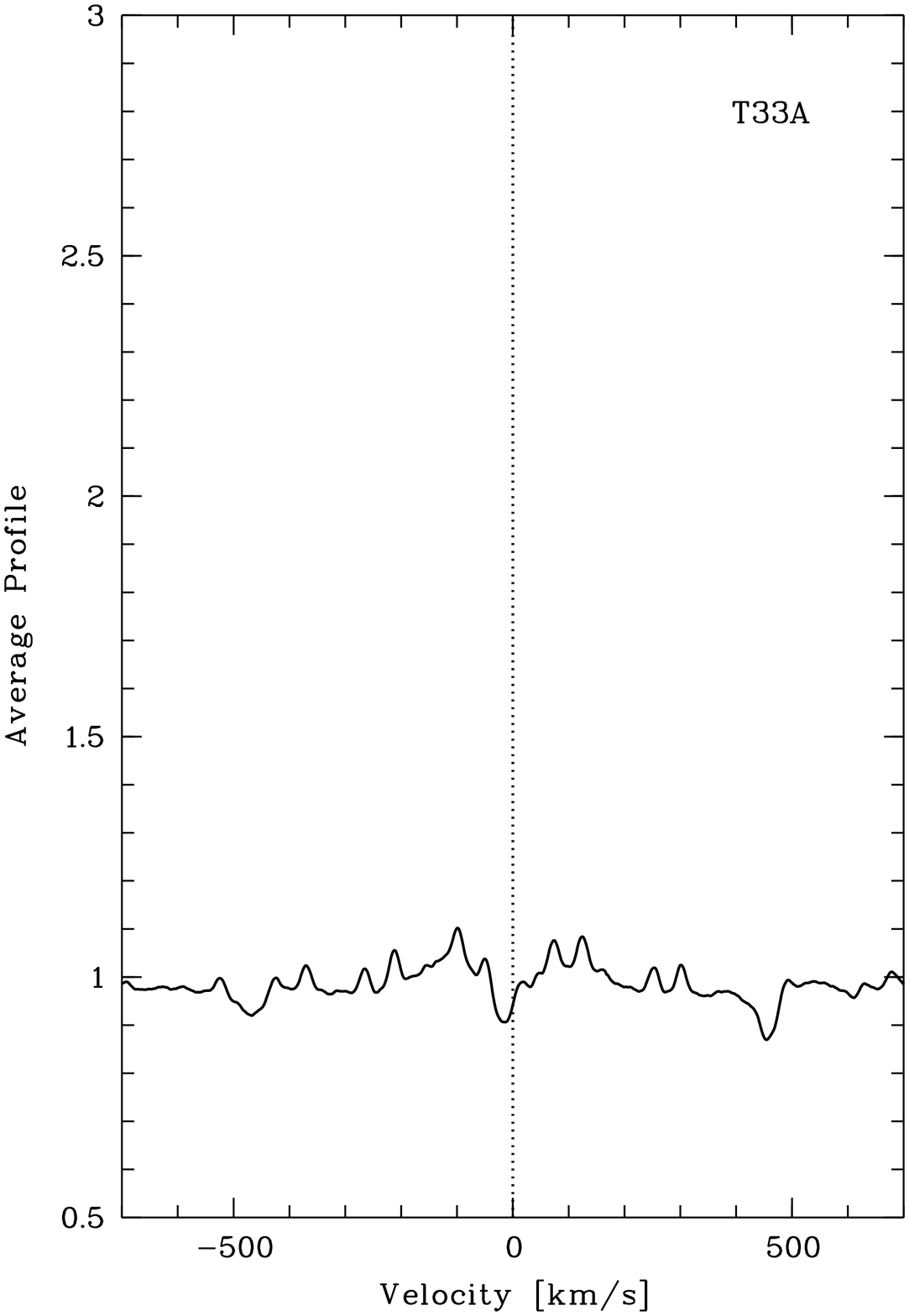} 
\includegraphics[scale=0.180]{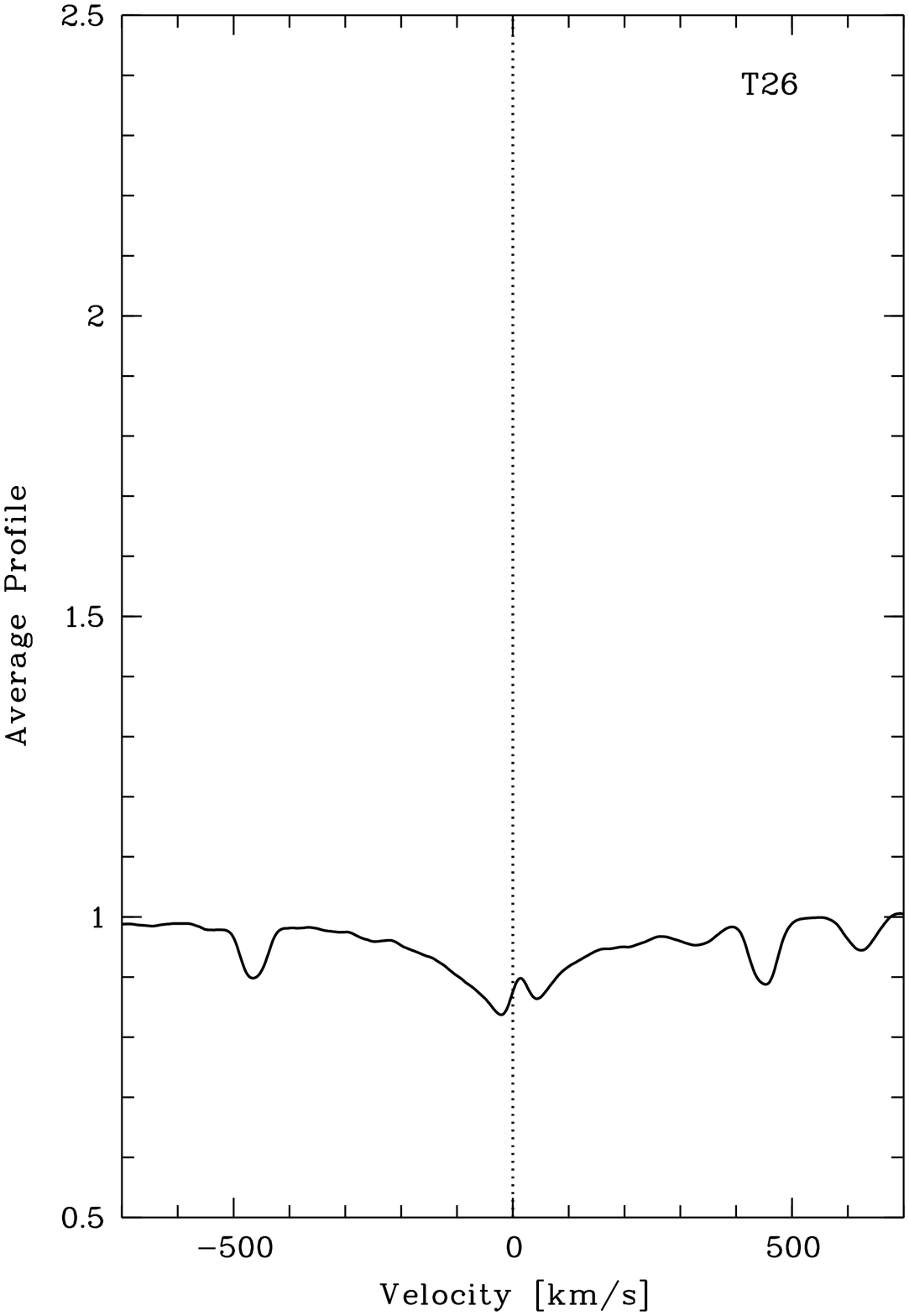}\\
\end{tabular}
\caption{Average and Variance profiles for Ca\,II emission in the 10 accretors. Variance profiles are only shown for objects with strong Ca\,II emission. Note for T33A, the average profile does not show any Ca\,II emission, but this object does show weak Ca\,II emission in four epochs. }
\label{fig:Ca_Profiles}
\end{figure*}

\begin{figure}
\begin{tabular}{ccccc}
\includegraphics[scale=0.2]{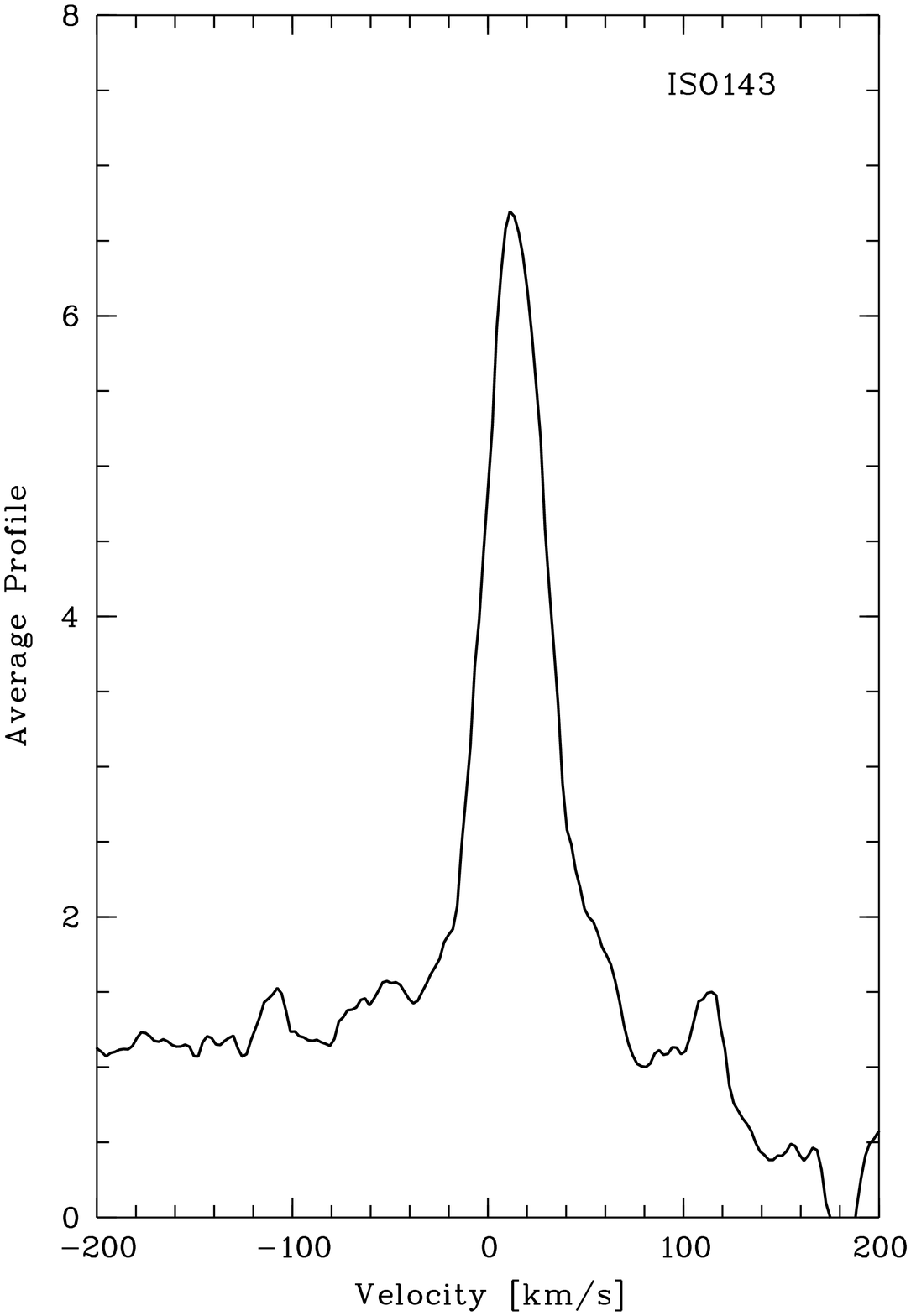}
\includegraphics[scale=0.2]{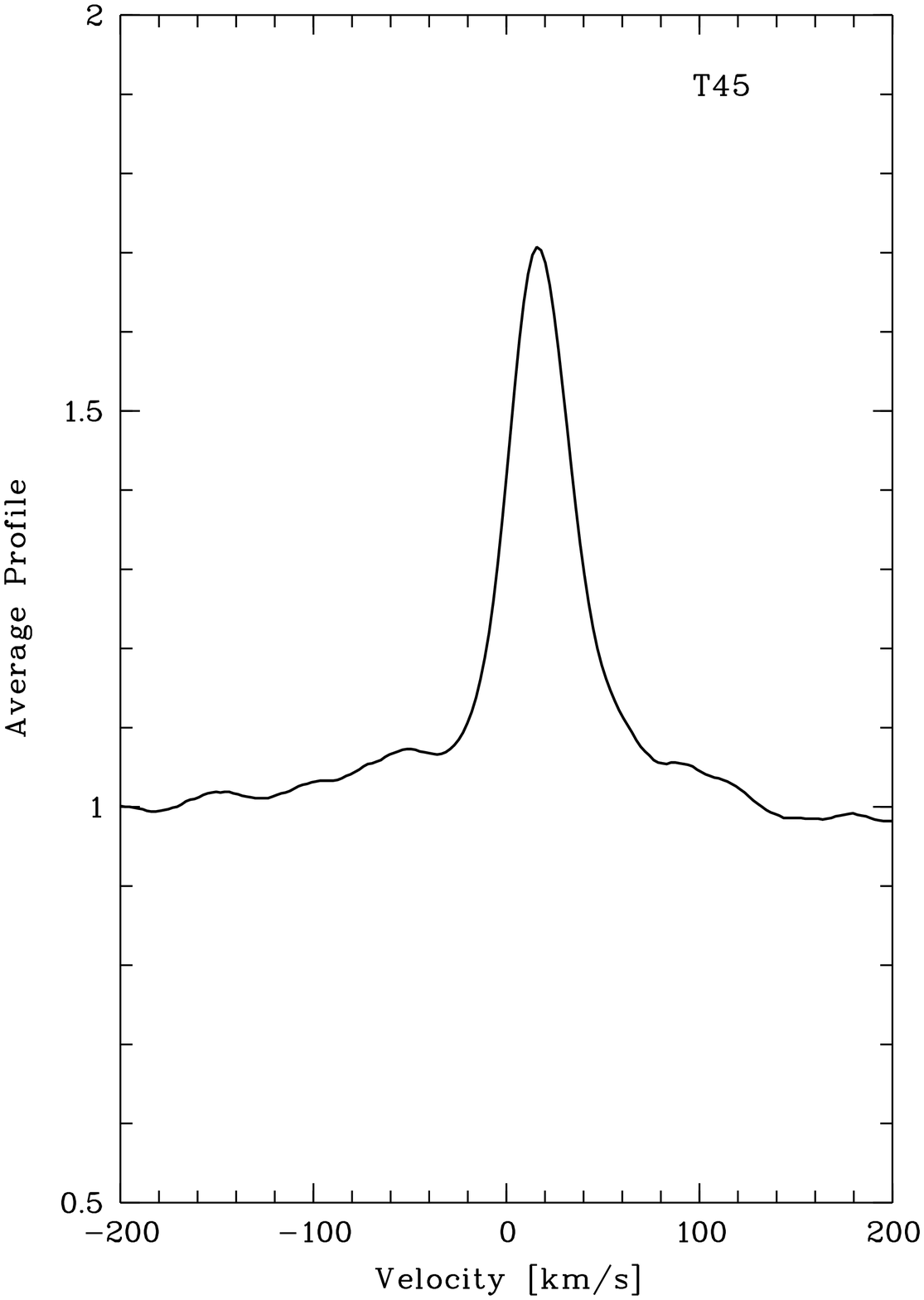}\\
\includegraphics[scale=0.2]{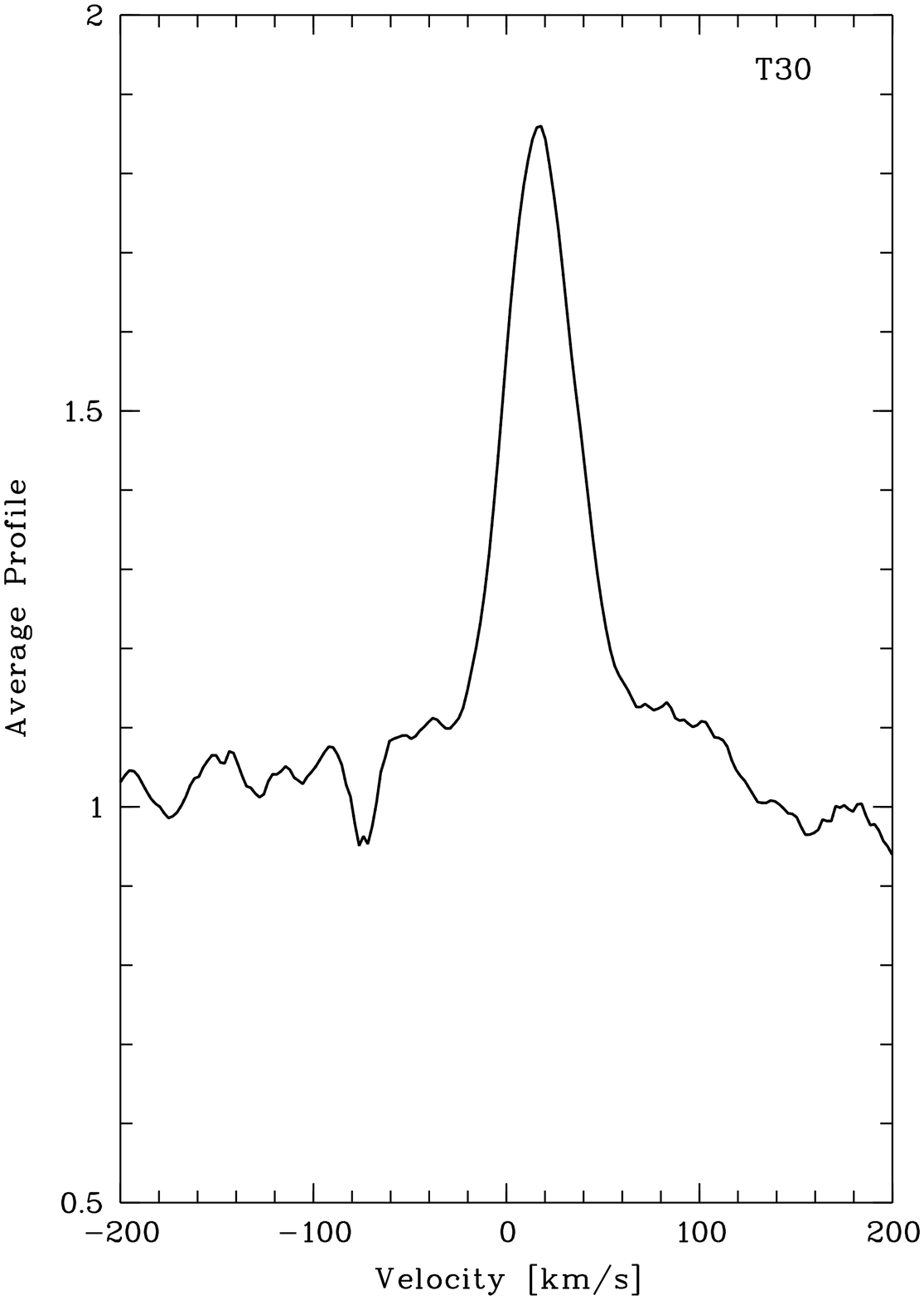}
\includegraphics[scale=0.2]{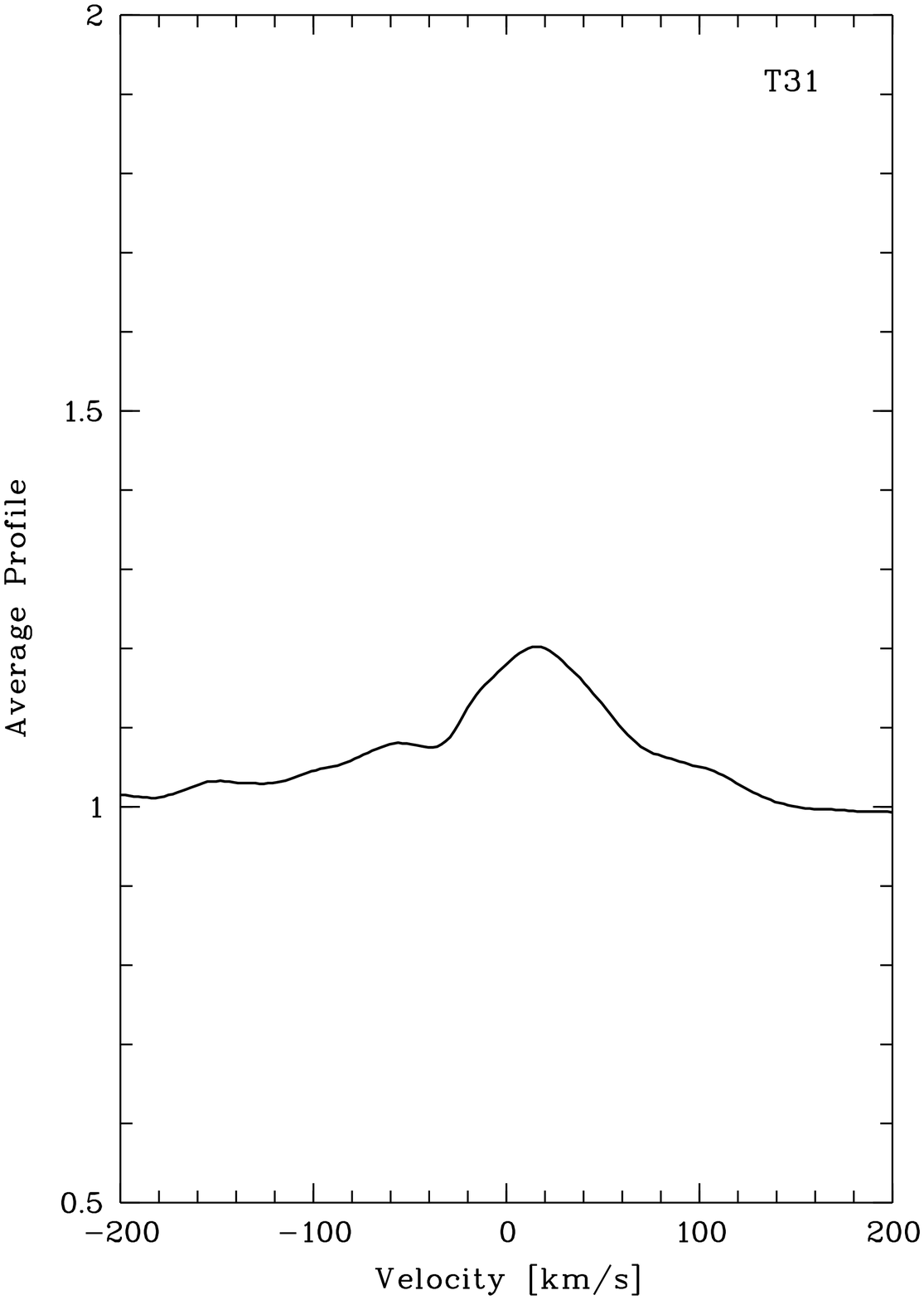}\\

\includegraphics[scale=0.2]{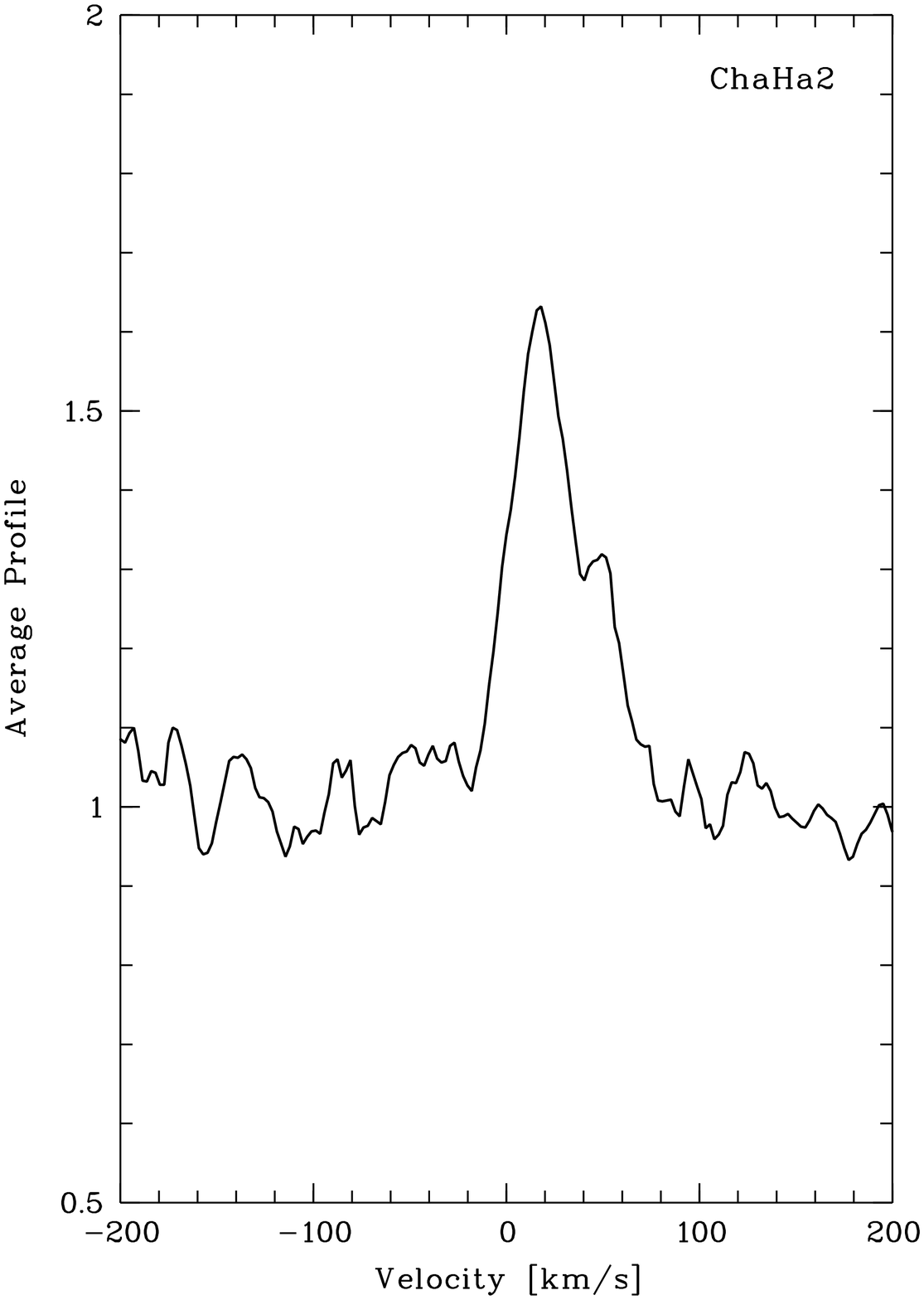} 
\includegraphics[scale=0.2]{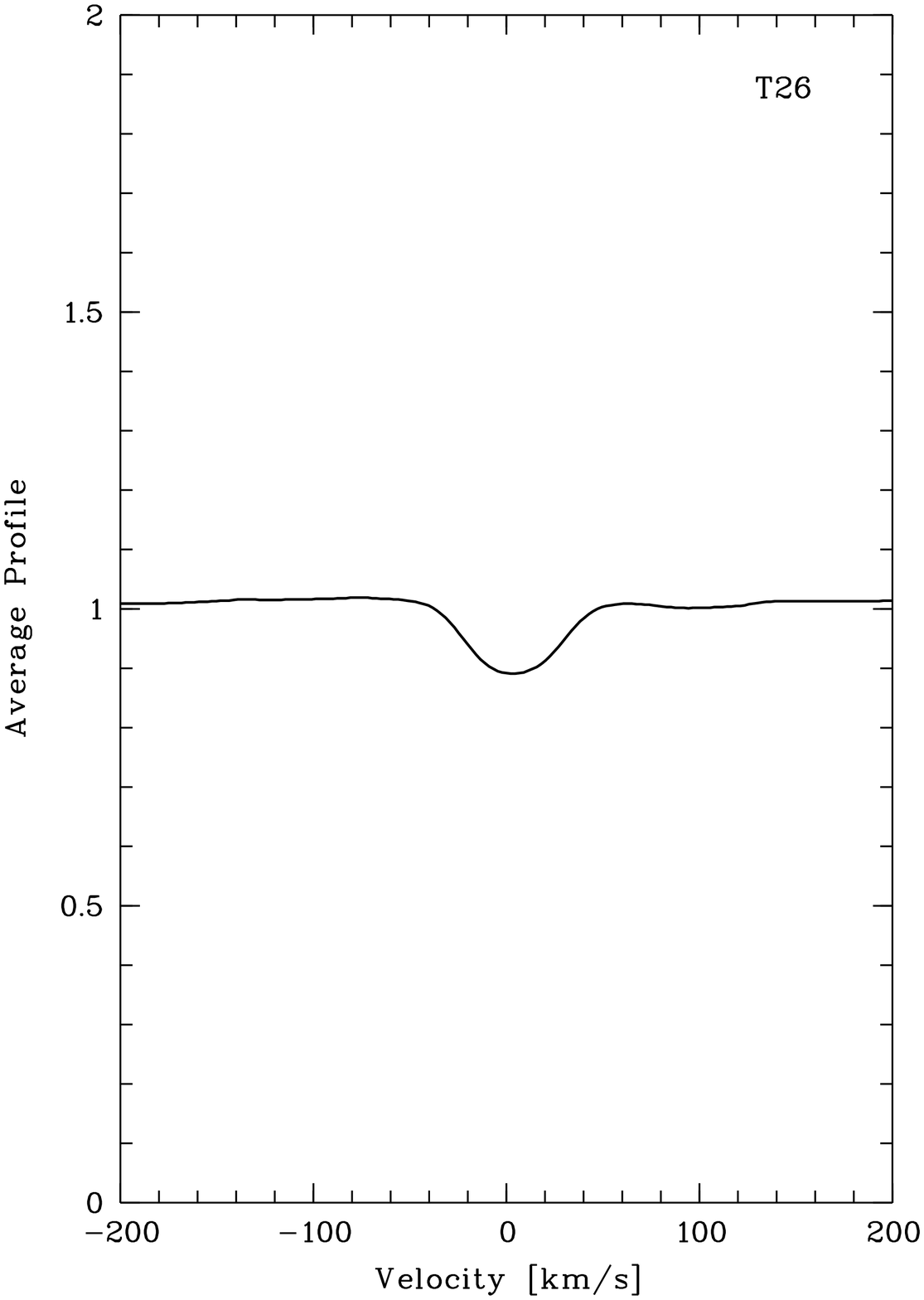} \\                                
\end{tabular}
\caption{Average profiles for 5 objects with He\,I emission and one without. }
\label{fig:He_Profiles}
\end{figure}

\indent In the lower left panel of Fig. \ref{fig:10EW}, we have marked the objects with Ca\,II or He\,I emission with squares and triangles respectively. It clearly shows that these emission lines are found along with strong and variable H$\alpha$ emission. 
Calcium absorption increases in earlier types, correspondingly, the two earliest type stars in our sample, T33A and T26, show Ca\,II in absorption. (For four epochs in the first observation period T33A does show weak Ca\,II emission). Individual epoch Ca\,II EW measurements for the accretors are given in Table \ref{tab:All_obs_Ca_values}.  Fig. \ref{fig:Ca_Profiles} shows the average Ca\,II profiles for all 10 accretors, Fig. \ref{fig:He_Profiles} shows the average He\,I profiles for the five objects that show He\,I emission and one which does not.

\section{Origin of Emission}\label{sec:OriginEmission}
\subsection{Accretion}\label{sec:subsection_accretion}
Based on the emission lines in the spectra, our sample shows a clear division into two groups (see Sect.\ref{section:emissionlines}). In this section we argue that 10 objects in our sample are accreting from a disc for the following reasons: (i) Strength and width of H$\alpha$ emission lines (ii) Shape of H$\alpha$ profiles (iii) Presence of Calcium and Helium and (iv) Presence of circumstellar discs in the sample. 

\begin{figure}
    \includegraphics[scale=0.33,angle=270]{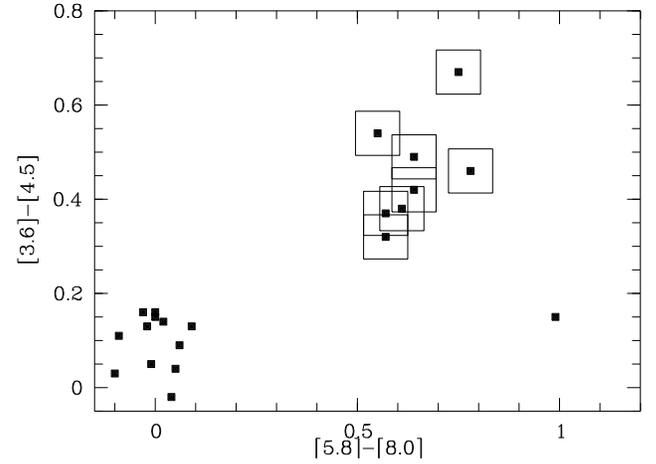}
 \caption{Colour plot for the sample based on Spitzer photometry from \citet{2008ApJ...675.1375L}. Here the accreting objects are indicated with a box. Four objects are not shown as not all of the 4 IRAC magnitudes were available for them (T26, T33A, ESOHa560 and CHXR21).}
\label{fig:Spitzer}
\end{figure}

\begin{enumerate}
 \item H$\alpha$ 10\% versus EW: Eight objects show H$\alpha$ emission with average EW $>20$\,\AA~and 10\% width $>340$\,km\,s$^{-1}$ (see Fig. \ref{fig:10EW}) (Two further objects, B43 and ChaH$\alpha$2, which we classify as accretors have strong H$\alpha$ emission with 10\% width $>340$\,km\,s$^{-1}$, but do not have a continuum for measurement of the H$\alpha$ EW). In the case of accretion we would expect the line velocities to correspond to the typical velocities in an accretion flow. Taking the mass range in our sample to be around~0.1$M_{\odot}$-1$M_{\odot}$, free fall velocities should be in the region of $\thicksim$160\,-\,300\,km\,s$^{-1}$, which would correspond to a range in 10\% widths of $\thicksim$320\,-\,600\,km\,s$^{-1}$. The apparent free-fall velocities will be smaller than this range in most cases due to the inclination of the stellar system with respect to us. For example taking an inclination of 45$\degree$ this 10\% width range becomes $\sim$ 225\,-\,425 km\,s$^{-1}$. The line widths for the 10 accretors are consistent with these ranges (see Fig. \ref{fig:10EW} and Table \ref{tab:Objects}). In contrast, the rotational velocities of M-type stars in Cha-I have on average $v$sin$i$ $\thicksim$11\,km\,s$^{-1}$ \citep{2009ApJ...695.1648N}, which cannot account for the broadening that we see.

 \item Shape of the H$\alpha$ profiles: The accretors' H$\alpha$ profiles are almost all asymmetric, often with more than one peak and an overlying absorption feature (see Fig. \ref{fig:Accretion_Profiles}). Therefore the emission is likely to originate from a structured flow. One possible explanation of the asymmetry is absorption by stellar wind or material in the inner disc \citep{1994AJ....108.1056E,1994ApJ...426..669H}.

 \item H$\alpha$ Variability: The accretors show distinctively different variations across the emission line compared to the non-accretors. The variance profiles seen in Fig. \ref{fig:Accretion_Profiles} show multiple peaks of variations across the emission line and are much stronger than those of the non-accretors (see Fig. \ref{fig:Non-Accretor_Prof}).

\item Presence of Calcium and Helium emission: Strong emission of these lines is thought to originate in the infall region of a magnetosphere \citep{1998AJ....116..455M,2003ApJ...592..282J}. As seen in the lower right panel of Fig. \ref{fig:10EW} and from Table \ref{tab:HaHeCa}, 8 out of 10 of our accretors show signs of either He\,I or Ca\,II emission. None of the non-accretors show the presence of Ca\,II or He\,I. Both of these emission lines can be produced in the chromospheres of active M-type objects, however it is either at low levels compared with our accretors or in transitory flare events \citep{2006ApJ...648.1206J,1998ASPC..154.1508M,2002AJ....123.3356G}.

\item Discs: The lower right panel of Fig. \ref{fig:10EW} shows a plot of 10\% width versus EW of the H$\alpha$ emission in our sample, where the objects with discs are marked with a diamond. This is based on the SED classification by \cite{2008ApJ...675.1375L} from Spitzer photometry (see Fig. \ref{fig:Spitzer}). All of the objects in our sample are either Class II, which are considered to have inner discs, or Class III objects. All 10 accretors are Class II objects.                                                                                                                                                                               

\end{enumerate}

\indent The above arguments strongly suggest that these 10 objects are actively accreting. This accretion is further discussed in Sect. \ref{sec:accretion_rate} and \ref{sec:discusion}. We define the 15 remaining objects as non-accretors (further discussed below). There is a possibility that some are weakly accreting objects that fall below our detection limits. However, the distinctive difference in H$\alpha$ emission between the accretors and non-accretors is strong evidence that they are two separate populations, with different physical processes determining the H$\alpha$ emission.

There are four objects that do show H$\alpha$ EW of $\sim$ 10\,\AA~and have large H$\alpha$ wings (these are ringed in Fig.\ref{fig:10EW}). Apart from these broad wings, they do not share the other characteristics with the accretors. Most notably, only one object has any evidence of having a disc and none have Ca\,II or He\,I in emission. The H$alpha$ emission profiles are quite different to the accretors, they are centrally peaked and show with no overlying absorption or other structures within the profile. Their H$\alpha$ EWs are also below the cut-offs for accretors \citep{2003AJ....126.2997B}, and they do not show the same amplitude of variations in EW (4.8\,\AA~on average) as we see in the accretors (36\,\AA). This can also be seen by comparing the variance profiles of the non-accretors, Fig.\ref{fig:Non-Accretor_Prof}, with the strength of those of the accretors, Fig. \ref{fig:Accretion_Profiles}. The origin of this emission remains unclear, however a full discussion of these objects is beyond the scope of this paper.

\subsection{Magnetic Activity}

The 15 non-accretors in the sample all show an average H$\alpha$ EW of $\lesssim$ 13\,\AA, which is consistent with chromospheric emission from young M-type stars caused by magnetic activity \citep{2007ApJ...662.1254S}. They mostly show narrow, symmetric H$\alpha$ profiles (see Fig. \ref{fig:Non-Accretor_Prof}), which is expected for chromospheric emission.

\section{Accretion Rate Estimates}\label{sec:accretion_rate}

The Ca\,II EW (at 8662.1\,\AA), H$\alpha$ 10\% width and the H$\alpha$ EW were used to estimate mass accretion rates. The He\,I emission was not further examined due to the fact that the line was very weak and noisy or else not observed at all. 

\begin{enumerate}
 \item Ca\,II: To derive $\dot{M}$ we follow the method given by \cite{2005ApJ...626..498M}. Firstly, we determine
the underlying continuum flux around the Ca\,II emission line from the AMES-Dusty synthetic spectra by \cite{2001ApJ...556..357A}, which are available for a range of effective temperatures, gravities and metallicities. The appropriate model spectrum is selected using the effective temperature from \cite{2007ApJS..173..104L}. In each case we used the lowest $\log{g}$ available, which for $\mathrm{T}_{\mathrm{eff}}$ \textless~3100K was $\log{g}$ = 3.5, $\mathrm{T}_{\mathrm{eff}}$ \textless~4000K was $\log{g}$ = 4 and $\mathrm{T}_{\mathrm{eff}}$ \textgreater~4000K was $\log{g}$ = 5. Considering a small range of $\log{g}$ values will not adversely affect our results as the continuum flux is only negligibly dependent on $\log{g}$, and primarily determined by $\mathrm{T}_{\mathrm{eff}}$ \citep{2005ApJ...626..498M}. The average continuum flux ($\mathrm{F}_{\mathrm{cont}}$) was then computed from the model over the range 8600\,-\,8700\,\AA. 

The Calcium flux $\mathrm{F}_{\mathrm{CaII}}$ (per unit area on the stellar surface [ergs sec$^{-1}$cm$^{-2}$]) was calculated from the EW, using the following equation:
\begin{equation}\label{eq:ca_flux}
 \mathrm{F}_{\mathrm{CaII}} = \mathrm{F}_{\mathrm{cont}}\cdot (1 + r_{\lambda(\mathrm{CaII})} )\cdot \mathrm{EW}_{\mathrm{CaII}}
\end{equation}
where $r_{\lambda(\mathrm{CaII})}$ is a measure of excess continuum emission produced by the accretion shock relative to the photospheric emission, and is given by $r_{\lambda(\mathrm{CaII})} = \frac{ F_{\mathrm{excess}}(\lambda)}{F_{\mathrm{cont}}(\lambda)}$. Following \cite{2005ApJ...626..498M} we assume $r_{\lambda(\mathrm{CaII})} \sim 0$, which is a reasonable approximation for the red part of the spectrum and at low accretion rates. If this is not the case, and $r_{\lambda(\mathrm{CaII})}$ is not close to zero, we will have underestimated the accretion rates. (See Appendix \ref{sec:appendix} for a test of this statement).

Using accretion rates derived from UV veiling measurements and detailed modeling of the H$\alpha$ profiles in a sample with a mass range of $\sim$0.02\,$M_{\sun}$ to $\sim$2\,$M_{\sun}$), \cite{2005ApJ...626..498M} found the following relation between accretion rates and Ca\,II fluxes. 
\begin{equation} \label{eq:CaAccretion}
 log(\dot{M}) = 1.06 \cdot \log(F_{\mathrm{CaII}}) - 15.40
\end{equation}
\noindent where the units of $\dot{M}$ are $M_{\sun} yr^{-1}$. (Note \cite{2008ApJ...681..594H} found a very similar relation $log(\dot{M}) = 1.03 \cdot \log(F_{\mathrm{CaII}}) - 15.2$ ).

\item H$\alpha$ 10\% Width: Using veiling measurements for calibration purposes and H$\alpha$ line modeling from a few different studies with a sample mass range of 0.04\,$M_{\sun}$ to 0.8\,$M_{\sun}$, \cite{2004A&A...424..603N} found that the 10\% width of the H$\alpha$ emission line correlates with the mass accretion rate: 
\begin{equation}
\log(\mathrm{\dot{M}}) = \mathrm{A} + \mathrm{B}\cdot (\mathrm{H}\alpha ~ 10\% ~ \mathrm{width})
\end{equation}
\noindent where $\mathrm{A} = -12.89 \pm 0.03 $ and $\mathrm{B} = 0.0097 \pm0.0007$. Units of $\mathrm{\dot{M}}$ are $M_{\sun} yr^{-1}$. We do not propagate the spread in the coefficients through our calculations, as we are interested in the variations rather than the absolute accretion rates.

\item H$\alpha$ EW: To estimate the accretion rate using the H$\alpha$ EW, we begin by converting the H$\alpha$ EW a flux per unit area on the stellar surface [ergs sec$^{-1}$cm$^{-2}$]. As with Ca\,II and following Eq. \ref{eq:ca_flux}, we use the AMES-Dusty models to compute the continuum flux in the range 6554\,-\,6572\,\AA. Note that the AMES-Dusty models do not go to high enough temperatures to cover the $\mathrm{T}_{\mathrm{eff}}$ of two objects T26 and T33A. In these two cases we used the \cite{1993yCat.6039....0K} atmosphere models to convert the H$\alpha$ EW to a flux. Veiling is not taken into account (see Appendix \ref{sec:appendix} for further details). Using the published stellar luminosities and effective temperatures from \cite{2007ApJS..173..104L} we determine estimates for the stellar radii (R). The H$\alpha$ fluxes are then converted to luminosities by multiplying by $4\pi R^2$. 

By fitting models to accretion continuum emission to objects in the range $0.02 \mathrm{M}_{\sun}\lesssim \mathrm{M} \lesssim 1\mathrm{M}_{\sun}$, \cite{2008ApJ...681..594H} derived a series of relations between accretion luminosity and the luminosity of certain emission lines associated with accretion. They derived the following relation for H$\alpha$: 
\begin{equation}
 \log(L_{\mathrm{acc}}) = \mathrm{A} + \mathrm{B}\cdot \log(L_{\mathrm{H\alpha}})
\end{equation}
\noindent where $\mathrm{A} = 2.0\pm0.4$ and $\mathrm{B} = 1.20\pm0.11$. $L_{\mathrm{H\alpha}},L_{\mathrm{acc}}$ are in units of $L_{\sun}$. We use this equation to determine accretion luminosities but we do not take into account the spread in the coefficients. Assuming that all the gravitational energy from the accretion is converted into luminosity, the accretion luminosity is related to the accretion rate as follows \citep{2008ApJ...681..594H}:
\begin{equation}
 \dot{M} = \left(1 - \frac{R_{*}}{R_{\mathrm{in}}}\right)^{-1} \cdot \frac{L_{\mathrm{acc}}R_{*}}{G M_{*}}
\end{equation}
Here $R_{*}$ is the stellar radius, $R_{\mathrm{in}}$ is the infall radius and $M_{*}$ is the stellar mass. Following \cite{2008ApJ...681..594H}, we approximate the factor $(1 - \frac{R_{*}}{R_{\mathrm{in}}})^{-1} \sim 1.25$, assuming $R_{\mathrm{in}} \sim 5R_{*}$ \citep{1998ApJ...492..323G}.
 
\indent To get an estimate of the stellar mass, we compare the published effective temperatures of our objects \citep{2007ApJS..173..104L} with a 2\,Myr isochrone from the BCAH98 evolutionary models by \citet{baraffe98}. Two objects in our sample have temperatures that are higher than those covered by the Baraffe models (T26 and T33A); for these objects we estimate $M \sim 2.5\,M_{\odot}$ based on the
\cite{1997MmSAI..68..807D} isochrones. The \citet{baraffe98} models have been found to systematically overestimate masses compared to the \cite{1997MmSAI..68..807D} models \citep{2003ApJ...592..266M}. Therefore we have underestimated the H$\alpha$ EW derived accretion rates of T26 and T33A compared to those of the other accretors. This means that although these accretion rates are not fully comparable, the relative spread in accretion rates are. 

\end{enumerate}

\indent The time-average accretion rates and the spread (max - min) in accretion rates for all three diagnostic are given in Tables \ref{tab:accretion_rates} and \ref{tab:accretion_obs1_obs2}. The parameters such as $F_{\mathrm{cont}}$, $M_{*}$~and $R_{*}$~that were used in the derivation of these accretion rates are given in Table \ref{tab:model_parameters}.

\begin{table}
 \caption{Parameters used in the derivation of accretion rates. Fluxes are in units of [ergs sec$^{-1}$cm$^{-2}$\AA$^{-1}$]x$10^{5}$}
 \begin{tabular}{@{}lccccc@{}}
 \hline
Object & F$_{\mathrm{cont}}$($\mathrm{H}\alpha$) & F$_{\mathrm{cont}}$($\mathrm{Ca}$) &  M$_{*}$  & R$_{*}$ \\
       &                                 &                            & [M$_{\sun}$] & [R$_{\sun}$] \\

\hline
ISO143   & 1.369  &   3.600   & 0.175 & 1.02  \\  
T33A     & 58.89$^{*}$ & -   & 2.5   & 1.77 \\  
ChaH$\alpha$2   & -      &  -   & 0.15  & 1.17  \\  
B43      & -      &  5.731   & 0.35  & 1.38  \\  
T45      & 4.383  &  8.073   & 0.75  & 2.29  \\  
ChaH$\alpha$6   & 0.879  &  2.797   & 0.11  & 0.87  \\ 
ISO126   & 4.383  &  8.073   & 0.75  & 1.43  \\ 
T30      & 3.055  &  6.467   & 0.50  & 1.07  \\ 
T26      & 70.16$^{*}$ & -   & 2.5   & 3.39  \\ 
T31      & 10.295 &  11.567  & 1.15  & 3.71  \\ 
\hline  
\multicolumn{4}{|l|}{$^{*}$ Estimates from \cite{1993yCat.6039....0K} models.}\\
\end{tabular}
\label{tab:model_parameters}
\end{table}

 \begin{table*}
   \caption{ Derived time-averaged accretion rates for the sample in Table \ref{tab:model_parameters}. Following each average is the max\,-\,min spread in accretion rates over the 12 epochs of observations. An estimate of measurement errors are give in parentheses. Units: Log($M_{\sun}yr^{-1}$). A 2$\sigma$ upper limit is given for the object with no discernible Ca\,II emission in it's spectrum. No Ca\,II accretion rate estimates are given for the 3 objects in our sample with Ca\,II in absorption. For B43 and ChaH$\alpha$2 no H$\alpha$ EW estimates are given as they have no measurable continuum at that wavelength.  }
   \begin{tabular}{@{}lcccccccccc@{}}                                                                                                                                                                                                                                                                                                                         
   \hline
Object & Log($\dot{\mathrm{M}}$)&  Error & Log($\dot{\mathrm{M}}$) & Error & Log($\dot{\mathrm{M}}$) & Error \\ 
        & Ca EW & & H$\alpha$ EW & & H$\alpha$ 10\% &   \\
\hline
 ISO143      &   -8.97 $\pm$ 0.67 & (0.07) &   -8.82 $\pm$  0.29 & (0.10)  &     -9.06 $\pm$ 0.64 & (0.10)  \\
 T33A        &   -                &        &   -8.09 $\pm$  0.42 & (0.01)  &     -8.68 $\pm$ 1.53 & (0.01)  \\
 ChaH$\alpha$2                    &   -                &        &                  -  &         &     -9.54 $\pm$ 0.68 & (0.12)  \\ 
 B43         &   -8.39 $\pm$ 1.04 & (0.09) &   -                 &         &     -9.02 $\pm$ 0.70 & (0.09)  \\
 T45         &   -7.85 $\pm$ 0.68 & (0.003)&   -7.82 $\pm$  0.34 & (0.01)  &     -7.98 $\pm$ 0.66 & (0.01)  \\
 ChaH$\alpha$6                    &    \textless-9.16$\dagger$  & (0.08) &   -9.37 $\pm$  0.40 & (0.05)  &     -9.49 $\pm$ 1.14 & (0.12) \\
 ISO126      &   -8.39 $\pm$ 0.73 & (0.04) &   -8.65 $\pm$  0.46 & (0.06)  &     -9.14 $\pm$ 0.66 & (0.01)   \\
 T30         &   -8.39 $\pm$ 1.11 & (0.03) &   -9.24 $\pm$  0.59 & (0.04)  &     -7.45 $\pm$ 2.07 & (0.02)  \\
 T26         &   -                &        &   -7.16 $\pm$  0.19 & (0.01)  &     -8.71 $\pm$ 0.88 & (0.01)  \\
 T31         &   -8.20 $\pm$ 0.76 & (0.03) &   -7.06 $\pm$  0.27 & (0.01)  &     -8.32 $\pm$ 2.08 & (0.01)  \\
\hline                                        
Av. Spread &      $\pm$ 0.83      &    &    $\pm$ 0.37      	 &  &	$\pm$ 1.11  \\
                                              
\hline                                        
\multicolumn{2}{|l|}{$\dagger$: 2$\sigma$ upper limit}

 \end{tabular}
 \label{tab:accretion_rates}
 \end{table*}

\begin{table*}
  \caption{Derived average accretion rates shown separately for observation periods 1 and 2. Units: Log($M_{\sun}yr^{-1}$). The 2$\sigma$~upper limits are given for the object with no discernible Ca\,II emission in it's spectrum. For B43 and ChaH$\alpha$2 no H$\alpha$ EW estimates are given as they have no measurable continuum at that wavelength. No Ca\,II accretion rate estimates are given for the 3 objects in our sample with Ca\,II in absorption.}
  \begin{tabular}{@{}lcccccc@{}}
  \hline
&\multicolumn{3}{|c|}{Observation Period 1} & \multicolumn{3}{|c|}{Observation Period 2} \\
& Log($\dot{\mathrm{M}}$) &  Log($\dot{\mathrm{M}}$) &  Log($\dot{\mathrm{M}}$)  &  Log($\dot{\mathrm{M}}$) & Log($\dot{\mathrm{M}}$) & Log($\dot{\mathrm{M}}$)  \\ 
 & Ca EW  & H$\alpha$ EW &  H$\alpha$ 10\% & Ca EW  & H$\alpha$ EW  & H$\alpha$ 10\%   \\

\hline
 ISO143     &    -8.89 $\pm$ 0.58  &   -8.80 $\pm$ 0.18    &   -9.06 $\pm$ 0.50  &  -9.04 $\pm$  0.44    &   -8.96 $\pm$  0.01  &    -9.05 $\pm$  0.64      \\
 T33A       &       -              &   -8.02 $\pm$ 0.25    &   -9.00 $\pm$ 0.52  &  -                    &   -8.20 $\pm$  0.34  &    -8.31 $\pm$  1.10       \\
 ChaH$\alpha$2&     -    	   &   -                   &   -9.65 $\pm$ 0.42  &  -                    &   -                  &    -9.45 $\pm$  0.47       \\
 B43        &    -8.24 $\pm$ 0.64  &                  -    &   -8.91 $\pm$ 0.23  &  -8.60 $\pm$ 0.50     &         -            &    -9.26 $\pm$  0.65        \\
 T45        &    -7.78 $\pm$ 0.68  &   -7.82 $\pm$ 0.34    &  -7.84 $\pm$  0.51  &  -7.88 $\pm$ 0.54     &   -7.83 $\pm$ 0.14   &    -8.11 $\pm$  0.39       \\
 ChaH$\alpha$6&\textless -9.26 $\dagger$  &  -9.36 $\pm$ 0.40 &  -9.44 $\pm$  1.12  &\textless  -9.11 $\dagger$      &   -9.38 $\pm$ 0.24   &    -9.55 $\pm$ 1.05        \\
 ISO126     &    -8.45 $\pm$ 0.64  &   -8.70 $\pm$ 0.30    &  -9.14 $\pm$  0.66  &  -8.35 $\pm$ 0.64     &   -8.61 $\pm$ 0.36   &    -9.13 $\pm$ 0.33        \\
 T30        &    -8.28 $\pm$ 1.10  &   -9.28 $\pm$ 0.59    &  -7.07 $\pm$ 0.94   &  -8.49 $\pm$ 1.10     &   -9.23 $\pm$ 0.31   &    -7.83 $\pm$ 2.01        \\
 T26        &        -             &   -7.13 $\pm$ 0.10    &  -8.81 $\pm$ 0.69   &   -                   &   -7.20 $\pm$  0.11  &    -8.62 $\pm$ 0.64        \\
 T31        &    -8.36 $\pm$ 0.50  &   -7.10 $\pm$ 0.16    &  -8.45 $\pm$ 1.31   &  -8.06 $\pm$ 0.59     &   -7.02 $\pm$ 0.24   &    -8.20 $\pm$ 2.08         \\
\hline
Av. Spread &	$\pm$ 0.69        & $\pm$ 0.29      &   $\pm$ 0.69    &     $\pm$ 0.54        &  $\pm$ 0.22      &   $\pm$ 0.94      \\                       
\hline
$\dagger$: 2$\sigma$ upper limit
                       
\end{tabular}          
\label{tab:accretion_obs1_obs2}
\end{table*}

\section{Discussion}\label{sec:discusion}

\begin{figure}
\begin{tabular}{l}
\includegraphics[width=0.32\textwidth,angle=270]{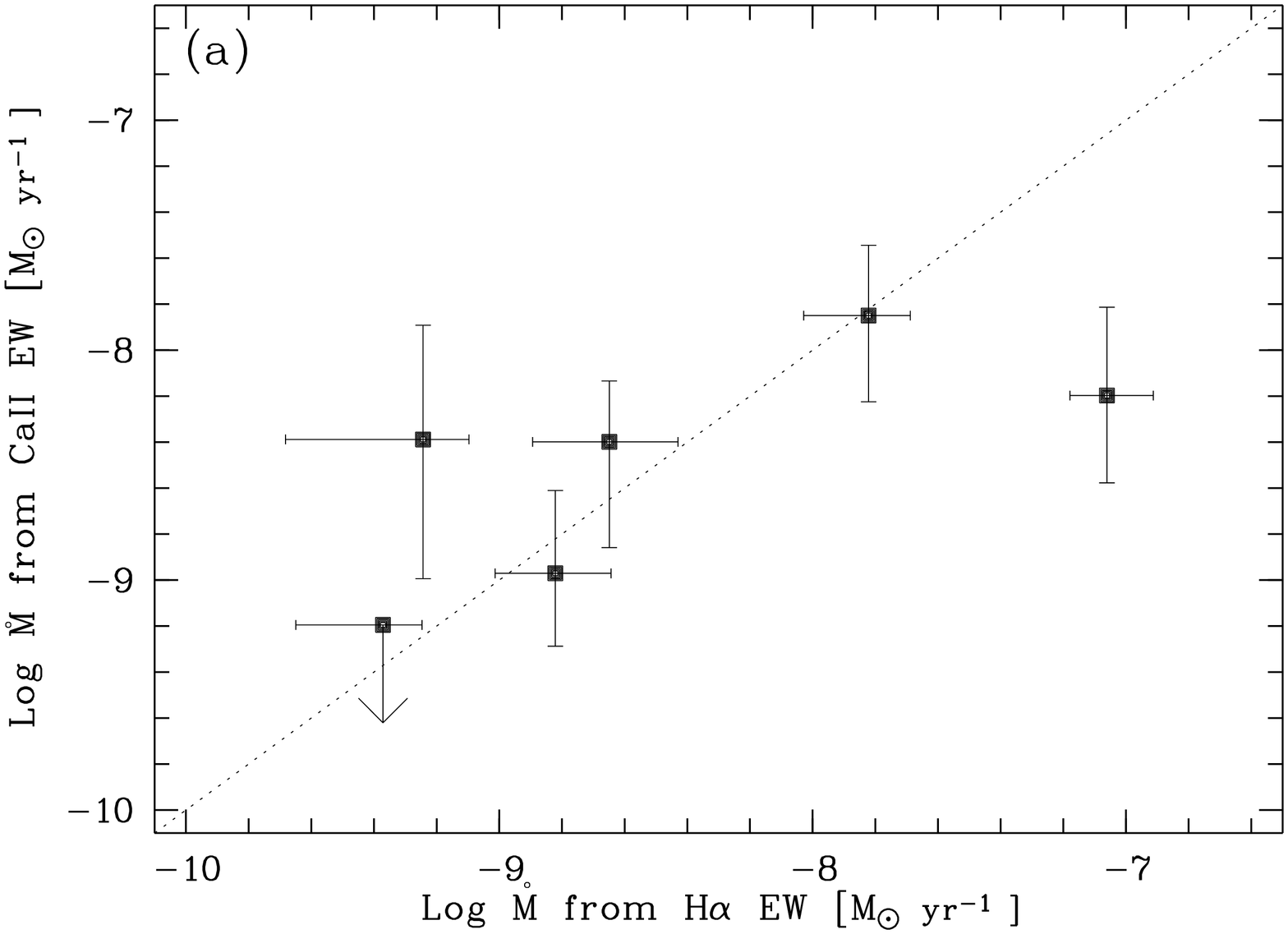}\\
\includegraphics[width=0.32\textwidth,angle=270]{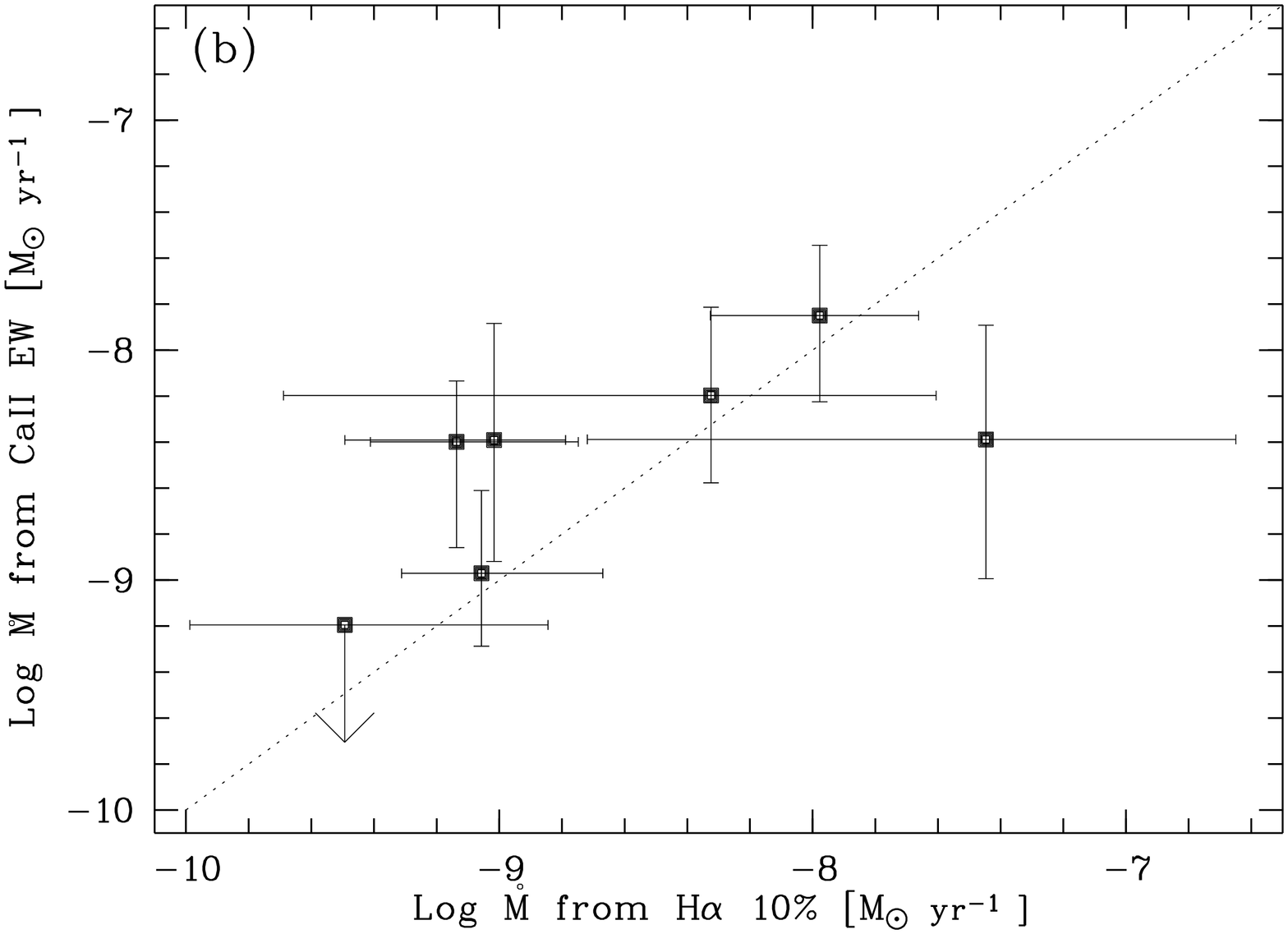}\\
\includegraphics[width=0.32\textwidth,angle=270]{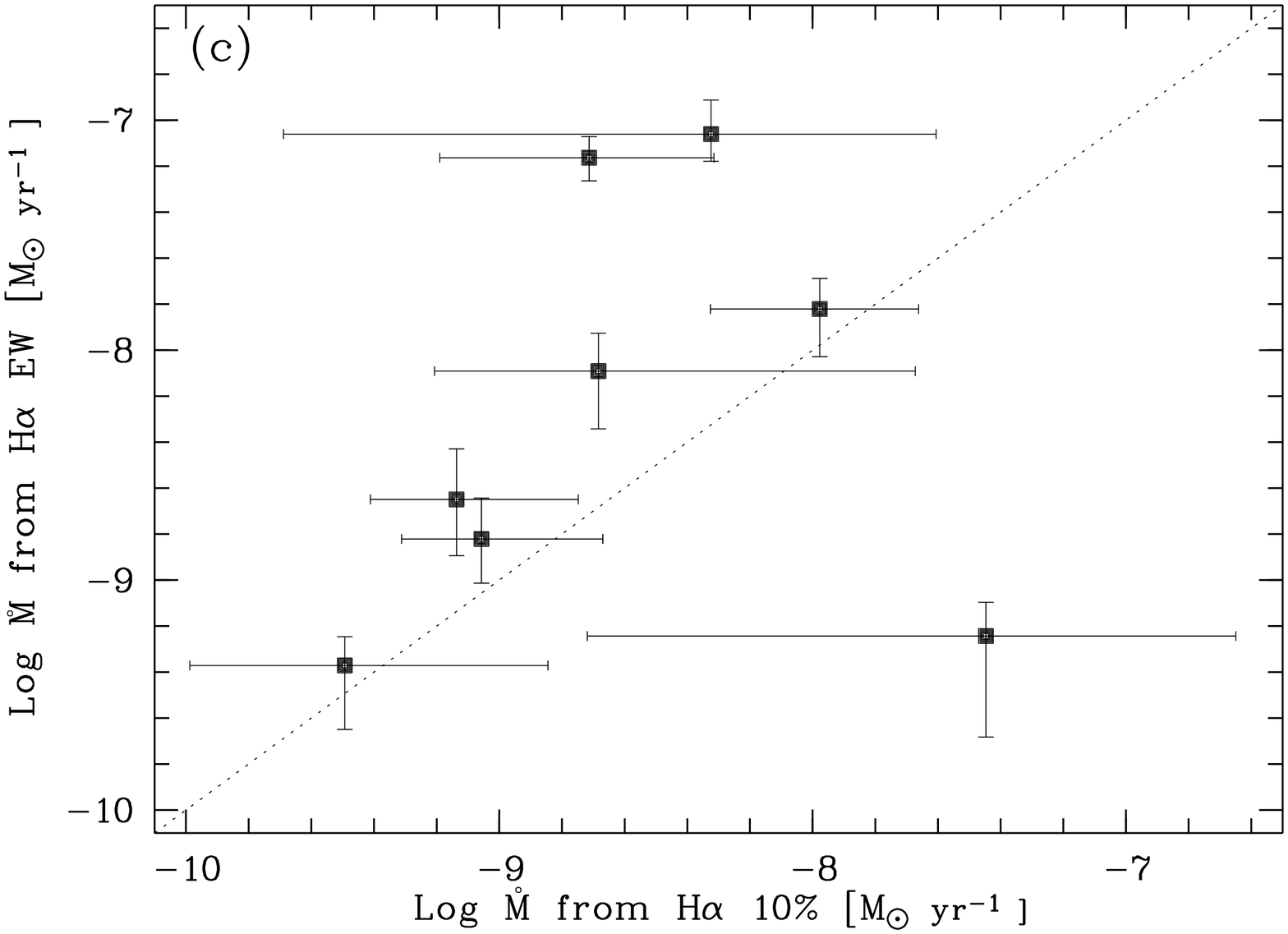}\\
\end{tabular}
\caption{(a) Calculated accretion rates from Ca EW and H$\alpha$ EW. Over-plotted horizontal and vertical bars indicate spread in measurements for all 12 epochs. The object with a downwards pointing arrow is the 2$\sigma$ upper limit for the Ca EW accretion rates as this object has no discernible calcium in its spectrum. (b) Calculated accretion rates from H$\alpha$ EW and H$\alpha$ 10\% width. (c) Calculated accretion rates from Ca EW and H$\alpha$ 10\% width. (see Sect. \ref{sec:accretion_rate}). The dotted line in these three plots indicates the 1-1 relation between the two indicators.}

\label{accretion_rate_plot}
\end{figure}

\subsection{Comparing accretion rate indicators}
In Fig. \ref{accretion_rate_plot} a graphical comparison of the three accretion rate estimates can be seen, with the 1-1 relation indicated by the dotted line. The over-plotted horizontal and vertical bars do not show the errors in measurements, which are substantially smaller (see Table \ref{tab:accretion_rates}. These bars denote the max - min spread in derived accretion rates from the 12 epochs of observations. The data points with arrows mark the 2$\sigma$ upper limit for the object with no discernible calcium in its spectrum. Three further objects had Ca\,II in absorption, and so we have no Ca\,II measurements for these three objects. Taking into account the spread in derived accretion rates of individual objects, there is a reasonable agreement between the three accretion rate estimates.  Some more additional trends can be seen by comparing the average accretion rates and the amplitude of variations.

\indent On average the Ca EW and the H$\alpha$ EW give the lowest variations in accretion rates over the course of our observations (see top panel of Fig. \ref{accretion_rate_plot}). For a given object the Ca EW estimates show greater variations than the H$\alpha$ EW estimates do. This could possibly be explained by the presence of telluric emission lines in the wavelength domain of Ca\,II, which could contaminate the EWs \citep{1996PASP..108..277O}. The telluric lines are drowned out by the Ca\,II emission when it is strong (e.g. in the case of T31 and T45, see Fig. \ref{fig:Ca_Profiles}), but they can clearly be seen in objects with weak Ca\,II emission. These telluric lines will change over the course of our observations, and increase the variability that we see in the Ca EW. This does not account for all of the variability though, as some of our objects with the largest spread in Ca EW (e.g. T45 and T31) show no sign of telluric lines in their spectra. 

The variations in individual object's accretion rates are larger when derived from the H$\alpha$ 10\% width ($\sim$1.11 dex) than they are when derived from the other two signatures. (Note: 1 dex means there is an order of magnitude spread in measurements). These results suggest that the H$\alpha$ 10\% width does not give reliable estimates of average accretion rates, especially with single epoch observations.  

The fact that the amplitude of variations for individual accretion rates derived from H$\alpha$ 10\% width is on average greater than what it is for the other two accretion rate estimates implies there is another contributor to the 10\% width changes. It has been shown that outflows do contribute to the H$\alpha$ emission in the wings in some cases \citep{2009ApJ...691L.106W,2001AJ....122.3335A}. Looking at the average and variance profiles in Fig \ref{fig:Accretion_Profiles}, we can see that much of the variability in the H$\alpha$ line is found in the wings. For the calcium emission lines we see a different behaviour: we found that the Ca line emission varies all across the profile and not just in the wings (see Fig. \ref{fig:Ca_Profiles}).
 
Most of the accretors show blue- or redshifted absorption in their H$\alpha$ profiles. The blue absorption is unlikely to be associated with accretion and more likely to be a consequence of a stellar wind. Over the course of our observations we see these absorption features change in strength and wavelength. This will contribute to the variability in the 10\% width, but it could also reduce the peak height of the H$\alpha$ emission line, which would lead us to overestimate the 10\% width and the accretion rate. Hence, the absorption features could explain the fact that accretion rates derived from the 10\% width are more variable. Furthermore, in a number of profiles there is evidence of high velocity absorption in the wings, which the 10\% width measurements are far more sensitive to than the EW measurements. From these results we consider the Ca EW and H$\alpha$ EW estimates to reflect the accretion rates more accurately. In the following we focus the discussion on these two quantities and do not consider the H$\alpha$ 10\% width estimates.

The variability cannot entirely be due to our measurement errors. Estimate of the measurements errors are given in Tables \ref{tab:Objects} and \ref{tab:HaHeCa}. These have been carried through to the accretion rate estimates and are given in Table \ref{tab:accretion_rates}. In all cases these errors are less than the variations in accretion rates.

\subsection{Spread in $\dot{M} \propto M_{*}$ relation}

The average spread in accretion rates is $\sim$ 0.83 dex for the values derived from the Ca EW and $\sim$ 0.37 dex for the H$\alpha$ EW. We can take this as the maximum spread in accretion rates for young objects ($\sim$ 2 Myr) over a time period of 15 months. This spread is an upper limit as it is likely that emission from the wind and the chromosphere both contribute to these lines.

\indent This spread is relevant in the context of the $\dot{M} - M_{*}$ relation (see Sect. \ref{s1}). As the size of our sample is small, we do not aim to derive such a relation from our dataset. However, it is found that when these two stellar properties are plotted against each other the dominant feature is the large spread in accretion rates for any given mass. Our multi-epoch observations are designed to probe this significant attribute of the $\dot{M} - M_{*}$ relation.

\indent \cite{2005ApJ...626..498M} derived a correlation  $\dot{M} \propto M_{*}^2$ for a sample with a mass range of 0.15\,$M_\odot$ to $\sim$ 2\,$M_\odot$ and age range from $\gtrsim$ 1\,Myr to 10\,Myr. Apart from the correlation, there is a $\pm$1.5 order of magnitude spread in accretion rates at any given mass. \cite{2006A&A...452..245N} found a similar trend for a sample of (0.03\,-\,3$M_\odot $) in Ophiuchus, which has an age of about ($\gtrsim$0.5\,Myr - 1\,Myr). For this single region there is still a spread of $\sim$2 dex at least for any given mass. Our results show that this spread cannot be due to variations of individual objects on the time-scale of about a year. This statement was also checked using the standard deviation of the accretion rates within the sample, which on average for the H$\alpha$ EW derived accretion rates is 0.16, for the H$\alpha$ 10\% width derived accretion rates it is 0.61 and for the Ca\,II EW derived accretion rate it is 0.25. 

\begin{figure}
\begin{tabular}{l}
 \includegraphics[width=0.30\textwidth,angle=270]{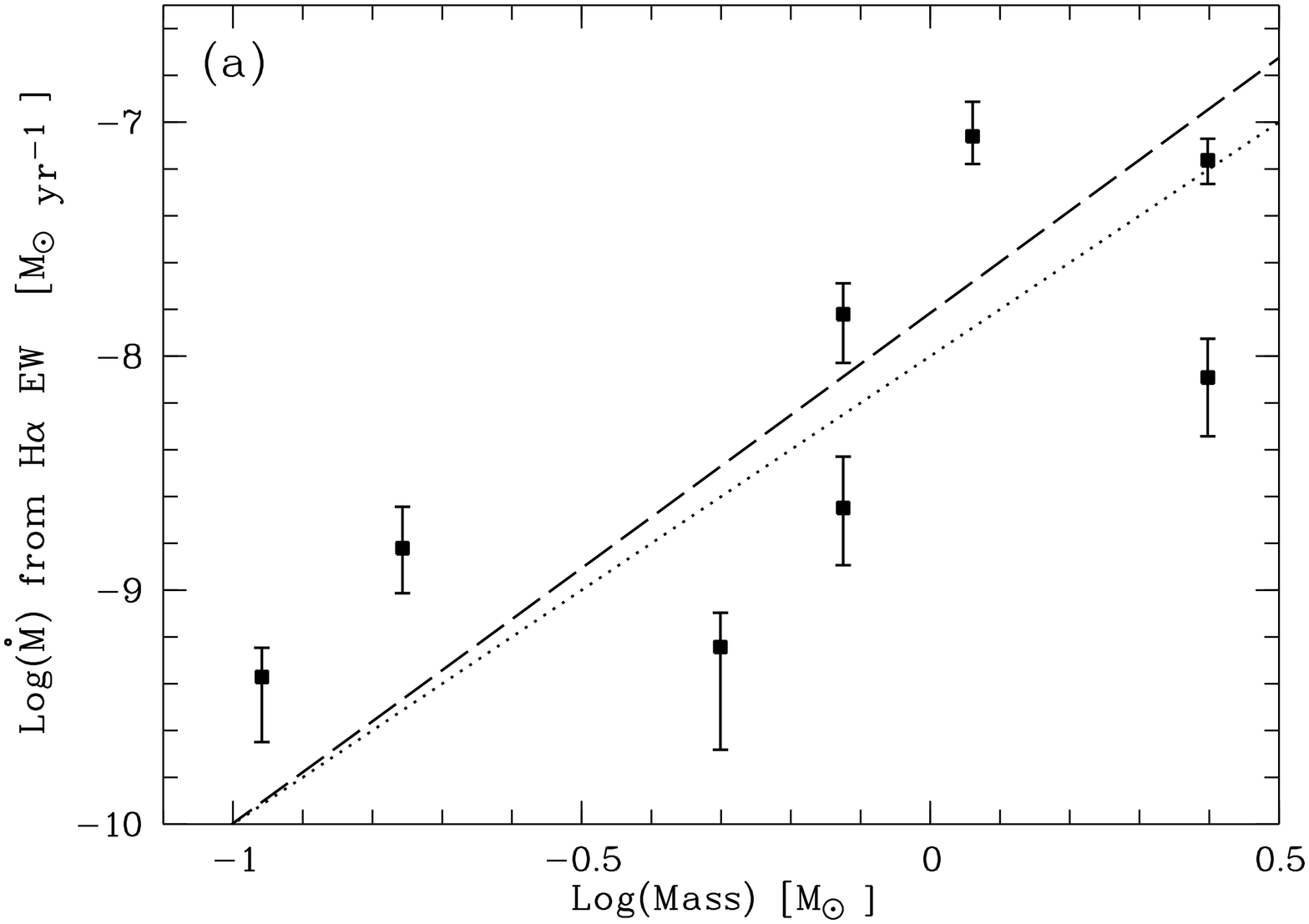}\\
 \includegraphics[width=0.30\textwidth,angle=270]{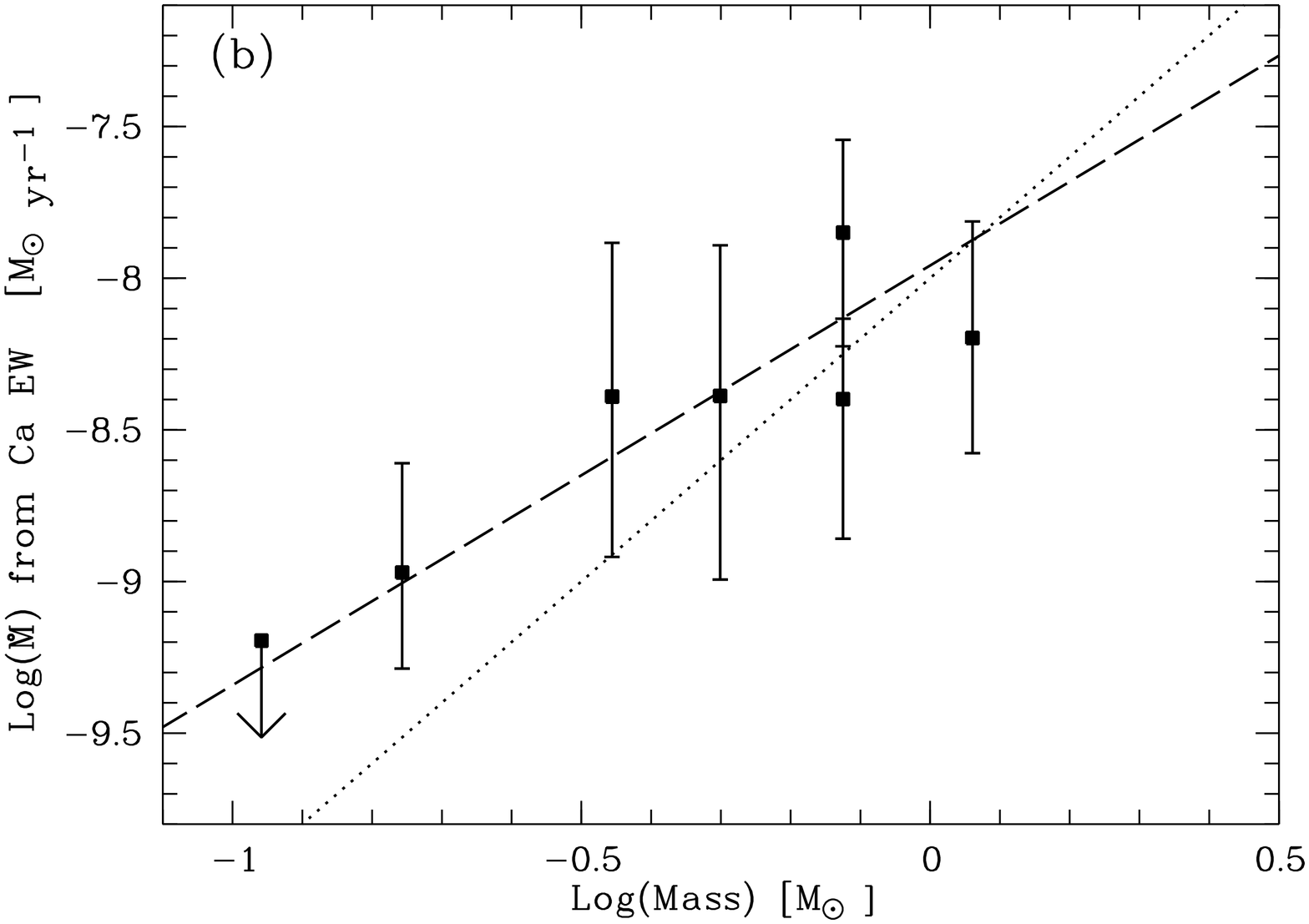}
\end{tabular}
\caption{(a) H$\alpha$ EW derived accretion rates versus object mass. (b) Ca EW derived accretion rates versus object mass. The dotted line represents the $\dot{M}$-$M_{*}^{2}$ relation found by \citet{2005ApJ...626..498M}. The dashed line represents the linear fit derived from our measurements. The over-plotted vertical bars indicate max\,-\,min spread in measurements. Upper limits are indicated by arrows.}
\label{fig:AccretionRate_v_mass} 
\end{figure}

To illustrate this, in Fig. \ref{fig:AccretionRate_v_mass} we have plotted the mean H$\alpha$ EW and Ca EW derived accretion rates versus mass for our sample. The over-plotted vertical bars in this plot indicate the spread in measurements over the 12 epochs of observations. The dotted line on both of these plots is the $\dot{M}$-$M_{*}^{2}$ relation found by \citet{2005ApJ...626..498M}. Also plotted as a dashed line is the linear fit derived from our data set using standard linear regression. The relation derived from our sample is much shallower for the Ca EW derived accretion rates than the H$\alpha$ EW derived accretion rate. Significantly, our plot shows that the spread around the relation is much larger than any spread we see in accretion rates over the observations.

Several alternative explanations for the scatter in $\dot{M}$ have been suggested in the literature. It could be accounted for by evolutionary differences between sources, as $\dot{M}$ is expected to change with time. However as \cite{2004A&A...424..603N} still find large spreads in accretion rates within single star formation regions, it suggests that evolutionary effects are unlikely to explain the full spread. Differences in initial conditions or environmental conditions between objects could also result in a variety of accretion rates \citep{2006ApJ...645L..69D}. For instance, as a result of different initial disc conditions \cite{2006ApJ...639L..83A} have predicted a 2\,dex spread in accretion rates for stars in the mass range 0.02\,-\,0.2 $M_\odot$ and a 2.8\,dex range for masses 0.4\,-\,4.0\,$M_\odot$. Observations such as those by \cite{2011yCat..35259047R} have provided evidence for a mass dependent evolution in the accretion rates, with very low mass stars and brown dwarfs evolving faster, steepening the slope of the $\dot{M} \propto M_{*}$ relation at the low mass end. Furthermore, we cannot rule out that a significant portion of the scatter in the $\dot{M}$\,-\,$M_{*}$ plots could be due to uncertainties in the determination of the stellar masses \citep{2011A&A...534A..32A}, stellar radius, infall radius or from the veiling calibrations themselves.

The exponent in this relation is also poorly defined and there has been a range of values found, from $\dot{M}$\,-\,$M_{*}^{1.1}$ \citep{2011MNRAS.415..103B}, $\dot{M}$\,-\,$M_{*}^{1.8}$ \citep{2006A&A...452..245N} and up to $\dot{M}$\,-\,$M_{*}^{3.1}$ \citep{2009A&A...504..461F}. There are also suggestions that a close to quadratic relation is a result of limited sample size and detectability \citep{2006MNRAS.370L..10C,2011MNRAS.415..103B}, although some relation between mass and accretion rate is expected. However all these relations have a large spread of at least 1 order of magnitude in accretion rates for a given mass.

\subsection{Time-scales of Variations}
 
In the first observation period, the mean amplitude of variations in $\dot{M}$ is $\sim$ 0.69 dex when derived from Ca\,II and $\sim$ 0.29 when derived from H$\alpha$ EW. For the second observation period they are $\sim$ 0.54 and $\sim$ 0.22 dex respectively. The total variations over 15 months (0.83 and 0.37 dex, see above) are not significantly different. The slight increase in the amplitude of variations in the full sample shows there is some variability on the longer time-scales. However, the fact that it does not increase by much means the dominant variations occur in a single observation period i.e. it is shorter than 8 weeks. 

\begin{figure}
\includegraphics[width=0.33\textwidth,angle=270]{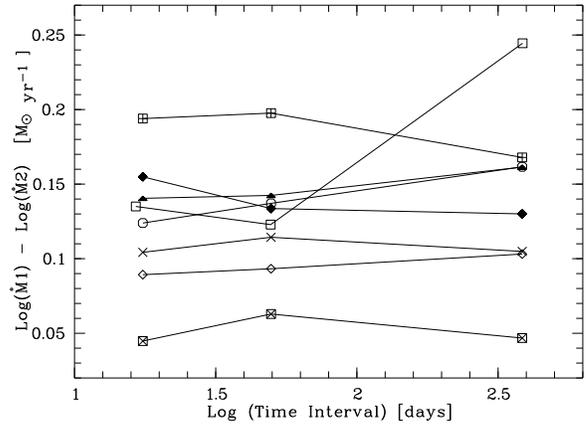}
\caption{Average spread in accretion rate over different time-scales for accretors in the sample, excluding B43 and ChaH$\alpha$2, as these accretion rates  were derived from H$\alpha$ EW. Three different time bins were used (i)8\,d(ays) \textless~ T \textless~25 d (ii) 40\,d \textless~T \textless~90\,d (iii) T \textgreater~1 yr. Seven out of the eight accretors here show very little increase in variations across the three time bins, demonstrating that the variations are dominated by short term variability. }
\label{fig:Variations}
\end{figure}

Fig. \ref{fig:Variations} shows the average difference in accretion rates on different time-scales within our sample. The accretion rates are those derived from the H$\alpha$ EW (B43 and ChaH$\alpha$2 are not included due to the lack of continuum around H$\alpha$) . In order to cover all the possible time-scales in our sample, every accretion rate estimate is compared to every other accretion rate estimate for that object. The differences in accretion rates ($\dot{M}_1$-$\dot{M}_2$) are then placed into three different time bins, depending on the time difference between measurements, (i) T \textless~25 d(ays) (ii) 40\,d \textless~T \textless~90\,d (iii) T \textgreater~1 yr. The mean accretion rate difference in each bin is then plotted for each object. For two objects the largest variations occur on the shortest timescales. Five more objects show a roughly even distribution over the three time bins, and one object shows the largest variations on the longest timescales, T33A. This could be a true increase in accretion rate, or it could also come from the fact that we have incomplete sampling of the variations. Although long term variations are seen in the sample, for all, excluding T33A, the amplitude of variations have reached a maximum or are within 70\% of the maximum after 25 days.

\indent \cite{Nguyen09} observed spectroscopically a number of low-mass pre-main sequence stars over a series of time-scales of days, weeks and months, with the majority of the observations taking place on shorter time-scales than our observations. They found that the amplitude of variations increases on time-scales from hours to several days, after which it saturates. Our observations are able to test if there are any larger variations when the period of observations is extended. Our results suggest that observations on time-scales of a few of weeks are sufficient to limit the majority of the extent of accretion rate variations in typical young stars over the time period of $\thicksim$~1 year.

A previous long term photometric monitoring program also found low levels of variability in a large sample 72 CTTs over the course of $\sim$20\,yrs  \citep{2007A&A...461..183G}. Only a fraction of their sample showed significant changes over time-scale of months to years, suggesting that T Tauri photometric variability is dominated by short term variations.

\indent A major proportion of the variations occur on the time-scale of 8-25 days, which is on the same order as the rotation period of young stellar objects, 1-10 days \citep{2007prpl.conf..297H}. Note that the rotation periods in our sample are unknown. This suggests the origin of the variations is close to the surface of the star, possibly within the co-rotation radius. If rotation is the underlying mechanism that causes the observed accretion variability, we could be tracing structural asymmetries in the accretion flow rather than actual accretion rate changes. These time-scales could also be in agreement with magnetic reconnection events \citep{1998AIPC..431..533G,1996ApJ...468L..37H}, but we do not have the temporal resolution to distinguish what is occurring on these shorter time-scales.

As can be seen in Fig. \ref{fig:10EW} our accreting objects are steadily accreting through all epochs in our observations. We do not see a radical change in the emission lines of the non-accretors over the course of our observations, they always have much weaker H$\alpha$ emission than the accretors. This shows that if they are accreting it is below our detection levels, and never accrete at the same levels as the 10 accretors in the sample. Extreme variations in accretion or episodic accretion is thought to be necessary for YSOs to gain the masses that we observe \citep{2009ApJS..181..321E}. There have now been a number of observations of FU Ori events where an object will suddenly jump from having an accretion rate of 10$^{-7}$\,M$_{\sun}$yr$^{-1}$ to 10$^{-4}$\,M$_{\sun}$yr$^{-1}$ in a very short amount of time, and then decay slowly over the course of 20-100\,yrs \citep{1996ARA&A..34..207H}. However these are rare events, only a handful have been found, and they are expected to occur at the early stages of stellar evolution when the infall is still occurring to the disc. Slightly less dramatic accretion changes occur in EXors type objects, which can show 1\,-\,3 magnitude flare ups every few years for a period of weeks \citep{2007AJ....133.2679H}. 

Highly variable, sporadic accretion rate changes have also been found in older objects such as in the 8\,Myr old $\eta$ Cha cluster \citep{2011MNRAS.411L..51M,2006ApJ...648.1206J}. \cite{2011MNRAS.411L..51M} suggest that such sporadic accretion could occur at `the critical age range of 5\,-\,10 Myr when inner discs are being cleared and giant planet formation takes place'. The T Tauri star, T Chamaeleontis, shows very large changes in H$\alpha$ EW, increasing from 0\,\AA~to 30\,\AA~ and falling again, \cite{2009A&A...501.1013S} infer these variations to be a result of an evolved, clumpy disc. Furthermore \cite{2006ApJ...648.1206J} have also found evidence to suggest that inner disc clearing and/or grain growth might indeed affect two accretors in $\eta$ Cha.

Our sample does not show signs of episodic or sporadic accretion like the cases discussed above. The fact that at 2\,Myr the younger stars in Cha\,I are continuously accreting, supports the results of \cite{2011MNRAS.411L..51M} that the sporadic accretion is likely related to the older age of the population and the evolutionary stage of the disc.

\section{Conclusions}\label{sec:conclusion}

We present in this paper a study designed to investigate the accretion variability of YSOs based on multi-epoch optical spectra of 25 targets in the Chameleon I region. Our observations took place over a total of 15 months. They were split into two separate periods, of 8 weeks each. In each of these periods 6 observations were made, evenly spread across the 8 weeks. Our high resolution spectra from the FLAMES/GIRAFFE spectrograph on the ESO-VLT covered the H$\alpha$~(6562.81\,\AA), He\,I (6678.2\,\AA) and Ca\,II (8662.1\,\AA) emission lines which we used to identify 10 accretors in our sample. We derived the accretion rates of these objects from the H$\alpha$ EW, H$\alpha$ 10\% width and the Ca EW.
 
The accretion rate variability in our sample reached a maximum or was within 70\% of the maximum within $\sim$ 8\,-\,25 days, for 7 out of 8 accretors (based on H$\alpha$ EW derived accretion rate, and hence excluding two accretors B43 and ChaH$\alpha$2). This result complements those from previous studies which covered shorter time-scales \citep{Nguyen09}. This period of variations roughly corresponds to the typical rotation period of these types of objects, suggesting the rotation could be modulating the observed accretion in these systems. Variability on these time-scales could be explained by several scenarios described in Sect. \ref{s1}. One such model is the asymmetric accretion flow hypothesis due to misaligned rotation and field axes or complex field topologies. In this case we would be probing the structural asymmetries in the accretion flow as different parts of the flow come into view. Differential rotation between the disc and the star has been shown to produce magnetic field line expansion and reconnection over the course of a few days \citep{1996ApJ...468L..37H}, and could also account for the variations. Further studies with greater temporal resolution are needed to discriminate between such hypotheses. 

When we considered the amplitude of variations in accretion rates from all three accretion rate estimators, we found that H$\alpha$ 10\% width accretion rate estimates of individual objects show much greater variability than the other two estimators. For this reasons we do not consider the H$\alpha$ 10\% width as a good quantitative accretion rate estimator.

Considering the full observation period we found the average spread in accretion rate to be 0.83\,dex for Ca EW estimates and 0.37\,dex for the H$\alpha$ EW estimates. These variations demonstrate that over $\lesssim$ 15 months, the accretion rate changes are 0.37\,-\,0.83\,dex or less. \cite{2005ApJ...626..498M}, using the Ca\,II (8662\AA) emission line, found the spread for a single mass in the $\dot{M}$-$M_{*}$ relation to be 1-2.5 orders of magnitude. Bearing in mind that our Ca\,II sample is small, our results show that the spread in the $\dot{M}$-$M_{*}$ relation is probably not due to the variability of individual objects on the time-scales of years. They also suggest that this spread is not entirely due to evolutionary differences within a sample, but is more likely due to environment or initial conditions or uncertainties in the derived parameters.

\begin{landscape}
\centering
\begin{table}
\centering
\caption{List of H$\alpha$ EW values for all observations. Here $\Delta$ EW is the Max EW - Min EW measurement. Also given are the Mean EW and RMS values which are calculated over the entire sample. B43 and ChaH$\alpha$2 H$\alpha$ EW measurements are not included here as they have no measurable continuum at that wavelength.}

  \begin{tabular}{@{}lcccccccccccccccccccccc@{}}
 \hline
 Object        & EW1  & EW2    &EW3    & EW4   &EW5     & EW6   &EW7   &EW8    & EW9   & EW10 & EW11  & EW12 &Mean& Max   &  Min  &   $\Delta$EW & RMS    \\
\hline
CHXR28   &   0.4 &   0.7 &  0.6 &   0.5 &   0.7 &   0.4 &   0.6 &   0.6 &  1.2 &   0.4 &   0.5 & 0.6 &   0.6 &   1.2 &  0.4 &  0.8 &   0.2 \\
CHXR20   &   0.1 &    -  & 0.7 &   0.6 &   0.2 &   0.7 &   0.8  &    -  &  0.2 &   0.2 &    -  & 0.5 &   0.3 &   0.7 &  0.1 &  0.7 &   0.3 \\
T22      &   3.1 &   2.2 &  3.1 &   3.4 &   3.3 &   3.2 &   3.2 &   2.7 &   -  &   3.1 &   2.7 & 3.5 &   3.0 &   3.5 &  2.2 &  1.3 &   0.4 \\
ISO143   & 172.2 & 129.2 &122.5 & 163.4 &    -  & 160.6 & 119.6 &    -  &   -  &    -  &    -  & 9.6 & 138.2 & 172.2 & 99.7 & 72.6 &  27.2 \\
T33A     &  35.7 &  22.1 & 26.0 &  30.9 &  30.8 &  34.8 &  17.0 &  30.6 & 21.1 &  19.2 &   3.1 & 6.1 &  25.9 &  35.7 & 16.1 & 19.7 &   9.4 \\
T39A     &   4.5 &   4.8 &  5.1 &   8.6 &   4.7 &   4.4 &   7.0 &   4.7 &  4.8 &   4.7 &   5.3 & 4.9 &   5.3 &   8.6 &  4.4 &  4.2 &   1.3 \\
T45      &  94.2 &  70.3 & 93.7 & 121.7 & 116.6 & 135.1 & 107.6 &  97.9 &101.1 & 124.8 &  94.6 & 5.1 & 104.4 & 135.1 & 70.3 & 64.8 &  17.7 \\
ChaH$\alpha$6   &  97.5 &  83.2 & 74.4 & 103.4 &  47.7 &  93.9 &  78.2 &  94.4 & 63.3 & 100.1 &    -  & 3.1 &  81.7 & 103.4 & 47.7 & 55.7 &  18.1 \\
ISO126   &  50.2 &  81.4 & 82.9 &  89.8 &  60.7 &  70.7 &  76.2 &  79.2 &122.4 &  61.7 & 104.8 & 8.8 &  79.9 & 122.4 & 50.2 & 72.1 &  19.6 \\
ChaH$\alpha$8   &   6.4 &   9.0 &   -  &    -  &   2.7 &   4.6 &    -  &    -  &   -  &    -  &   -   &  -  &   5.6 &   9.0 &  2.7 &  6.3 &   2.7 \\
ChaH$\alpha$5   &  10.2 &    -  & 8.2 &   7.1  &   6.7 &   5.8 &    -  &    -  &   -  &   5.8 &    -  &  -  &   7.3 &  10.2 &  5.8 &  4.5 &   1.7 \\
ESOHa566 &   9.3 &    -  &   - &   7.1  &   6.2 &   7.5 &    -  &    -  &   -  &    -  &    -  &  -  &   7.5 &   9.3 &  6.2 &  3.1 &   1.3 \\
T30      &  79.5 &  52.8 & 43.8 &  56.0 &  79.8 &  26.0 &  53.8 &  41.3 &   -  &  74.7 &  75.3 & 6.2 &  59.0 &  79.8 & 26.0 & 53.9 &  17.7 \\
T34      &   3.0 &   3.8 &  3.1 &   2.0 &   1.8 &   2.6 &   3.0 &   3.1 &   -  &   2.6 &   4.4 & 4.0 &   3.0 &   4.4 &  1.8 &  2.6 &   0.8 \\
ChaH$\alpha$3   &   5.4 &  10.5 &  6.4 &   4.4 &   2.1 &   5.5 &   5.7 &   4.0 & 12.2 &   2.7 &  12.7 & 0.8 &   6.9 &  12.7 &  2.1 & 10.6 &   3.7 \\
T26      &  20.4 &  24.7 & 22.0 &  20.3 &  22.3 &  22.6 &  17.2 &  17.1 & 20.9 &  19.4 &  20.2 & 1.2 &  20.7 &  24.7 & 17.1 &  7.6 &   2.2 \\
ESOHa560 &  11.4 &  14.6 & 13.2 &   9.9 &   7.9 &   8.9 &   6.6 &   8.3 & 11.4 &   9.0 &  11.4 & 0.1 &  10.2 &  14.6 &  6.6 &  8.0 &   2.3 \\
LM04 429 &  13.9 &  17.8 & 14.4 &  13.4 &   6.6 &   9.7 &  12.5 &    -  &   -  &   6.8 &  11.8 & 0.3 &  11.7 &  17.8 &  6.6 & 11.2 &   3.5 \\
CHXR22E  &  11.8 &   9.4 & 11.1 &   8.7 &   4.6 &   9.1 &   5.8 &  11.0 &   -  &   9.2 &  12.7 & 1.4 &   9.5 &  12.7 &  4.6 &  8.1 &   2.5 \\
CHXR21   &  12.3 &   7.3 & 12.2 &  10.0 &   4.8 &   9.2 &   5.7 &  11.9 &   -  &  10.2 &  11.2 & 5.6 &   9.1 &  12.3 &  4.8 &  7.5 &   2.8 \\
T31      &  74.7 &  59.1 & 62.2 &  55.1 &  67.4 &  62.1 &  79.2 &  72.9 & 91.7 &  58.0 &  60.6 & 3.4 &  68.9 &  91.7 & 55.1 & 36.7 &  11.5 \\
CHXR76   &  12.2 &  12.7 & 16.7 &  11.7 &  15.0 &  14.6 &  11.2 &  10.8 & 10.8 &  10.6 &  14.3 & 0.3 &  12.6 &  16.7 & 10.3 &  6.4 &   2.1 \\
CHXR74   &   7.4 &    -  & 10.1 &   9.3 &   7.3 &   8.6 &   10.0&   8.2 &  9.9 &   8.5 &   8.5 & 6.6 &   8.6 &  10.1 &  6.6 &  3.5 &   1.2 \\

\hline	
												
\end{tabular}

\label{tab:All_EW_values}                                  
\end{table}
\end{landscape}

\begin{landscape}
\begin{table}
 \centering
\caption{ List of H$\alpha$ 10\% width values for all observations. Here $\Delta$10\% is the Max - Min 10\% width measurement. Also given are the Mean 10\% width and RMS values which are calculated over the entire sample.}
  \begin{tabular}{@{}lcccccccccccccccccccccc@{}}

  \hline
Object   & \multicolumn{12}{|c|}{10\% width [km s$^{-1}$]}     &  &   &   &  \\
         &  1        & 2    & 3    & 4    & 5    & 6     & 7    & 8     & 9     &  10  &  11   & 12    & Mean  & Max & Min & $\Delta$10\% & RMS \\                                             
         \hline                                                                                                                                                
   CHXR28  &\textless115 &\textless219 &\textless130 &\textless 121 &\textless258 &\textless122 &     336 &     257 &\textless252 &\textless206 &\textless266 &\textless252 &\textless194 &   - &       -&     - &      - \\
    CHXR20 &\textless119 &\textless324 &\textless216 &\textless175 & \textless336 &\textless175 &\textless338 &\textless293 &\textless221 &\textless219 &\textless298 &\textless339 &\textless255       &   - &      - &       - &    - \\
       T22 &     393 &\textless492 &\textless370 &\textless362 &     384 & \textless400 &     407 &     438 &\textless145 &     415 &     416 &     414 &     418$^{*}$&     439 &      384 &     54 &   17 \\
    ISO143 &     381 &     369 &     391 &     387 &     421 &     416 &     409&     395&     368 &     435 &     390 &     374 &     395 &     435 &     368 &      66 &      21 \\
    T33A   &     430 &     391 &     392 &     432 &     379 &     381 &     430&     537&     475 &     424 &\textless380 &     492 &     433$^{*}$ &     537 &     379 &     158 &    50 \\
    T39A   &     174 &     212 &     177 &     261 &     182 &     187 &     213&     188&     193 &     188 &     178 &     204 &     196 &     261 &     174 &      86 &      24 \\
       T45 &     526 &     486 &     521 &     527 &     539 &     519 &     496&     510&     477 &     502 &     470 &     499 &     506 &     539 &     470 &      68 &      21 \\
    ChaH$\alpha$6 &     361 &     300 &     360 &     381 &     313 &     417 &     308&     385&     308 &     407 &     358 &     299 &     350 &     417 &     299 &     117 &      42 \\
    ISO126 &     427 &     403 &     380 &     358 &     379 &     370 &     412&     379&     378 &     385 &     380 &     390 &     386 &     427 &     358 &      68 &      18 \\
    ChaH$\alpha$8 &     203 &     178 &     195 &     192 &\textless221 &     202    &     159 &     186 &      95 &     187 &     161 &     112 &     170$^{*}$&     203 &      95 &     108 &    36 \\
    ChaH$\alpha$5 &     154 &     131 &     195 &     155 &     213 &     154 &     149 &     154 &     163 &     133 &     248 &\textless115 &     168$^{*}$&     248 &     131 &     116 &    36 \\
  ESOHa566 &     164 &\textless133 &     174 &     146 &     124 &     154 &     144 &     198 &-           &     162 &     176 &     140 &     158$^{*}$&     198 &     124 &      74 &      21 \\
       T30 &     612 &     623 &     643 &     610 &     546&     564 &     429 &     509 &     522 &     637 &     534&     499 &     561&     643&     429 &     213 &      65 \\
       T34 &     176 &\textless254 &     188 &     214 &     195 &     200 &     149 &     201 &     154 &     226 &     223&     209 &     194$^{*}$&     226&     149 &      77 &   26 \\
    ChaH$\alpha$3 &     265 &     217 &     267 &     269 &\textless245 &     122 &     218 &   241 &     166 &     207 &     247 &     216 &     222$^{*}$&     269 &     122 &     146 &   45 \\
       T26 &     408 &     452 &     437 &     439 &     381 &     405 &     405 &     451&     412 &     471 &     435 &     465 &     430 &     471 &     381 &      90 &      27 \\
  ESOHa560 &     517 &     645 &     578 &     497 &     516 &     555 &     465 &     478 &\textless535 &     515 &     491 &     533 &     526$^{*}$&     645 &     465 &     179 &      51 \\
  LM04 429 &\textless721 &     701 &\textless635 &\textless633 &\textless463 &\textless726 &\textless786 &\textless731 &\textless465 &\textless574 &\textless564 &\textless497 &\textless618 &     - &      - &    - &    - \\
   CHXR22E &     572 &     393 &     539&     427 &     214 &     469 &     274 &    542 &\textless168 &     410 &     444 &     270 &     412$^{*}$&     572 &     214&     357 &    199 \\
    CHXR21 &     494 &     371 &     485&     420 &     378 &     436 &     284 &    492 &\textless327 &     479 &     443 &     393 &     425$^{*}$&     494 &       211&     494 &      283 \\
       T31 &     478 &     436 &     477&     443 &     523 &     388 &     514 &    528 &     495 &     487 &     330 &    544 &     470 &     544 &     330 &     214 &      62 \\
    CHXR76 &     151 &     148 &     149&     171 &     167 &     156 &     146 &    148 &     125 &     148 &     142 &    184 &     153 &     184 &     125 &      58 &      15 \\
    CHXR74 &     186 &      -  &     155&     185 &     187 &     207 &     162 &    133 &     201 &     150 &     177 &    170 &     174 &     207 &     133 &      74 &      22 \\
       B43 &     403 &     -   &     422&     399 &      -  &     416 &        -&    417 &       - &     386 &     350 &      - &     399 &     422 &     350 &      72 &      25 \\
    ChaH$\alpha$2 &     317 &     339 &     336&     361 &     326 &     324 &     388 &    346 &     339 &     356 &     342 &    357 &     344 &     388 &     317 &      70 &      19 \\

   \hline
 \multicolumn{16}{|c|}{$^{*}$ Indicates the cases where the mean 10\% width is not based on all 12 epochs due to some measurements being upper limits.}     &  &   &   &  \\																	                                                                                                           \\                        
   \end{tabular}      
\label{tab:All_Obs_10width_values} 
   \end{table}
\end{landscape}

\begin{landscape}
\begin{table}                               
\centering
\caption{All Ca\,II EW measurements for the accretors with discernible Ca\,II emission. $\Delta$EW  is the Max - Min measurements. The RMS and the mean are calculated across the entire observation period.}                                   
                                             
 \begin{tabular}{@{}lcccccccccccccccccccc@{}}      
\hline
Object     &  EW1  & EW2     & EW3   & EW4   & EW5     & EW6    &  EW7  &   EW8  &  EW9    &  EW10 &  EW11  &  EW12 & Mean & Max & Min   & $\Delta$EW  & RMS    \\
\hline                                                                                                                            
    ISO143 &  4.0 &   2.6 &  2.0 &   4.2 &   4.8 &  7.1 &   3.6 &  1.6 &   4.6 &  4.1 &  2.0  &  2.3 &    3.6 &   7.1 &  1.6 &   5.5 &  1.6 \\
    T33A+B &  2.3 &   1.7 &  1.1 &   -   &  -    &  1.4 &  -    &      &  -    & -    &  -    &  -   &    1.6 &   2.3 &  1.1 &   1.2 &  1.0 \\
    ChaH$\alpha$2 &   -  &   -   &  -   &   -   &   -   &  -   &   -   &  -   &    -  &  -   &  -    &  -   &    -   &   -   &   -  &   -   &  -   \\
       B43 & 14.4 &   -   & 21.5 &   6.0 &   -   &  5.4 &   -   &  6.5 &   -   &  6.1 &   2.3 &  -   &    8.9 &  21.5 &  2.3 &  19.2 &  6.5 \\
       T45 & 16.3 &   7.3 & 12.1 &  22.3 &  31.9 & 26.9 &  21.7 & 21.6 &  23.9 & 17.2 &  10.2 &  7.3 &   18.2 &  32.0 &  7.3 &  24.6 &  7.8 \\
    ChaH$\alpha$6 &   -  &   -   &  -   &   -   &   -   &  -   &   -   &  -   &    -  &  -   &  -    &  -   &    -   &   -   &   -  &   -   &  -   \\
    ISO126 &  2.7 &   6.5 &  5.1 &   6.4 &   1.8 &  7.3 &   4.9 &  4.3 &   9.9 &  4.9 &   7.0 &  4.6 &    5.5 &   9.9 &  1.8 &   8.0 &  2.3 \\
       T30 & 18.7 &   5.8 &  8.2 &  11.3 &  15.1 &  1.7 &   3.7 &  2.2 &  19.3 & 15.4 &   5.3 &  3.0 &    9.2 &  19.3 &  1.7 &  17.6 &  6.8 \\
       T26 & -    &   -   &  -   &  -    &  -    & -    &   -   &   -  &   -   &  -   &   -   &  -   &    -   &   -   &  -   &   -   &  -   \\
       T31 &  6.9 &   2.7 &  3.1 &   2.4 &   5.7 &  -   &  12.4 & 10.8 &   6.0 &  3.5 &   5.2 & 9.8  &    6.2 &  12.4 &  2.4 &  10.1 &  3.0 \\
\hline
\end{tabular}
\label{tab:All_obs_Ca_values}                                                                                      
\end{table}                                                                                   
\end{landscape}

\appendix
\section{Test for Veiling} \label{sec:appendix}

As mentioned in Sect. \ref{s1}, accretion processes also emit continuum emission. It fills in and `veils' any photospheric absorption lines, making these lines appear shallower. One method of estimating the accretion rate is to compare a veiled spectrum of an accreting star to that of a non-accreting star of similar spectral type and age.

\begin{figure}
 \includegraphics[width=0.35\textwidth,angle=270]{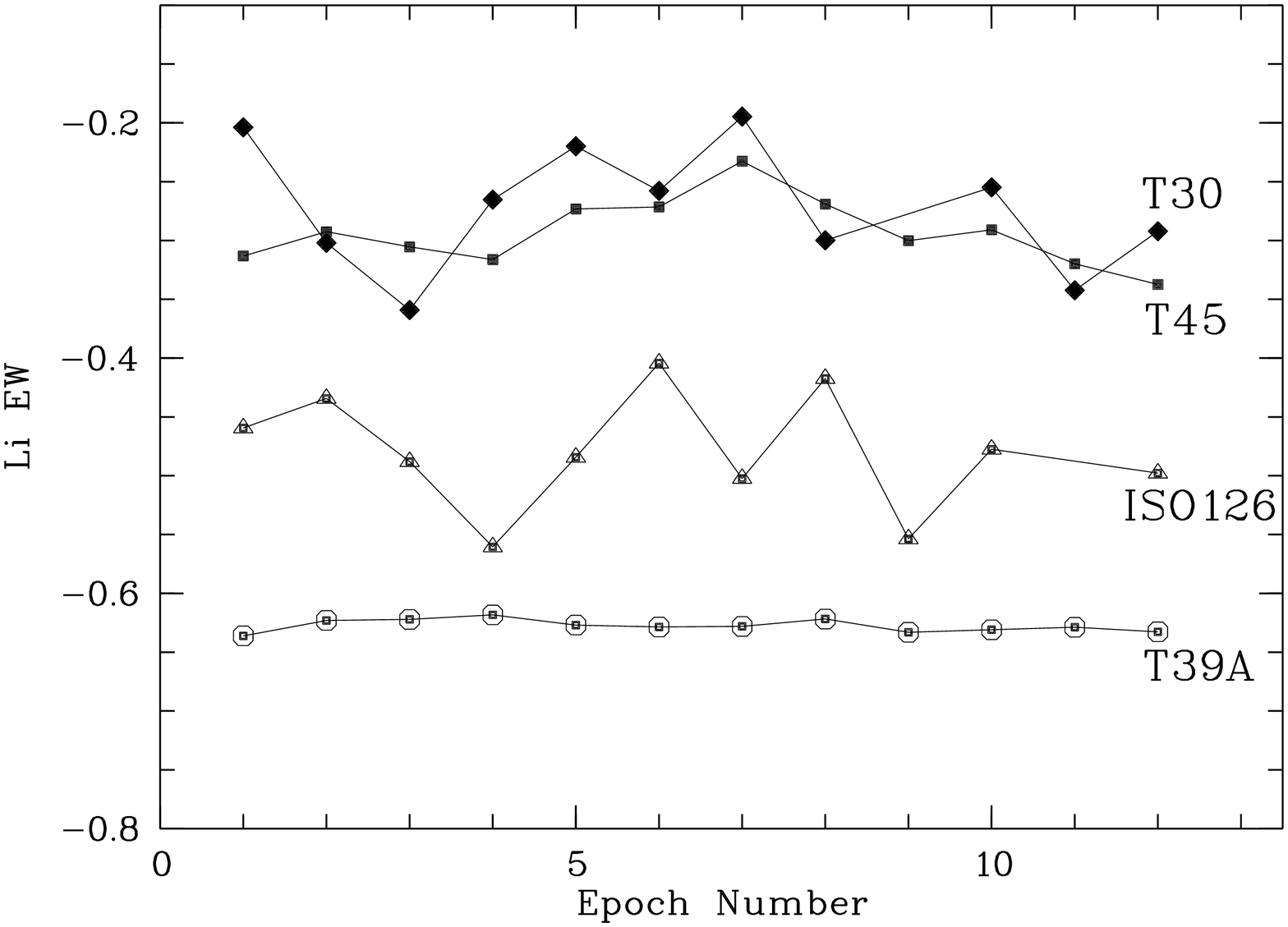}
\caption{Li EW measurements of three accretors and a non-accretor (T39A) used as a reference. Note: The measurements for the fourth accretor and its reference non-accretor are not shown in this plot, as they are of different spectral types.}
\label{fig:Li_EW}
\end{figure}

The strongest photospheric absorption line that lies in our spectral range is Li (6708.1\,\AA). In order to accurately determine the depth of the Li line we needed smooth, distinct and clearly visible absorption profiles. It is also necessary to have a spectral match for each of the accretors in order to estimate the amount of veiling. Due to these restrictions, our analysis was limited to four accretors with distinct Li absorption in their spectra for all 12 epochs, ISO126, T45, T30 with spectral types M1.25, M1.25, M2.5 respectively and T31 with a spectral type of K8. Two non-accretors with suitable spectral types and strong Li absorption were used for comparison, T39A (M2) and CHXR20 (K6). We measured the Li EW in all epochs for these six objects, and the measurement for three of these accretors and their non-accretor spectral match are shown in Fig. \ref{fig:Li_EW}. The non-accretor shows steady deep Li absorption, indicative of photospheric absorption. All three accretors show shallower and more variable Li absorption, which is what we would expect when there is variable veiling present in the continuum. (Note: only 11 epochs are considered in the case of ISO126 due to an increase in noise in the second last epoch.)

In our calculations of the accretion rates we assumed that veiling was negligible. Since the Li (6708.1\,\AA) absorption line is close to the H$\alpha$ (6562.81\,\AA) emission line, we can consider an estimate of the veiling derived at the wavelength of Li to be a good estimate for the veiling around the H$\alpha$ line. We then used this veiling estimate to correct our measurements for the H$\alpha$ EW, as the continuum we had measured is actually veiled continuum and not true continuum. To find the veiling factor we use the equation
\begin{equation}
r = \frac{EW_{P}}{EW_{V}}-1
\end{equation}
where $EW_{P}$ is the photospheric Li EW, and $EW_{V}$ is the veiled EW. (For definition of the veiling factor r see Sect. \ref{sec:accretion_rate}). In this case the $EW_{P}$ is the EW of the Li absorption line in the reference non-accretor, and $EW_{V}$ is the EW of the veiled Li line in the accretor. We then use the same equation with the calculated veiling factor, to derive an unveiled EW for H$\alpha$, and used it to recalculate the accretion rates. This was done for all epoch measurements, and so our new accretion rate estimates also take into account the variations in veiling. These measurements are presented in Table \ref{tab:veiling_accretion}.

\begin{table}
 \centering
\caption{Accretion rate estimates derived from H$\alpha$ EW with and without veiling correction for four accretors, and calculated average veiling factor, r. Units of $\dot{\mathrm{M}}$ are [$\mathrm{M}_{\sun} yr^{-1}$] }
\begin{tabular}{@{}lccccc@{}}
\hline
Object &  Log($\dot{\mathrm{M}}$) Corrected & Log($\dot{\mathrm{M}}$) Uncorrected & r\\
\hline
T45    & -7.42  $\pm$ 0.38 & -7.82 $\pm$ 0.34 & 1.16 \\
ISO126 & -8.53  $\pm$ 0.36 & -8.65 $\pm$ 0.46 & 0.32 \\
T30    & -8.78  $\pm$ 0.71 & -9.25 $\pm$ 0.59 & 1.39 \\
T31    & -6.86  $\pm$ 0.36 & -7.06 $\pm$ 0.27 & 0.47 \\

\hline
\end{tabular}
\label{tab:veiling_accretion}
\end{table}            

There are a number of sources of uncertainty using this method. The veiling factors were derived to give an estimate of how much veiling there is in the spectra and hence allow us to evaluate the affect of veiling on our results. Ideally to reduce the spread and errors in the veiling estimate, a number of lines are used, however since we just have a single one available, our calculations can only provide a rough estimate of the veiling. The abundance of lithium in YSOs decreases with age, so an older star will have shallower Li absorption. Previous studies have found stellar clusters, such as TW Hydrae and $\eta$ Chamaeleontis with ages of $\sim$ 12 Myr, to have a range of Li EW of 0.4 - 0.6\,\AA~\citep{2008ApJ...689.1127M}. However at $\sim$ 2\,Myr, ChaI is a younger region, where an even smaller range in Li EWs can be expected. Similarly, a fast rotator will have a broadened Li absorption line, making it appear shallower, and will lead to an overestimation of the veiling. In our sample all the accretors have narrow Li absorption, so this is unlikely to be a significant factor. 
                                                                                                                                                                                                                           
On average  when we take account of veiling within the continuum, the accretion rate for all four objects increases by 0.30 dex. However more importantly for our results is that the variations in the veiling have little effect on the amplitude of variations in the accretion rates, there is only a 0.09 dex increase on average in the spread. Using this figure we can hypothesize that on average, the spread in all the accretors will increase by the same amount. This will increase the average spread in accretion rates derived from H$\alpha$ EW to 0.46 dex. The same can most likely be said for the accretion rates derived from Ca\,II EW, however veiling tends to be less at longer wavelengths. Therefore, although the veiling present in the continuum may have caused us to underestimate the accretion rates and spread slightly, it does not affect our conclusions. 

\section{Error Estimation} \label{sec:appendix_errors}
In order to get an estimate of our measurement errors we varied both the sky removal and emission line measurement parameters. Specifically we used (i) three different sky continuum estimates, the spatial averaged sky continuum and the average sky continium $\pm$ standard deviation (ii) five different window sizes around the chosen window size given within the text, and (iii) five different window positions around the line centre. The ranges in parameters were chosen to allow for reasonable measurements of the emission line, while still differing from the correct parameters. We varied these three sets of parameters were seperately for each measurement of the H$\alpha$, Ca\,II and He\,I emission lines. 

The mean difference in the correct line measurement and the perturbed line measurement was found for each of the three parameter variations. After these mean differences were added in quadrature, the square root of the sum was taken to get an estimate of the total error in each line measurement (see Table \ref{tab:Objects} and \ref{tab:HaHeCa}).

These total error estimates were then carried through to the accretion rate estimates, giving the error esimtates in Table \ref{tab:accretion_rates}.

\section*{Acknowledgments}
We would like to thank  Antonella Natta, Pat Hartigan and Leonardo Testi for helpful comments and discussions. We also thank the anonymous referee for a constructive review.  Part of this work was funded by the Science Foundation of Ireland by grant no. 11/RFP/AST3331 to AS and grant no. 07/RFP/PHYF790 to TR. 
  
\newcommand\aj{AJ} %Astronomical Journal
\newcommand\actaa{AcA} %Acta Astronomica
\newcommand\araa{ARA\&A} %Annual Review of Astron and Astrophys
\newcommand\apj{ApJ} %Astrophysical Journal
\newcommand\apjl{ApJ} %Astrophysical Journal, Letters
\newcommand\apjs{ApJS} %Astrophysical Journal, Supplement
\newcommand\aap{A\&A} %Astronomy and Astrophysics
\newcommand\aapr{A\&A~Rev.} %Astronomy and Astrophysics Reviews
\newcommand\aaps{A\&AS} %Astronomy and Astrophysics, Supplement
\newcommand\mnras{MNRAS} %Monthly Notices of the RAS
\newcommand\pasa{PASA} %Publications of the Astron. Soc. of Australia
\newcommand\pasp{PASP} %Publications of the ASP
\newcommand\pasj{PASJ} %Publications of the ASJ
\newcommand\solphys{Sol.~Phys.} %Solar Physics
\newcommand\nat{Nature} %Nature
\newcommand\bain{Bulletin of the Astronomical Institutes of the Netherlands}
\newcommand\memsai{Mem. Societa Astronomica Italiana}

\bibliographystyle{mn2e}
\bibliography{lamp}

\label{lastpage}

\end{document}